\documentclass[]{aa}
\usepackage{graphicx,txfonts,amssymb}
\usepackage{natbib}

\renewcommand{\d}{\mathrm{d}}

\authorrunning{Pace et al.}
\titlerunning{Testing the reliability of weak lensing cluster detections}

\begin{document}

\title{Testing the reliability of weak lensing cluster detections}

\author{Francesco Pace\inst{1}, Matteo Maturi\inst{1}, Massimo
  Meneghetti\inst{2}, Matthias Bartelmann\inst{1}, Lauro Moscardini\inst{3,4},
  Klaus Dolag\inst{5}}

\institute{$^1$ ITA, Zentrum f\"ur Astronomie, Universit\"at Heidelberg,
  Albert \"Uberle Str. 2, 69120 Heidelberg, Germany \\ $^2$ INAF-Osservatorio
  Astronomico di Bologna, Via Ranzani 1, 40127 Bologna, Italy \\ $^3$
  Dipartimento di Astronomia, Universit\`a di Bologna, Via Ranzani 1, 40127
  Bologna, Italy \\ $^4$ INFN-National Institute for Nuclear Physics, Sezione
  di Bologna, Viale Berti Pichat 6/2, 40127 Bologna, Italy\\ $^5$
  Max-Planck-Institut f\"ur Astrophysik, Karl-Schwarzschild-Str. 1 85748,
  Garching bei Muenchen, Germany}

\date{\emph{Astronomy \& Astrophysics, submitted}}

\abstract{We study the reliability of dark-matter halo detections with
three different linear filters applied to weak-lensing data. We use
ray-tracing in the multiple lens-plane approximation through a large
cosmological simulation to construct realizations of cosmic lensing by
large-scale structures between redshifts zero and two. We apply the
filters mentioned above to detect peaks in the weak-lensing signal and
compare them with the true population of dark matter halos present in
the simulation. We confirm the stability and performance of a filter
optimised for suppressing the contamination by large-scale structure.
It allows the reliable detection of dark-matter halos with masses above
a few times $10^{13}\,h^{-1}\,M_\odot$ with a fraction of spurious
detections below $\sim10\%$. For sources at redshift two, 50\% of the
halos more massive than $\sim7\times10^{13}\,h^{-1}\,M_\odot$ are
detected, and completeness is reached at
$\sim2\times10^{14}\,h^{-1}\,M_\odot$.}

\maketitle

\section{Introduction}

How reliably can dark-matter halos be detected by means of weak lensing,
and what selection function in terms of mass and redshift can be
expected? This question is important in the context of the analysis of
current and upcoming wide-field weak-lensing surveys. This subject
touches upon a number of scientific questions, in particular as to how
the non-linear growth of sufficiently massive structures proceeds
throughout cosmic history, whether galaxy-cluster detection based on gas
physics agrees with or differs from lensing-based detection, whether
dark-matter concentrations exist which emit substantially less light
than usual or none at all, what cosmological information can be obtained
by counting dark-matter halos, and so forth.

As surveys proceed or approach which cover substantial fractions of the
sky, such as the CFHTLS survey, the upcoming Pan-STARRS surveys, or the
planned surveys with the DUNE or SNAP satellites, automatic searches
for dark-matter halos will routinely be carried out, see for example
\cite{ER00.1} and \cite{ER03.1}. It is important to study what they are
expected to find.

Several different methods for identifying dark-matter halos in weak-lensing
data have been proposed in recent years. They can all be considered as
variants of linear filtering techniques with different kernel
functions. Particular examples are the aperture mass with the radial filter
functions proposed by \cite{SC98.2} and modified by \cite{SC04.2} and
\cite{HE05.1}, and the filter optimised for separating the weak-lensing signal
of dark-matter halos from that of the large-scale structures (LSS) they are
embedded in \citep{MA05.1}.
The non-negligible contamination by the large-scale structure was
already noted by \cite{RE99.1} and \cite{WH02.1}, and \cite{HO01.1} quantified
its impact on weak-lensing mass determinations. \cite{HE05.1} showed that the
redshift of background galaxies can be used to improve the number of reliable
detections. An approach alternative to matched filters is based on
the peak statistics of convergence maps
\citep{JA00.1}, e.g.\ obtained with the Kaiser-Squires inversion technique
\citep{KA93.1,KA95.1} or variants thereof.

In this paper, we evaluate three halo-detection filters in
terms of their performance on simulated large-scale structure data in which
the dark-matter halos are of course known. 
One of the filters is specifically designed to optimally suppress the LSS
contamination \citep{MA05.1}.
This allows us to quantify the completeness of the resulting halo
catalogues, the fraction of spurious detections they contain, and the
halo selection function they achieve. In particular, we compare the
performance of the three filters mentioned in order to test and
compare their reliability under a variety of conditions.

We summarise the required aspects of lensing theory in Sect.~2 and
describe the numerical simulation in Sect.~3. The weak-lensing filters
are discussed in Sect.~4, and results are presented in Sect.~5. We
compare suitably adapted simulation results to the GaBoDS data in
Sect.~6, and present our conclusions in Sect.~7.

\section{Lensing theory}

In this section, we briefly summarise those aspects of gravitational
lensing that are relevant for the present study. For a more detailed
discussion on the theory of gravitational lensing we refer the reader to
the review by \cite{BA01.1}. 

\subsection{Lensing by a single plane of matter}

We start with the deflection of light rays by thin structures in the
universe. This \emph{thin-screen} approximation applies when the
physical size of the gravitational lens is small compared to the
distances between the observer and the lens and between the lens and the
sources. Accordingly, we project the three-dimensional matter
distribution of the lens $\rho(\vec\theta,w)$ on the lens plane and
obtain the surface mass density
\begin{equation}
\label{sigma}
  \Sigma(\vec{\theta})=\int\rho(\vec{\theta},w)\d w,
\end{equation}
where $\vec\theta$ is the angular position vector on the lens plane and
$w$ is the radial coordinate distance along the line-of-sight.

The {\em critical surface density} is defined as 
\begin{equation}\label{sigma_crit}
  \Sigma_{crit}=\frac{c^2}{4\pi G}\frac{D_{ds}}{D_{d}D_{s}}\;, 
\end{equation}
where $D_{s}$, $D_{d}$ and $D_{ds}$ are the angular-diameter distances
between the observer and the source, between the observer and the lens,
and between the lens and the source, respectively. Rescaling the surface
density with the critical surface density we obtain the {\em
convergence}
\begin{equation}
  \kappa(\vec\theta)=\frac{\Sigma(\vec{\theta})}{\Sigma_{crit}}\;.
\end{equation}

A thin lens is fully described by the {\em lensing potential} $\psi$,
which is related to the convergence through the two-dimensional Poisson
equation
\begin{equation}\label{angpoisson}
  \nabla^2_{\vec{\theta}} \psi=2\kappa\;.
\end{equation}
The deflection angle is the gradient of the lensing potential,
\begin{equation}
  \hat{\vec{\alpha}}=\vec\nabla\psi\;.
\end{equation}

The point $\vec\theta$ on the lens plane is mapped onto the point
$\vec\beta$ on the source plane given by the {\em lens equation},
\begin{equation}\label{lensequation}
  \vec{\beta}=\vec{\theta}-\vec{\alpha}(\vec{\theta}),
\end{equation}
where $\vec\alpha(\vec\theta)=D_{ds}/D_{s}\hat{\vec\alpha}(\vec\theta)$
is the {\em reduced deflection angle}. 

The complex {\em shear}, $\gamma=(\gamma_1+i\gamma_2)$, is also obtained
from the lensing potential. Its components $\gamma_1$ and $\gamma_2$ are
\begin{equation}
  \gamma_1=\frac{1}{2}\left(\frac{\partial^2\psi}{\partial\theta_1^2}-
  \frac{\partial^2\psi}{\partial\theta_2^2}\right)\;,\quad
  \gamma_2=\frac{\partial^2\psi}{\partial\theta_1\partial\theta_2}\;.
\end{equation}

\subsection{Multiple lens-plane theory}
\label{sect:multi}
We now generalise the previous formalism to the case of a continuous
distribution of matter filling the volume between the observer and the
sources. We can divide this volume into a sequence of equidistant
sub-volumes whose depths along the line-of-sight are small compared to
the cosmological distances between the observer and the centres of the
sub-volumes, and between those and the sources. Doing so, the thin screen
approximation remains locally valid. If the light cone is narrow enough,
the matter distribution within each sub-volume can be replaced by a
two-dimensional matter sheet. Thus, light rays are approximated by a
sequence of straight lines between the $N$ planes between the observer
and the sources, where $N$ is the number of sub-volumes into which the
space between observer and sources has been divided. In the following
discussion, we assume that the sources lie on the $(N+1)$-th plane.

The deflection angles on each plane can be computed by spatially
differencing the corresponding lensing potential. Following
\cite{HA01.1}, the gravitational potential of the matter within each
sub-volume is decomposed into a background and a perturbing potential.
The equation relating the lensing potential to the mass distribution
responsible for lensing on the $i$-th plane is very similar to
Eq.~\ref{angpoisson}, but the surface density is substituted by the
density contrast of the projected matter
\begin{equation}\label{delta}
  \delta^{proj}_i=\int\d w\delta_i(\vec\theta,w)=
  \int\d w\left(\frac{\rho_i}{\bar{\rho}}-1\right)\;,
\end{equation}
where $\rho_i$ is the three-dimensional density field in the $i$-th
sub-volume, $\bar{\rho}$ is the mean density of the universe, and
$\delta_i=(\rho_i-\bar\rho)/\bar\rho$ is the density contrast of the
three-dimensional mass distribution in the $i$-th sub-volume. The Poisson
equation (\ref{angpoisson}) is thus
\begin{equation}\label{pospoisson}
  \nabla^2\psi_i(\vec{\theta})=
  \frac{8\pi G\bar{\rho}}{3c^2}\delta^{proj}_i(\vec{\theta})=
  3\Omega\left(\frac{H_0}{c}\right)^2\delta^{proj}_i(\vec{\theta})\;,
\end{equation}
and the deflection angle is
\begin{equation}
  \hat{\vec\alpha}_i=\vec\nabla \psi_i\;.
\end{equation}

A light ray propagating from the source to the observer is thus
deflected on each plane by the amount $\vec\alpha_i(\vec\theta_i)$,
where $\vec\theta_i$ indicates the position where the light ray
intercepts the $i$-th lens plane. The lens equation, relating the
positions of a light ray on the $N$-th and the first planes can be
constructed iteratively by summing the deflections on all intermediate
planes. It becomes
\begin{equation}
  \vec{\theta_N} = \vec{\theta}_1-\sum_{i=1}^{n-1}\,\frac{D_{in}D_s}
      {D_nD_{is}}\vec{\alpha}_i(\vec{\theta}_i),
      \label{defangle}
\end{equation}
which represents the generalisation of Eq.~\ref{lensequation} to the
multi-plane case. In the previous equation, we introduced $D_{in}$ and
$D_{is}$ to indicate the angular diameter distances between the $i$-th
and the $N$-th planes, and between the $i$-th plane and the sources,
respectively.

The lens mapping between the $i$-th and the first planes is described by
the Jacobian matrix
\begin{eqnarray}
  A_i & \equiv & \frac{\partial{\vec{\theta}_i}}{\partial{\vec{\theta}_1}}\;.
  \label{jacdef}
\end{eqnarray}
The matrix describing the mapping through a sequence of $N$ planes is
obtained by recursion. It is given by
\begin{equation}
  A_N=I-\sum_{i=1}^{n-1}\,
  \frac{D_{in}D_{s}}{D_{n}D_{is}}  U_i A_i\;,
  \label{jacobian}
\end{equation}
where
 \begin{eqnarray}
  U_i & \equiv & \frac{\partial{\vec{\alpha}_i}}{\partial{\vec{\theta}_i}}\;.
\end{eqnarray}

On the source plane, we define the Jacobian matrix $A_{N+1}$ by
\begin{equation}
  A_{N+1} = \left(
  \begin{array}{cc}
    1-\kappa-\gamma_1 & -\gamma_2+\omega \\
    -\gamma_2-\omega & 1-\kappa+\gamma_1
  \end{array}
  \right).
  \label{sourcejacobian}
\end{equation}
This is not necessarily symmetric since it is the product of two
symmetric matrices. The asymmetry is given by the rotation term
$\omega$. This term only appears in the multiple lens-plane theory
because the Jacobian matrix is symmetric for a single lens plane. The
terms $\kappa$ and $\gamma$ appearing in Eq.~\ref{sourcejacobian} are
now the {\em effective convergence} and the {\em effective shear},
respectively.

\section{The numerical simulation}

\subsection{The cosmological box}
The cosmological simulation used in this study is the result of a
hydrodynamical, $N$-body simulation, carried out with the code {\small
GADGET-2} \citep{SP05.1}. It has been described and used in several
previous studies \citep{MU04.1,RO06.1}. We only briefly summarise
here some of its characteristics. A more detailed discussion can be found in
the paper by \cite{BO04.1}.

The simulation represents a concordance $\Lambda$CDM model, with matter
density parameter $\Omega_m=0.3$ and a contribution from the
cosmological constant $\Omega_{\Lambda}=0.7$. The Hubble parameter is
$h=H_0/100=0.7$ and a baryon density parameter $\Omega_{bar}=0.04$ is
assumed. The normalisation of the power spectrum of the initial density
fluctuations, given in terms of the \emph{rms} density fluctuations
in spheres of $8\,h^{-1}$Mpc, is $\sigma_8=0.8$, in agreement with the
most recent constraints from weak lensing and from the observations of
the Cosmic Microwave Background \citep[e.g.][]{HO06.1,SP06.1}.

The simulated box is a cube with a side length of $192\,h^{-1}$Mpc. It
contains $480^3$ particles of dark matter and an equivalent number of
gas particles. The Plummer-equivalent gravitational softening is set to
$\epsilon_{Pl}=7.5 kpc/h$ comoving between redshifts two and zero, and
chosen fixed in physical units at higher redshift. 

The evolution of the gas component is studied including radiative
cooling, star formation and supernova feedback, assuming zero
metalicity. The treatment of radiative cooling assumes an optically
thin gas composed of $76\%$ hydrogen and $24\%$ of helium by mass, plus
a time-dependent, photoionising uniform UV background given by quasars
reionising the Universe at $z\approx 6$. Star formation is implemented
using the hybrid multiphase model for the interstellar medium introduced
by \cite{SP03.2}, according to which the ISM is parameterised as a
two-phase fluid consisting of cold clouds and hot medium.

The mass resolution is $6.6\times10^9 M_{\odot}/h$ for the cold dark
matter particles, and $8.9\times10^8 M_{\odot}/h$ for the gas particles.
This allows resolving halos of mass $10^{13}\,h^{-1}M_\odot$ with
several thousands of particles.

Several snapshots are obtained from the simulation at scale factors
which are logarithmically equidistant between $a_{ini}=0.1$ and
$a_{fin}=1$. Such snapshots are used to construct light-cones for the
following ray-tracing analysis.

\subsection{Construction of the light-cones}

Aiming at studying light propagation through an inhomogeneous universe,
we construct light-cones by stacking snapshots of our cosmological
simulation at different redshifts. Each snapshot consists of a cubic
volume containing one realization of the matter distribution in the
$\Lambda$CDM model at a given redshift. However, since they are all
obtained from the same initial conditions, these volumes contain the
same cosmic structures in different stages of their evolution. Such
structures are approximately at the same positions in each box. Hence,
if we want to stack snapshots in order to build a light-cone
encompassing the matter distribution of the universe between an initial
and a final redshift, we cannot simply create a sequence of consecutive
snapshots. Instead, they must be randomly rotated and shifted in order
to avoid repetitions of the same cosmic structures along one
line-of-sight. This is achieved by applying transformations to the
coordinates of the particles in each cube. When doing so, we consider
periodic boundary conditions such that a particle exiting the cube on
one side re-enters on the opposite side.

One additional problem in stacking the cubes is caused by the fact that,
as they were written at logarithmically spaced scale factors,
consecutive snapshots overlap with each other by up to two-thirds of
their comoving side-length (at the lowest redshift). Thus, we have to
make sure to count the matter in the overlapping regions only once. For
doing so, we chose to remove particles from the later snapshot. The
choice of the particles to remove from the light cone is not critical,
since snapshots are relatively close in cosmic time. Several tests have
confirmed this expectation.

Hence, the light-cone to a given source redshift $z_s$ is constructed by
filling the space between the observer and the sources with a sequence
of randomly rotated and shifted volumes. As explained in
Sect.~\ref{sect:multi}, if the size of the volumes is small enough, we
can approximate the three-dimensional mass distribution in each volume
by a two-dimensional mass distribution. This is done by projecting the
particle positions on the mid-plane through each volume perpendicular to
the line-of-sight. Such planes will be used as lens planes in the
following ray-tracing simulations.

The opening angle of the light-cone is defined by the angle subtending
the physical side-length of the last plane before the source plane. For
sources at $z_s=1$ and $z_s=2$, this corresponds to opening angles of
$4.9$ and $3.1$ degrees, respectively. In principle, tiling
snapshots at constant cosmic time allows the creation light-cones of
arbitrary opening angles. However, this is not necessary for the
purposes of the present study.

\begin{figure}[!ht]
  \includegraphics[width=\hsize]{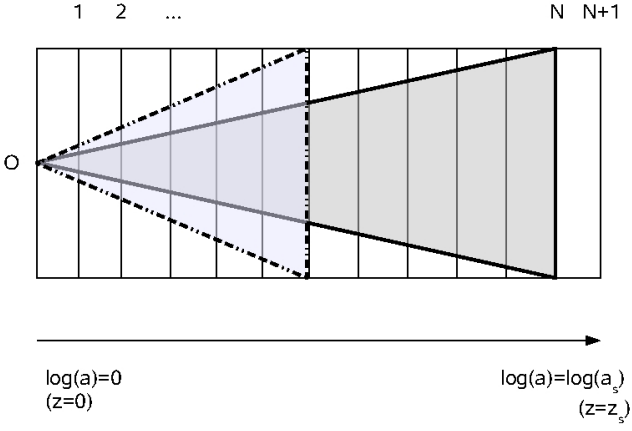}
\caption{Sketch illustrating the construction of the light cones. A
sequence of $N$ lens planes (vertical lines) is used to fill the space
between the observer (O) and the sources on the $(N+1)$-th plane. The
aperture of the light cone depends on the distance to the last lens
plane. At low redshifts, only a small fraction of the lens planes enters
the light-cone (dark-gray shaded region). This fraction increases by
reducing the redshift of the sources, increasing the aperture of the
light cone (light-gray shaded region).}
\label{fig:1}
\end{figure}

If the size of the light-cone is given by the last lens plane,
increasingly smaller fractions of the remaining lens planes will enter
the light cone as it approaches $z=0$ ($a=1$, see Fig.\ref{fig:1}).

\subsection{Halo catalogues}

Each simulation box contains a large number of dark-matter halos. For
our analysis, it is fundamental to know the location of the halos as
well as some of their properties, such as their masses and virial radii.
Thus, we construct a catalog of halos for each snapshot. The procedure
is as follows. We first run a friends-of-friends algorithm to identify
the particles belonging to a same group. The chosen linking length is
$0.15$ times the mean particle separation. Then, within each group of
linked particles, we identify the particle with the smallest value of
the gravitational potential. This is taken to be the centre of the halo.
Finally, we calculate the matter overdensity in spheres around the halo
centre and measure the radius that enclosing an average density equal to
the virial density for the adopted cosmological model, $\rho_{\rm
vir}=\Delta_c(z)\rho_{\rm crit}(z)$, where $\rho_{\rm crit}(z)$ is the
critical density of the universe at redshift $z$, and the overdensity
$\Delta_c(z)$ is calculated as described in \cite{EK01.1}.

We end up with a catalogue containing the positions, the virial masses and
radii, and the redshifts of all halos in each snapshot. The positions are
given in comoving units in the coordinate system of the numerical simulation.
They are rotated and shifted in the same way as the particles during the
construction of the light-cone. The positions of the halos in the cone
are finally projected on the corresponding lens plane.

\begin{figure}[!ht]
  \includegraphics[angle=-90,width=\hsize]{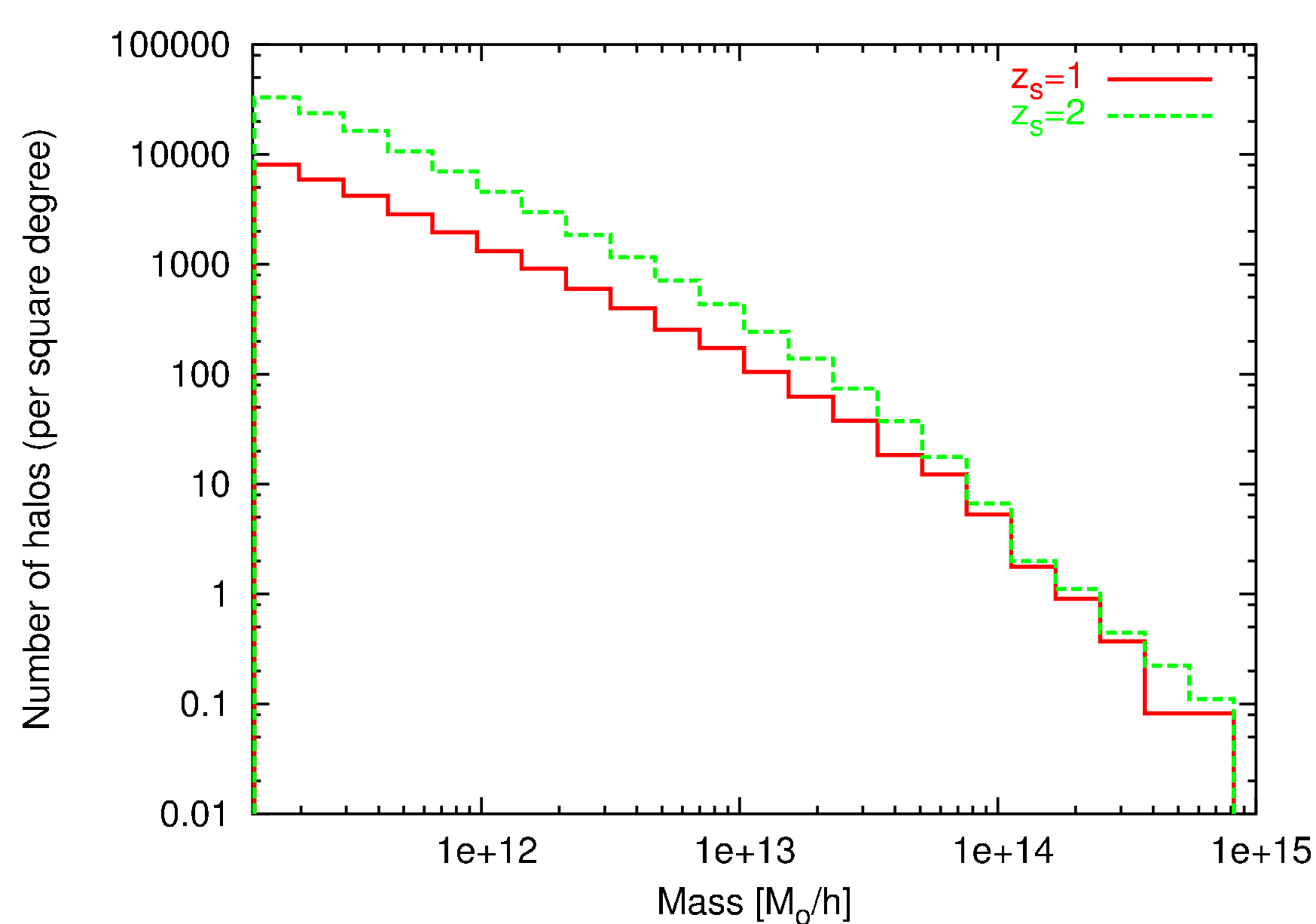}
  \caption{Number of halos per mass bin per square degree. The red and
    green curves show the halo mass distribution for sources at
    $z_\mathrm{s}=1$ and $z_\mathrm{s}=2$, respectively.}
\label{fig:2}
\end{figure}

In Fig.\ref{fig:2}, we show the mass functions of the dark-matter
halos normalised to one square degree and contained in the light-cones
corresponding to $z_\mathrm{s}=1$ (solid line) and $z_\mathrm{s}=2$
(dashed line). Obviously the light-cones contain a large number of
low-mass halos ($\sim 10^{11}-10^{13}\,h^{-1}M_\odot$) which are
expected to be undetectable through weak lensing. On the other hand, a
much lower number of halos with mass $M\gtrsim 10^{13}\,h^{-1}M_\odot$
are potential lenses.
We note that the numbers of haloes with masses larger than $5\times
10^{13}M_\odot$ are approximately equal in both light cones, because such
haloes are mainly contained in the low-redshift portion of the volume which
is common to both light cones.

We ignore the intracluster gas here because it contributes about one order of
magnitude less mass than the dark matter and therefore does not significantly
affect the weak-lensing quantities.

\subsection{Ray-tracing simulations}

The lensing simulations are carried out using standard ray-tracing
techniques. Briefly, starting from the observer, we trace a bundle of
$2048\times2048$ light rays through a regular grid covering the first
lens plane. Then, we follow the light paths towards the sources, taking
the deflections on each lens plane into account.

In order to calculate the deflection angles, we proceed as follows. On
each lens plane, the particle positions are interpolated on regular
grids of $2048\times 2048$ cells using the triangular-shaped-cloud (TSC)
scheme \citep{HO88.1}. This allows to avoid sudden discontinuities in
the lensing mass distributions, that would lead to anomalous deflections
of the light rays \citep{ME00.1,HA01.1}. The resulting projected mass
maps, $M^i_{lm}$, where $l,m=1,...,2048$ and $i=1,...,N$, are then converted
into maps of the projected density contrast,
\begin{equation}
  \delta^{proj,i}_{lm}=\frac{M^i_{lm}}{A_i\bar\rho}-L_i \;,
\end{equation} 
where $A_i$ and $L_i$ are the area of the grid cells on the $i$-th plane
and the depth of the $i$-th volume used to build the light cone,
respectively. 

The lensing potential at each grid point, $\psi^i_{lm}$, is then
calculated using Eq.~\ref{pospoisson}. Owing to the periodic boundary
conditions of the density-contrast maps, this is easily solvable using
fast-Fourier techniques. Indeed, Eq.~\ref{pospoisson} becomes linear in
Fourier space,
\begin{equation}
  \hat\psi(\vec k)=
  -3\Omega\left(\frac{H_0}{c}\right)^2
  \frac{\hat\delta^{proj}(\vec k)}{k^2} \;,
\end{equation}
where $\vec k$ is the wave vector and $\hat\psi$ and $\hat\delta^{proj}$
are the Fourier transforms of the lensing potential and of the projected
density contrast, respectively. Using finite-differencing schemes, we
finally obtain maps of the deflection angles on each plane,
$\alpha^i_{lm}$ \citep{PR98.1}.

The arrival position of each light ray on the source plane is computed
using Eq.~\ref{defangle} which incorporates the deflections on all
preceding $N$ lens planes. However, the ray path intercepts the lens
plane at arbitrary points, while the deflection angles are known on
regular grids. Thus, the deflection angles at the ray position are
calculated by bi-linear interpolation of the deflection angle maps.

Again using finite differencing schemes, we employ Eqs.~\ref{jacdef} to
\ref{sourcejacobian} to obtain maps of the effective convergence and
shear.

\subsection{Testing the ray-tracing code}

We test the reliability of the ray-tracing code by comparing the statistical
properties of several ray-tracing simulations with the theoretical
expectations for a $\Lambda$CDM cosmology. In these tests, we assume that all
source redshifts are $z_s=1.5$. For this source redshift, the light cone spans
a solid angle of roughly $3.6^2\simeq13$ square degrees on the sky. We perform
ray-tracing through 60 different light-cones in total.

\begin{figure}[!ht]
  \includegraphics[height=\hsize,angle=-90]{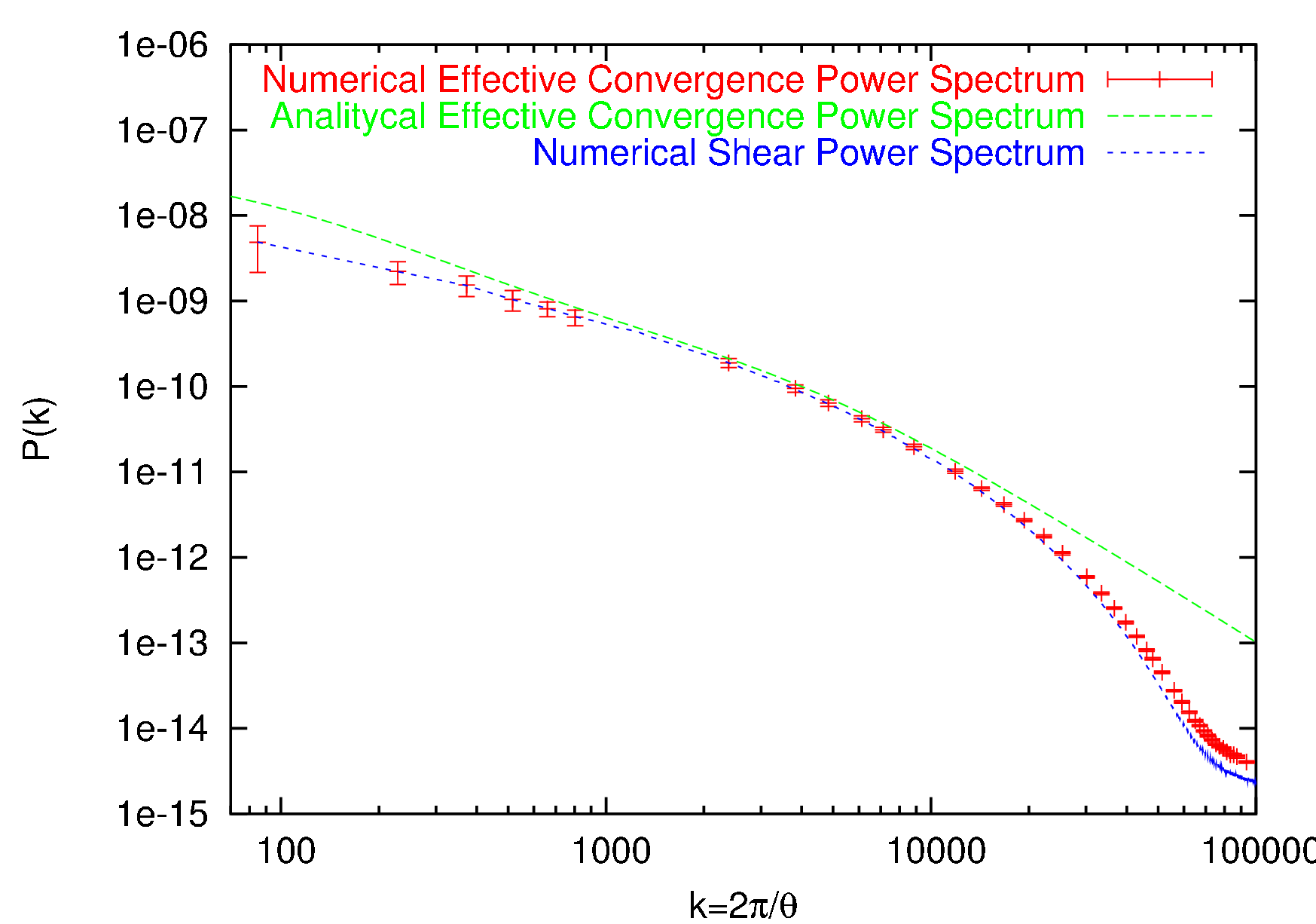}
\caption{Numerical power spectra of the effective convergence (solid
line) and of the shear (dotted line) obtained by averaging over $60$
different light cones corresponding to a solid angle of $\sim 13$ square
degrees. The power spectrum expected for a $\Lambda$CDM model with the
same cosmological parameters as the simulation is given by the dashed
line. The errorbars are shown only for the effective convergence power
spectrum, but are of equivalent size for the shear power spectrum.}
\label{fig:3}
\end{figure}

In Fig.~\ref{fig:3}, we show the power spectra of the effective
convergence and the shear, obtained by averaging over all different
realizations of the light-cone. These are given by the solid and by the
dotted lines, respectively. The theoretically expected power spectrum is
shown as the dashed line. As expected, the convergence and the shear
power spectra are equal. We note that they agree with the theoretical
expectation over a limited range of wave numbers. Indeed, they deviate
from the theoretical power spectrum for $k\lesssim200$ and for $k\gtrsim
20000$. These two values of the wave vector define the reliability range
of these simulations and are both determined by numerical issues. On
angular scales $\gtrsim1^\circ$, we miss power because of the small size
of the simulation box, while on angular scales smaller than $\lesssim1'$
we suffer from resolution problems due to the finite resolution of the
ray and the mass grids.

\subsection{Lensing of distant galaxies}

Using the effective convergence and shear maps obtained from the
ray-tracing simulations, we are now able to apply the lensing distortion
to the images of a population of background sources. 

In the weak-lensing regime, a sufficiently small source with intrinsic
ellipticity $\epsilon_s$ is imaged to have an ellipticity  
\begin{equation}
  \epsilon=\frac{\epsilon_s+g}{1+g^*\epsilon_s}\;,
  \label{eq:epstr}
\end{equation}
where $g$ is the complex reduced shear, and the asterisk denotes complex
conjugation. We adopt here the standard definition of ellipticity,
$\epsilon=\epsilon_1+i\epsilon_2=(a-b)/(a+b)e^{i2\phi}$, were $a$ and
$b$ are the semi-major and the semi-minor axes of an ellipse fitting the
object's surface-brightness distribution. The position angle of the
ellipse's major axis is $\phi$.

The complex reduced shear is defined as
\begin{equation}
  g(\vec{\theta})=\frac{\gamma(\vec{\theta})}{1-\kappa(\vec{\theta})}\;.
\end{equation}
In the weak-lensing regime, $\kappa \ll 1$ and $\gamma \approx g$.
Equation~\ref{eq:epstr} illustrates that the lensing distortion is
determined by the reduced shear. Since the background galaxies are
expected to be randomly oriented, the expectation value of the intrinsic
source ellipticity $\epsilon_s$ is assumed to vanish. Thus, the
ellipticity induced by lensing can be measured by averaging over a
sufficiently large number of galaxies and the expectation value of the
ellipticity is equal to the reduced shear, $\langle\epsilon\rangle=g$.

In order to generate a mock catalogue of lensed sources, galaxies are
randomly placed and oriented on the source plane. Their intrinsic
ellipticities are drawn from the distribution
\begin{equation}
  p(|\epsilon_s|)=\frac{\exp{|(1-|\epsilon_s|^2)/\sigma_{\epsilon_s}^2|}}
  {\pi\sigma_{\epsilon_s}^2|\exp{(1/\sigma_{\epsilon_s}^2)}-1|},
\end{equation}
where $\sigma_{\epsilon_s}=0.3$. We assume a background galaxy number
density of $n_g=30$ arcmin$^{-2}$. Observed ellipticities are obtained
from Eq.~\ref{eq:epstr} by interpolating the effective convergence and
shear at the galaxy positions. This procedure results in catalogues of
lensed galaxies for each source redshift chosen, where galaxy positions
and ellipticities are stored.

\section{Weak lensing estimators}

We investigate the performances of three weak-lensing estimators which
have been used so far for detecting dark-matter halos through weak
lensing. These are the classical aperture mass \citep{SC96.2,SC98.2}, an
optimised version of it \citep{SC04.2}, and the recently developed,
optimal weak-lensing halo filter \citep{MA05.1}. More details on these
three estimators are given below.

All of them measure the amplitude of the lensing signal $A$ within
circular apertures of size $\bar{\theta}$ around a centre $\vec{\theta}$.
Generalisations are possible to apertures of different shapes. In
general, $A$ is expressed by a weighted integral of the tangential
component of the shear relative to the point $\vec{\theta}$, $\gamma_t$.
The weight is provided by a filter function $\Psi$, such that
\begin{equation}
  A(\vec\theta,\bar\theta)=\int\d^2\theta'\gamma_t(\vec{\theta'},\vec{\theta})
  \Psi(|\vec{\theta'}-\vec{\theta}|)\;,
  \label{eq:wlest}
\end{equation}
and the integral extends over the chosen aperture. The variance of the
weak-lensing estimator is given by
\begin{equation}
  \sigma^2=\frac{1}{(2\pi)^2}\int|\hat\Psi(\vec{k})|^2P_N(k)\d^2k,
  \label{sigma_am}
\end{equation}
where $\hat\Psi(\vec{k})$ is the Fourier transform of the filter, and
$P_N(k)$ the power spectrum of the noise.

\subsection{Aperture Mass}\label{sec:apm}

The aperture mass was originally proposed by \citep{SC96.2} for
measuring the projected mass of dark-matter concentrations via weak
lensing. It represents a weighted integral of the convergence,
\begin{eqnarray}
  M_{APT}(\vec\theta) & = & \int\d^2\vec{\theta'}\kappa(\vec\theta')
  U(|\vec\theta'-\vec\theta|)\;.
\end{eqnarray}
The weight function $U(\theta)$ is symmetric if the aperture is chosen
to be circular, and it is compensated, i.e.
\begin{equation}
  \int_0^{\theta}\d\theta'\theta'U(\theta')=0\;.
\end{equation}

Since the convergence is not an observable, the aperture mass is more
conveniently written as a weighted integral of the tangential shear,
\begin{eqnarray}
  M_{APT}(\theta) & = & \int
  \d^2\vec(\theta')\gamma_t(\vec\theta',\vec\theta)
  \Psi_{APT}(|\vec\theta'-\vec\theta|)\;,
\end{eqnarray}
where the function $\Psi_{APT}$ is related to the filter function $U$ by
the equation
\begin{equation}
  \Psi(\theta) =
  \frac{2}{\theta^2}\int_0^{\theta}\d\theta'\theta'U(\theta')-U(\theta) \;.
\end{equation}

The variance $\sigma^2_{M_{APT}}$ of $M_{APT}$ is defined as
\begin{equation}
  \sigma^2_{M_{APT}} =
  \frac{\pi\sigma^2_{\epsilon_s}}{n_g}\int_0^{\theta}\d\theta'\theta'
  \Psi^2_{APT}(|\vec\theta'-\vec\theta|)\;,
\end{equation}
which takes into account the shot noise due to the finite number and the
intrinsic ellipticities of the sources \citep{BA01.1}.

The shape of the filter function $\Psi$ is usually chosen to have a
compact support and to suppress the halo centre because the lensing
measurements are more problematic there. Indeed, the weak-lensing
approximation may break down and the cluster galaxies may prevent the
ellipticity of background galaxies to be accurately measured.

\cite{SC98.2} propose the polynomial function 
\begin{equation}\label{PSIAPT}
  \Psi_{APT}=\frac{(1+l)(2+l)}{\pi\theta^2_{max}}x^2(1-x^2)^lH(1-x)\;,
\end{equation}
where $H(x)$ is the Heaviside step function, and $x=\theta/\theta_{max}$
is the radial angular coordinate in units of the radius, $\theta_{max}$,
where $\Psi_{APT}$ vanishes. $l$ is a free parameter which is usually
set to $l=1$. Note that this filter function was designed especially for
measuring cosmic shear. However, several authors have used it for
searches for dark matter halos \citep{ER00.1,SC04.2}.

More recently, other filter functions $\Psi$ have been proposed which
maximise the signal-to-noise ratio $M_{APT}/\sigma_{M_{APT}}$.
\cite{SC98.2} show that this is the case if $Q$ mimics the shear profile
of the lens. For example, \cite{SC04.2} propose a fitting formula that
approximates the shear profile of a Navarro-Frenk-White (NFW) halo
\citep{NA96.1}. Their filter function is
\begin{equation}\label{eq:oapt}
  \Psi_{OAPT}(x)=\frac{1}{1+e^{6-150x}+e^{-47+50x}}\frac{\tanh{x/x_c}}{x/x_c},
\end{equation}
where $x_c$ is a parameter controlling the shape of the filter
\citep[see also][]{PA03.2,HE05.2}. In the rest of the paper, we will
refer to this implementation of the aperture mass as to the
``optimised aperture mass''.

\cite{HE05.1} included the photometric redshifts of background sources,
increasing the halo-detection sensitivity at higher redshifts and for smaller
masses. 
Aiming at a comparison of different filters, we neglect this
additional information here. 
We can therefore not apply their tomographic approach, which is based
on an NFW fitting formula. They also suggested using a Gaussian
profile which found application in actual weak-lensing surveys
\citep[see e.g.][]{MI02.1}, but here we focus on the filter proposed
by \cite{MA05.1} whose shape is statistically and physically well motivated.

\subsection{Optimal Filter}

\cite{MA05.1} have recently proposed a weak-lensing filter optimised for
an unbiased detection of the tangential shear pattern of dark-matter
halos. Unlike the optimised aperture mass, the shape of optimal filter
is determined not only by the shear profile of the lens, but also by the
properties of the noise affecting the weak lensing measurements. 

The measured data $D$ is composed of the signal from the lens $S$ and by
the noise $N$, and can be written as
\begin{equation}
  D(\vec{\theta})=S(\vec\theta)+N(\vec\theta)=A\tau(\vec{\theta})+
  N(\vec{\theta})\;, 
\end{equation}
where $A$ is the total amplitude of the tangential shear and
$\tau(\vec\theta)$ is its angular shape. The noise $N$ comprises several
contributions that can be suitably modeled. 

The optimal filter accounts for the noise contributions because it is
constructed such as to satisfy two conditions. First, it has to be
unbiased, i.e.~the average error on the estimate of the lensing
amplitude,
\begin{equation}
  A_{\rm est}(\vec\theta)=\int\d^2 \theta'
  D(\vec\theta')\Psi(|\vec\theta'-\vec\theta|)
\end{equation}
has to vanish:
\begin{equation}
  b\equiv\langle A_{est}-A\rangle =
  A\left[\int\Psi(\vec{\theta})\tau(\vec{\theta})d^2\theta-1\right]=0\;.
\end{equation}
Second, the noise
\begin{equation}
  \sigma^2 \equiv\langle(A_{est}-A)^2\rangle
  =b^2+\frac{1}{(2\pi)^2}\int|\hat{\Psi}(\vec k)|^2 P_N(k)\d^2k
\end{equation}
has to be minimal with respect to the signal.

The filter function $\Psi$ satisfying these two conditions is found by
combining them with a Lagrangian multiplier $\lambda$. The variation
$L=\sigma^2+\lambda b$ is carried out, and the filter function $\Psi$ is
found by minimising $L$. In Fourier space, the solution of this
variational minimisation is
\begin{equation}\label{eq:opt_filter}
  \hat{\Psi}_{OPT}(\vec k)=
  \frac{1}{(2\pi)^2}\left[\int
  \frac{|\hat{\tau}(\vec k)|^2}{P_N(k)}\d^2k\right]^{-1}
  \frac{\hat{\tau}(\vec k)}{P_N(k)} \;,
\end{equation}
where the hats denote the Fourier transform. The last equation shows
that the shape of the optimal filter $\Psi$ is determined by the shape
of the signal, $\tau$, and by the power spectrum of the noise, $P_N$.

\cite{MA05.1} model the signal by assuming that clusters are on average
axially symmetric and their shear profile resembles that of an NFW halo
\citep[see e.g.][]{BA96.1,WR00.1,LI02.1,ME03.1}. Consequently, this
filter is optimised for searching for the same halo shape as the
optimised aperture mass, even if the filter profile is different.

The noise is assumed to be given by three contributions, namely the
noise contributions from the finite number of background sources, the
noise from their intrinsic ellipticities and orientations, and the
weak-lensing signal due to the large-scale structure of the universe.

The first two sources of noise are characterised by the power spectrum
\begin{equation}
  P_\epsilon(k)=\frac{1}{2}\frac{\sigma_{\epsilon_s}^2}{n_g} \;,
\end{equation}    
which depends on the dispersion of the intrinsic ellipticities of the
sources, $\sigma_{\epsilon_s}$, and on the number density of background
galaxies, $n_g$.

The statistical properties of the noise due to the lensing signal from
the large-scale structure of the universe are described by the
power-spectrum of the effective tangential shear. This is related to the
power-spectrum of the effective convergence by
\begin{equation}
  P_\gamma(k)=\frac{1}{2}P_\kappa(k) \;.
\end{equation}
Thus, the total noise power spectrum is
\begin{equation}
  P_N(k)=P_\gamma(k)+P_\epsilon(k) \;,
\end{equation}
where $P_\gamma$ is determined by the linear theory of structure growth.
Using the linear instead of the non-linear power spectrum avoids
suppressing a substantial fraction of the signal from the non-linear structures
we are searching for. To further reduce any loss of signal in the filtering
process, it would be possible to cut $P_\gamma$ off at angular
scales typical for galaxy clusters.
Doing so, we found that this approach has a negligible impact on the
final result.

\begin{figure}[!ht]
  \includegraphics[angle=-90,width=\hsize]{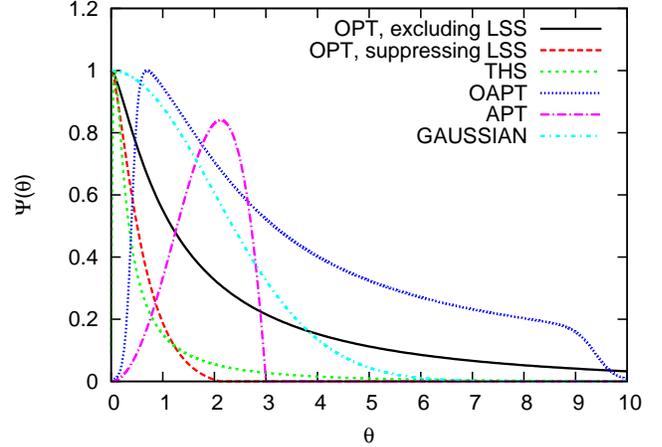}
  \caption{Comparison of the different filter shapes used here and in the
    literature. The filter scales $r_\mathrm{s}$ are those typically used in
    the literature. Note how the optimal filter (black solid curve) shrinks
    when the linear matter power spectrum is used to suppress the LSS
    contribution (red dashed curve).
    Interestingly, \cite{HE05.1} found experimentally that the
    truncated NFW-shaped filter (cyan curve) performs best when scaled
    to the green curve (THS), which approximates the optimal filter
    (OPT, red curve) almost precisely. The advantage of the optimal
    compared to the other filters is that its shape and scale are
    physically and statistically well motivated such that it needs not
    be experimentally rescaled.}
  \label{fig:14}
\end{figure}

In Fig. \ref{fig:14}, we compare the filters studied here and
in the literature. 
They are scaled in such a way as they are typically discussed or applied 
in the literature (see also the figure legend and caption for more
detail). At first sight, the scales are surprisingly different. When
the optimal filter is constructed including the linear matter power
spectrum such as to best suppress the LSS contribution, it shrinks
considerably. It is reassuring that the truncated NFW-shaped
filter (THS) proposed and heuristically scaled by \cite{HE05.1} to
yield best results almost exactly reproduces the optimal filter. They
are therefore expected to perform similarly well.
The optimised aperture-mass filter (OAPT) also peaks at fairly small
angular scales, but shows the long tail typical for the NFW
profile. The aperture mass has its maximum at comparatively large
radii, explaining why the APT filter yields results most severely
affected by the LSS.

\section{Results}

\subsection{Signal-to-noise maps}

We now use the above-mentioned weak-lensing estimators to analyse our
mock catalogues of lensed galaxies. 

In practice, the integral in Eq.~\ref{eq:wlest} is replaced by a sum
over galaxy images. Moreover, since the ellipticity $\epsilon$ is an
estimator for $\gamma$, we can write  
\begin{equation}
  A_{est}(\vec{\theta}) = \frac{1}{n_g}\sum_i\epsilon_{ti}(\vec{\theta_i})
  \Psi(|\vec{\theta_i}-\vec{\theta}|)\;,
\end{equation}
where $\epsilon_{ti}(\vec{\theta_i})$ is the tangential component of the
observed ellipticity of the galaxy at $\vec\theta_i$, with respect to
the point $\vec\theta$. Similarly, the noise estimate in $A_{est}$ is
given by
\begin{equation}
\sigma^2(A_{est})(\vec{\theta}) = \frac{1}{2n_g^2}\sum_i
|\epsilon_{i}(\vec{\theta_i})|^2 \Psi^2(|\vec{\theta_i}-\vec{\theta}|).
\end{equation}

Computing $A_{est}$ and $\sigma^2(A_{est})$ on a grid covering our
simulated sky, we produce maps of the signal-to-noise ratio for all the
weak lensing estimators. We use three different filter sizes for each
estimator in order to test the stability of the results achieved. These
have been calibrated among the different filters to allow the optimal
detection of similar objects. For the optimal filter, we used sizes of
$1'$, $2'$ and $4'$. These correspond to $2.75'$, $5.5'$ and $11'$ for
the aperture mass, for the optimised aperture mass we used the values $5'$,
$10'$ and $20'$ that are widely used in literature.

\begin{figure}[!ht]
  \includegraphics[width=0.49\hsize]{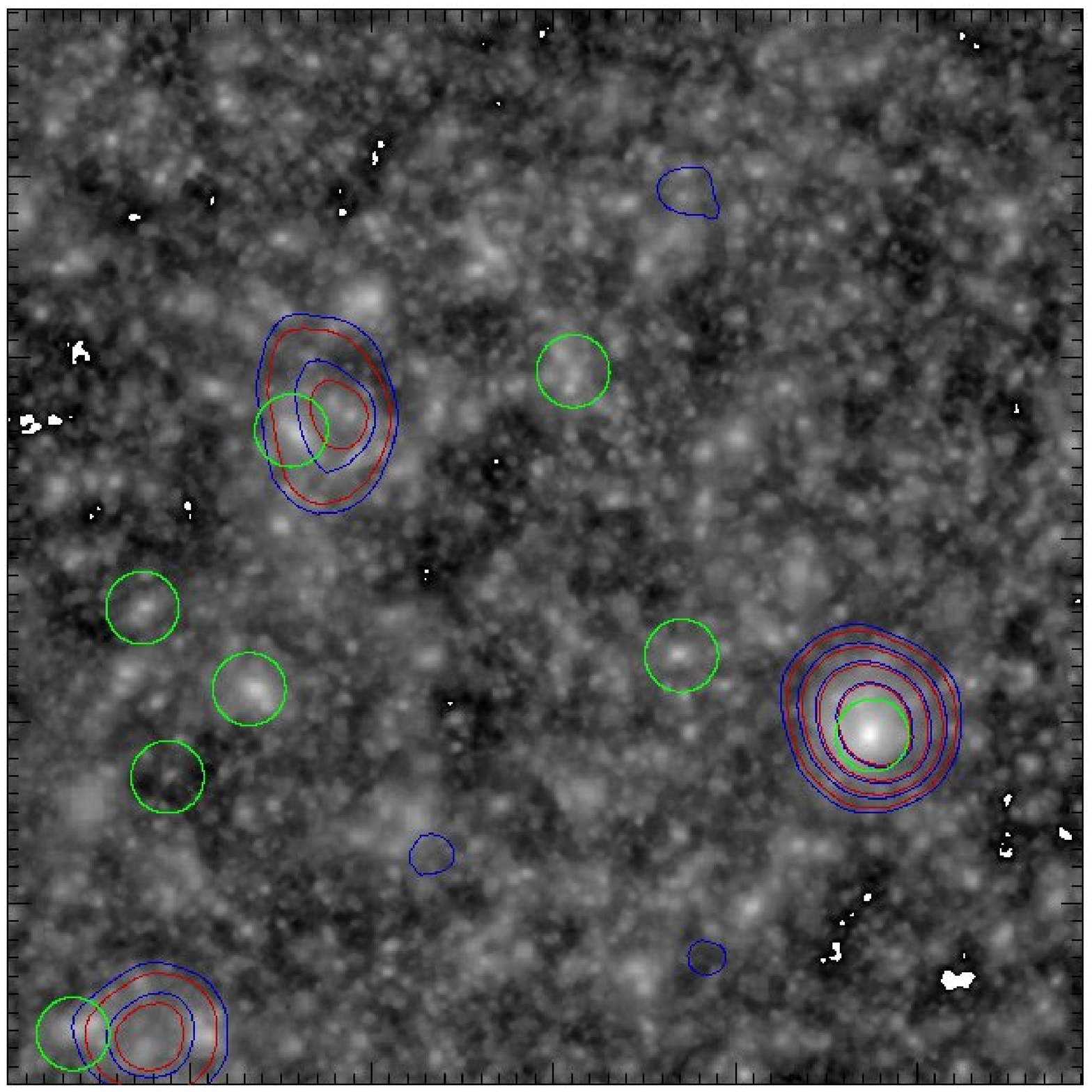}\hfill
  \includegraphics[width=0.49\hsize]{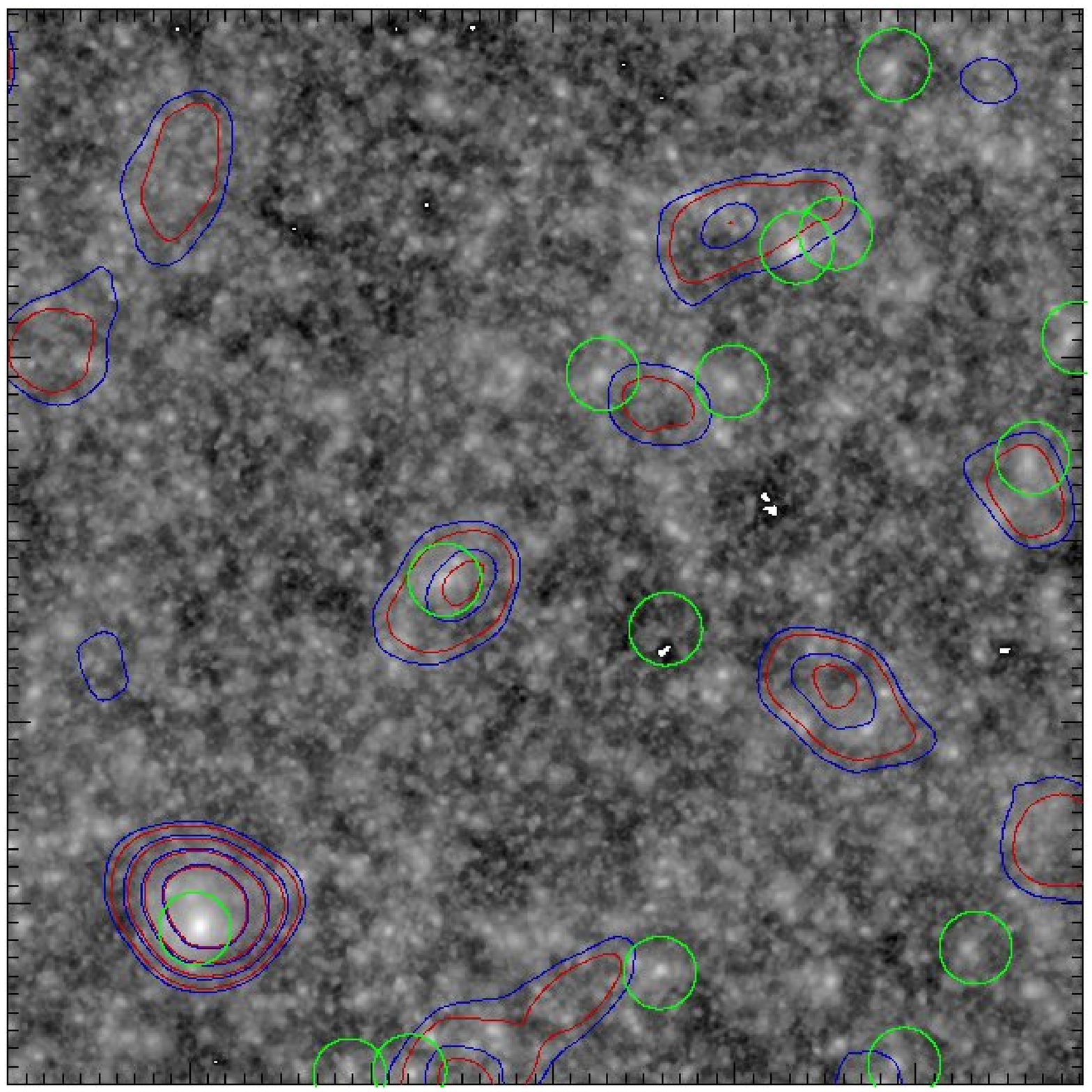}
  \includegraphics[width=0.49\hsize]{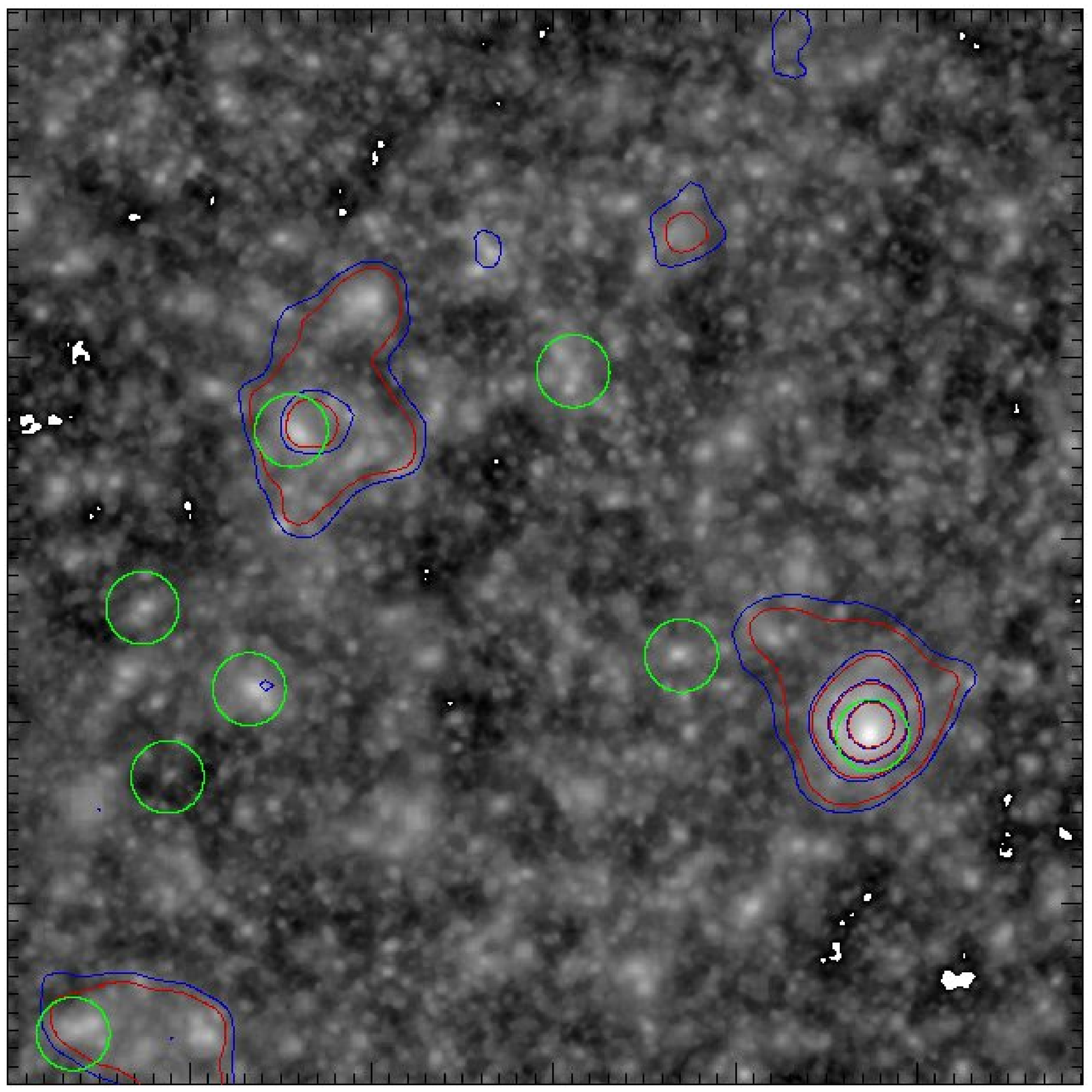}\hfill
  \includegraphics[width=0.49\hsize]{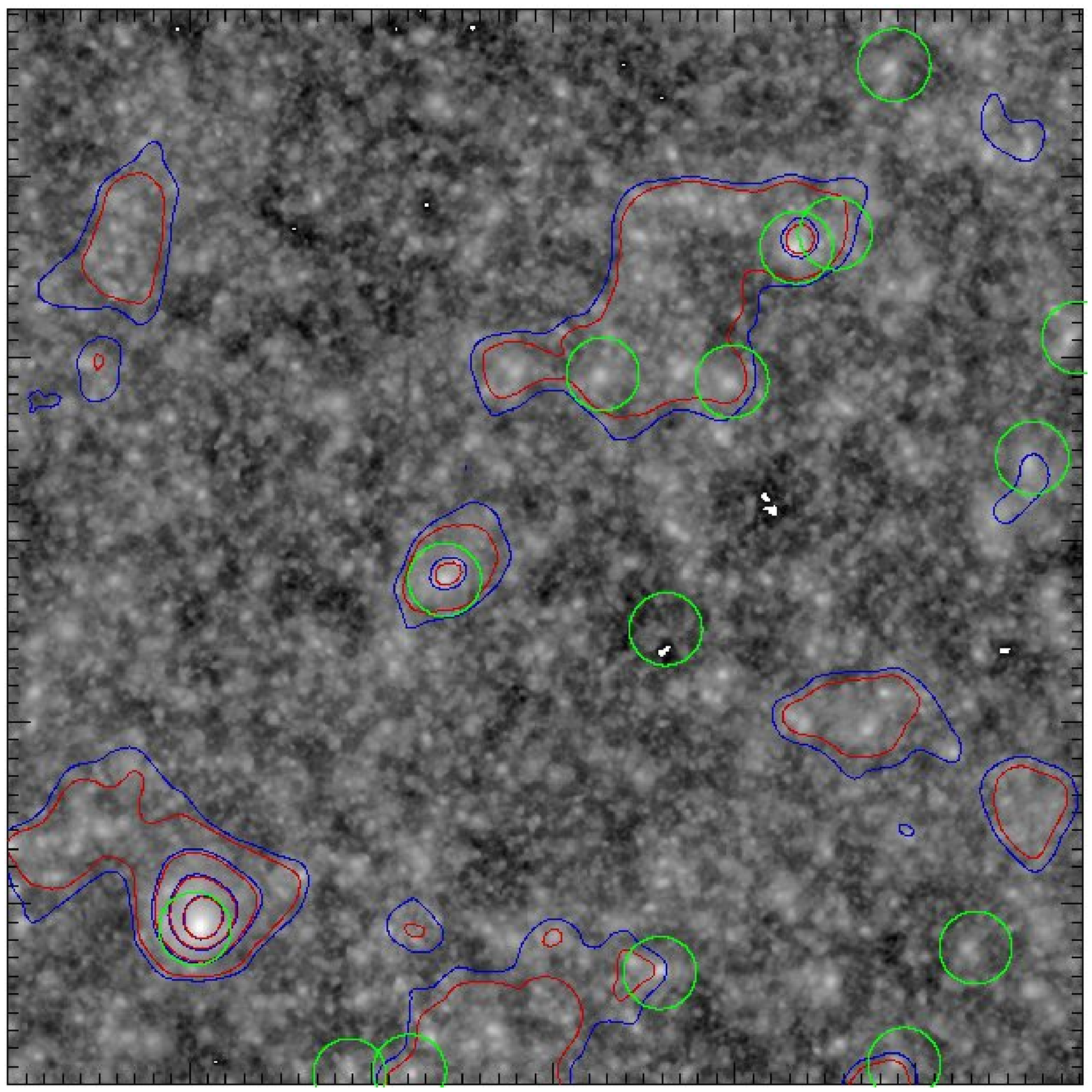}
  \includegraphics[width=0.49\hsize]{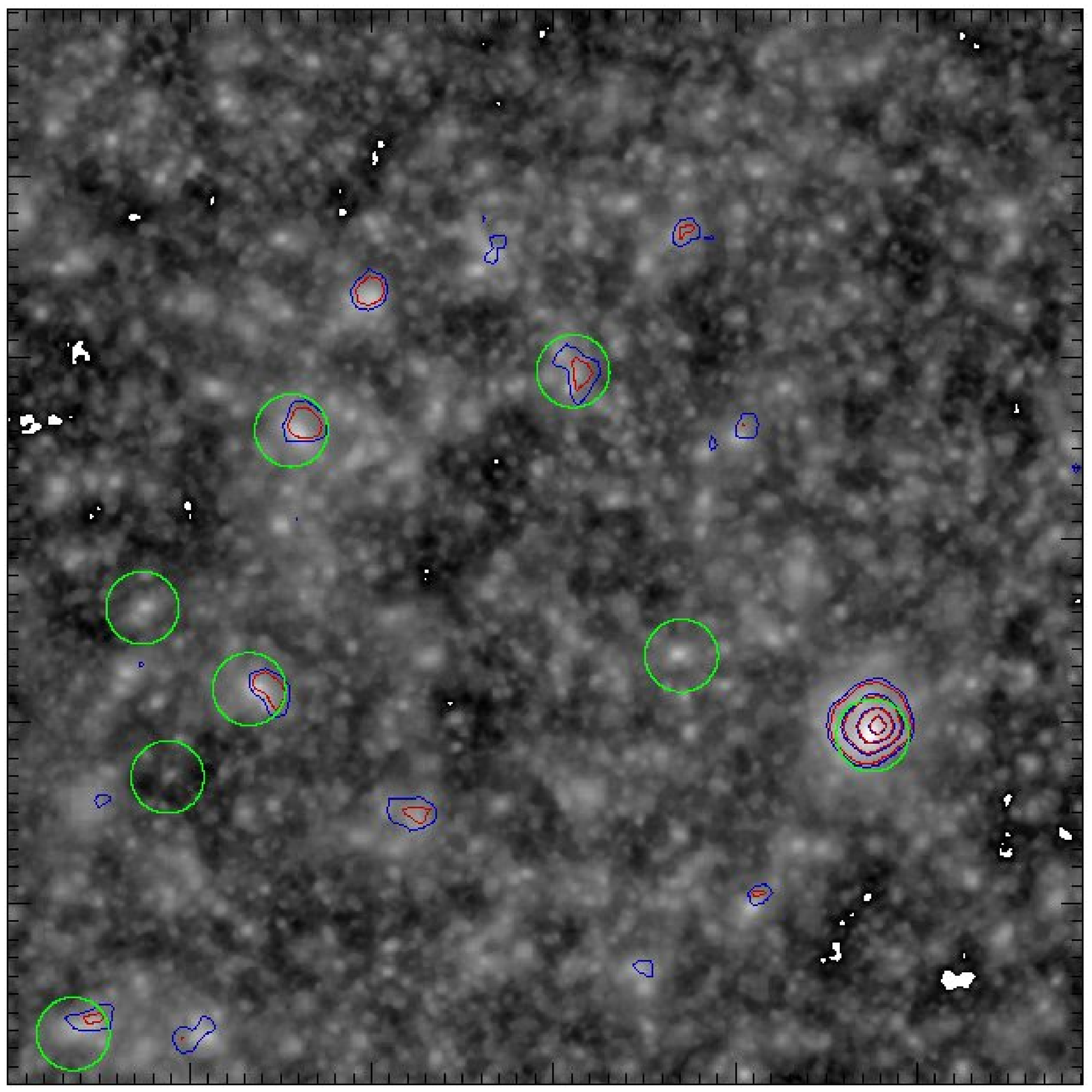}\hfill
  \includegraphics[width=0.49\hsize]{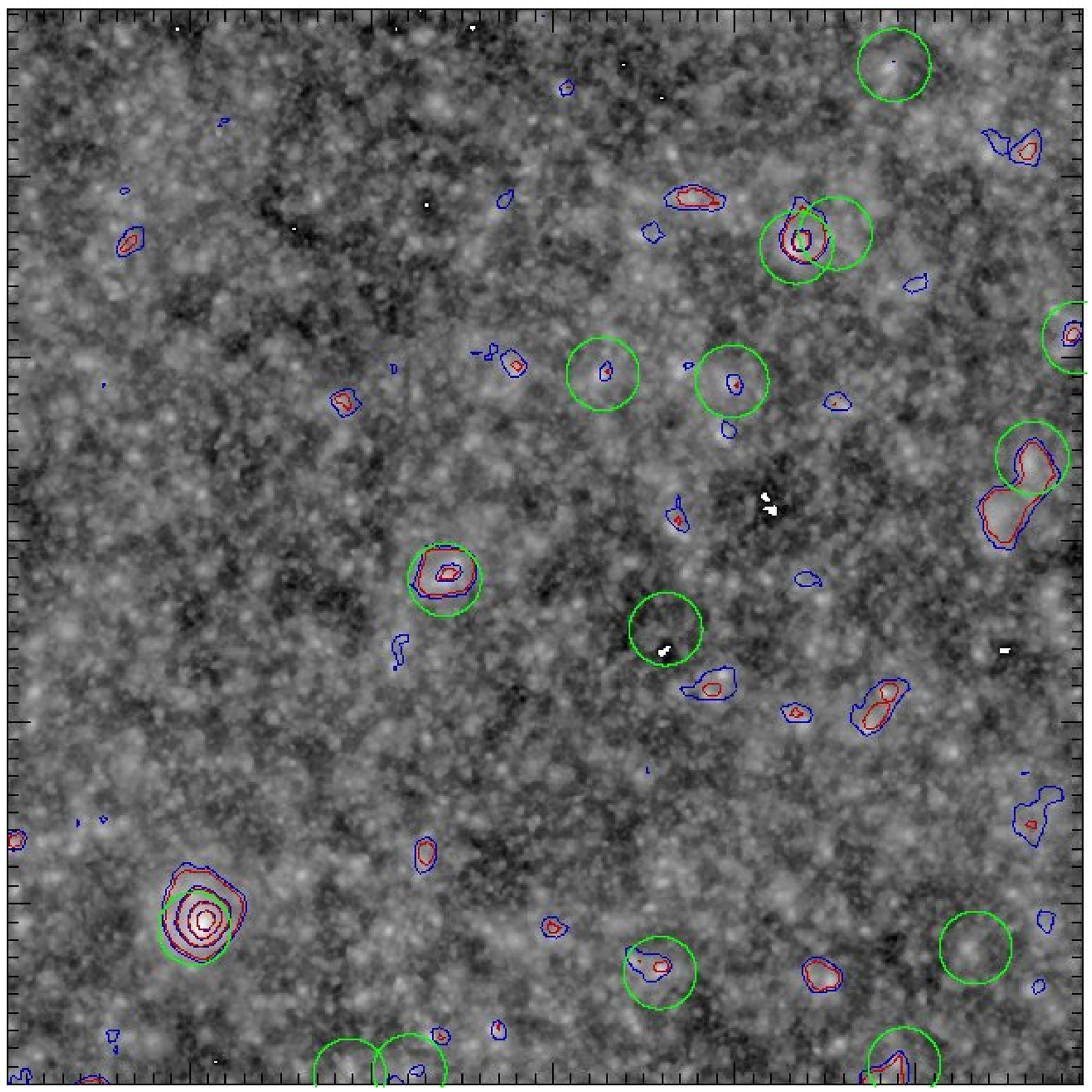}
\caption{Maps of the effective convergence for
sources at redshift $z_s=1$ (left panels) and $z_s=2$ (right panels) for
a region of simulated sky. Superimposed are the iso-contours of the
signal-to-noise ratio of the weak-lensing signal measured with three
estimators, namely the APT (top panels), the OAPT (middle panels) and the OPT 
(bottom panels). The iso-contours start from $S/N=4$ with a step of 3. The
positions of the halos contained in the field-of-view having mass
$M>7\times 10^{13}h^{-1}M_\odot$ are identified by circles. The filter sizes
are $11'$, $20'$ and $4'$ for the APT, the OAPT and the OPT, respectively.}
\label{fig:4}
\end{figure}

In Fig.~\ref{fig:4} we show examples of the signal-to-noise iso-contours
of the weak lensing signal, superimposed on the corresponding effective
convergence maps of the underlying projected matter distribution for
sources at redshift $z_s=1$ (left panels) and $z_s=2$ (right panels).
The iso-contours start at $S/N=4$ with a step of 3. From top to
bottom, the maps refer to the results obtained using the aperture mass
(APT), the optimised aperture mass (OAPT) and the optimal filter (OPT)
with sizes of $11'$, $20'$ and $4'$, respectively. The circles identify
halos with mass $M\ge 7 \times 10^{14} h^{-1} M_{\odot}$ present in the
field-of-view. The side length of each map is one degree.

The images show that, for sources at high redshift, all three estimators
can successfully detect the weak-lensing signal from clusters in the mass
range considered. However, spurious detections, corresponding to high
signal-to-noise peaks not associated with any halo, also appear. Their
significance and spatial extent is larger in the case of the APT and the
OAPT filters. This confirms the results of \cite{MA05.1}.

For lower-redshift sources, the OPT detects five out of the seven halos
present in the field, while the APT and the OAPT detect substantially
fewer halos. For the OPT, the number of spurious detections is roughly
the same or slightly smaller than for sources at higher redshift, while
it is strongly reduced for the APT and the OAPT. The natural explanation
of these results is that the detections with the APT and the OAPT are
strongly contaminated by the noise from large-scale structure lensing,
which becomes increasingly important for sources at higher redshift.
This noise is efficiently filtered out by the OPT.

\subsection{True and spurious detections}

In the following, we call a \emph{detection} a group of pixels in the
$S/N$ maps above a threshold $S/N$ ratio. Its position in the sky is
given by the most significant pixel, i.e.~that with the highest $S/N$
ratio.

A true detection is obviously a detection that can be associated with
some halo in the simulation. A spurious detection is instead mimicked by
noise, in particular by cosmic structures aligned along the
line-of-sight.

The association between weak-lensing detections and cluster halos is
established by comparing their projected positions on the sky. This
causes a problem, because the simulation boxes contain plenty of
low-mass halos that are not individually detectable through lensing but
happen to be projected near the line-of-sight towards a detection. Thus,
spurious detections could easily be erroneously associated with these
low-mass halos on the basis of the projected position only.

\begin{figure}[!ht]
  \includegraphics[width=0.49\hsize]{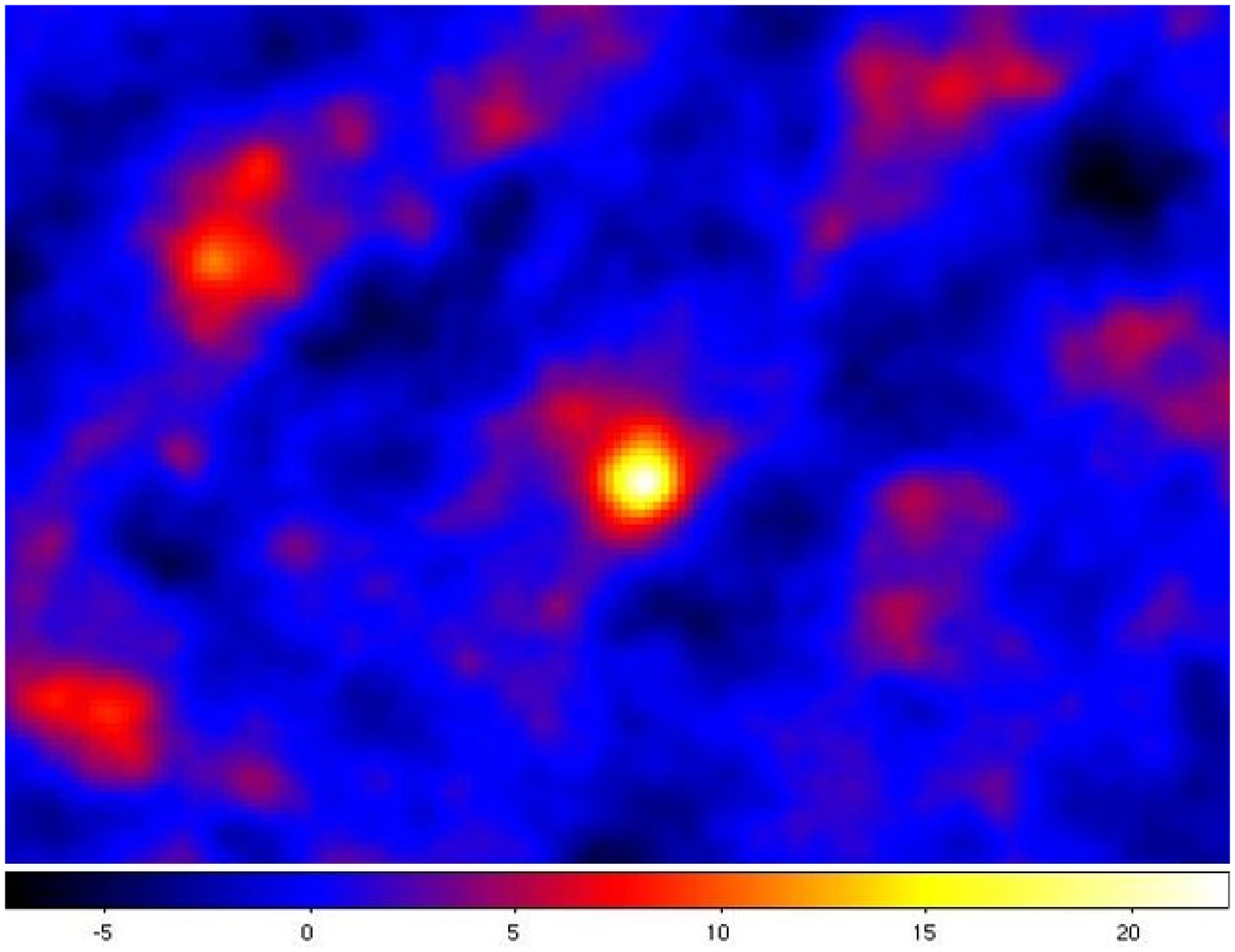}\hfill
  \includegraphics[width=0.49\hsize]{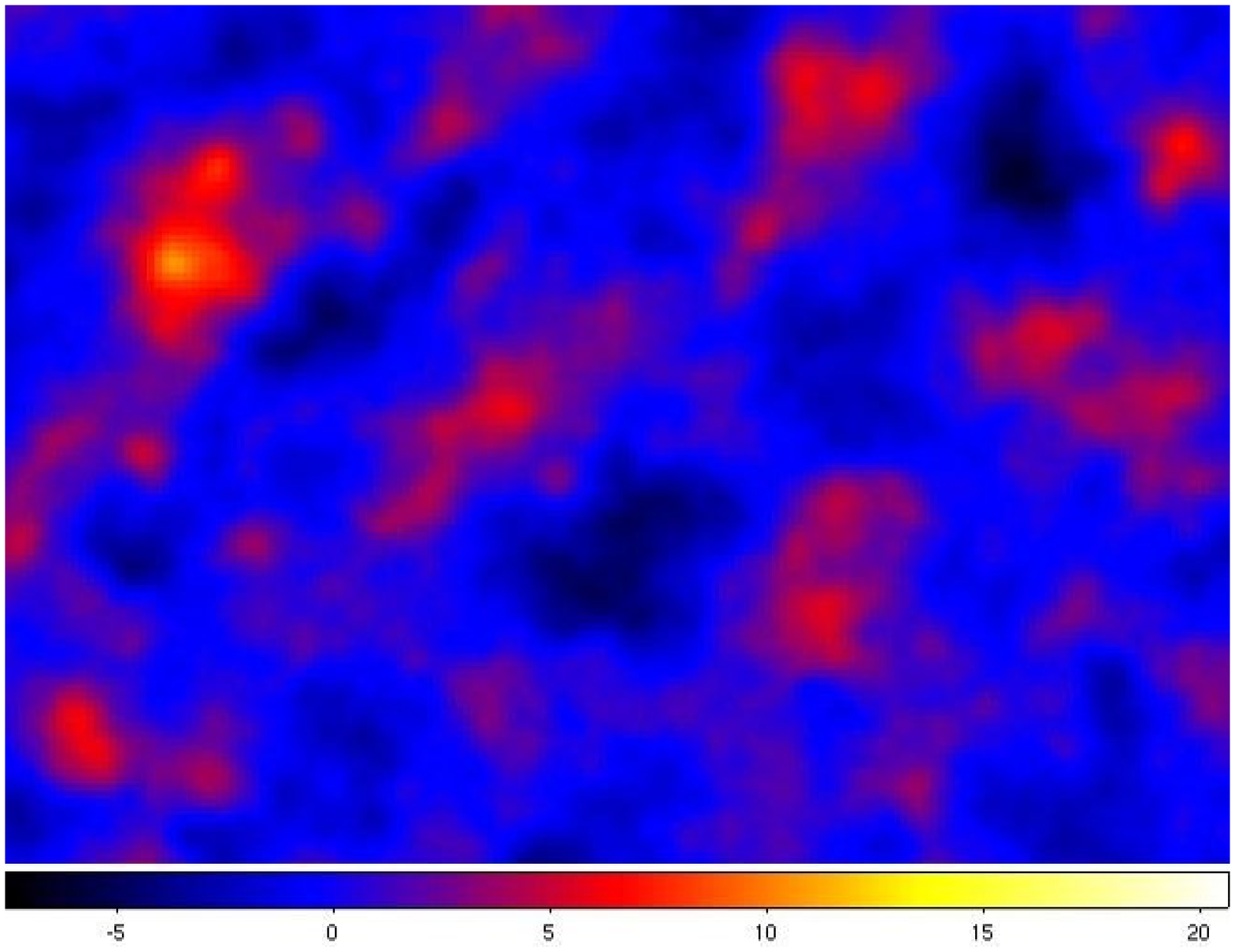}
\caption{Map of the $S/N$ ratio corresponding to a region of 3 square
degrees. The map was created using the OAPT estimator, with a filter scale of
$20'$ and assuming a source redshift of $z_s=1$. The left panel shows the
$S/N$ ratio map including all lens planes, while the right panel shows
the same map obtained after removing the lens plane containing the cluster
responsible for the highest $S/N$ peak in the left panel.}
\label{fig:5}
\end{figure}

\begin{figure}[!ht]
  \includegraphics[width=0.49\hsize]{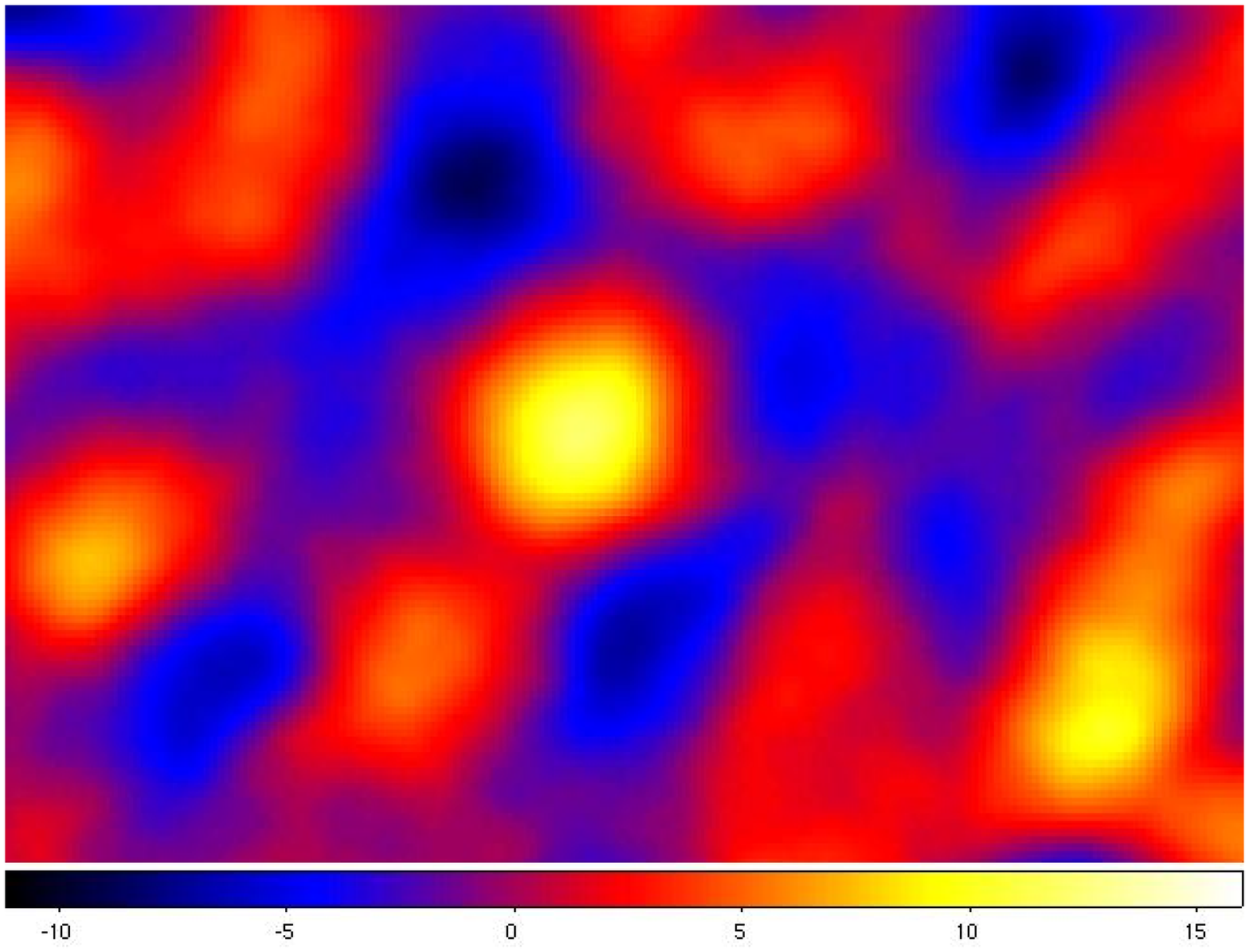}\hfill
  \includegraphics[width=0.49\hsize]{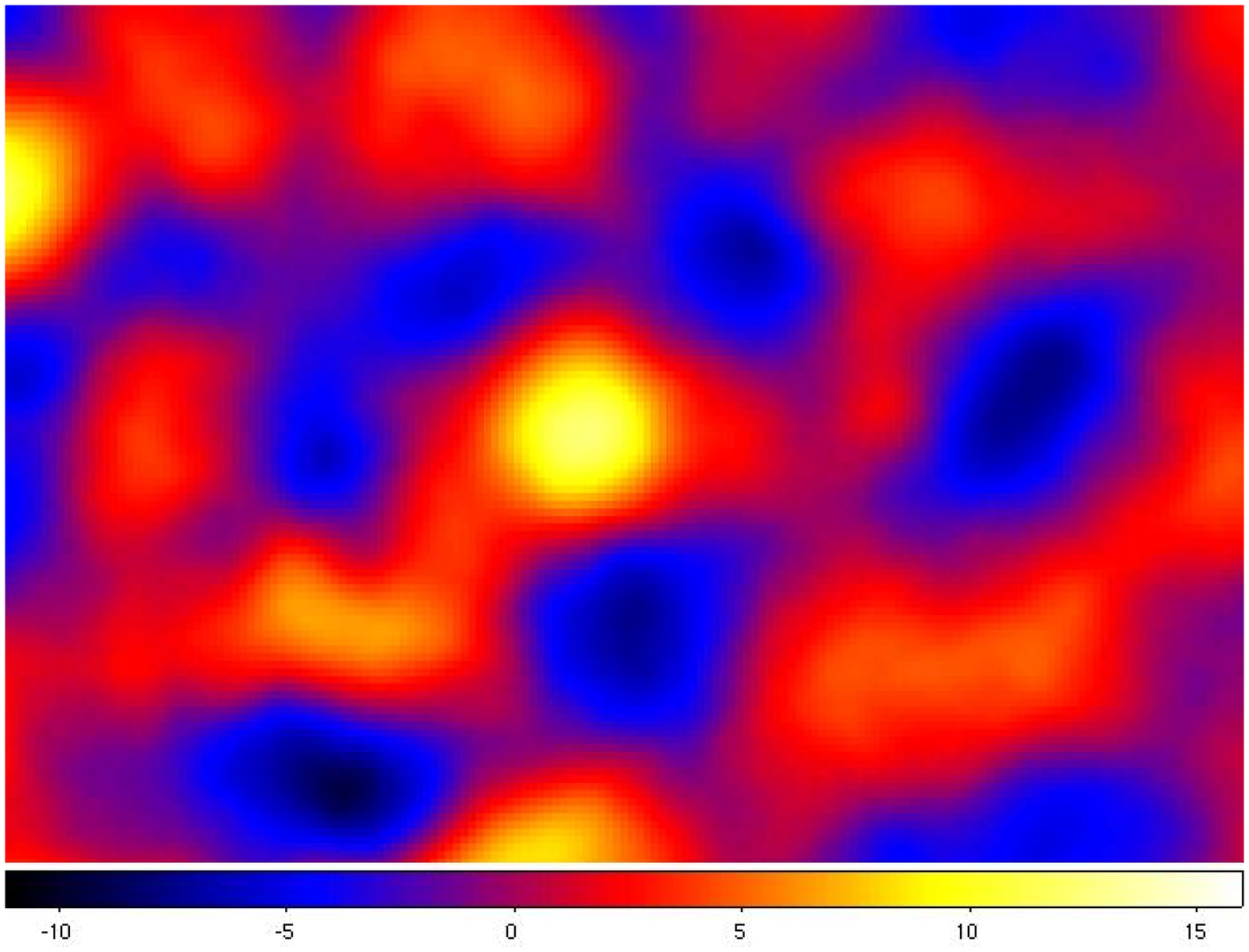}
\caption{Maps of the $S/N$ ratio corresponding to a region of 3 square
  degrees. The maps were created with the APT estimator, with a filter scale of
  $11'$ and assuming a source redshift of $z_s=2$. The left panel shows a true
  detection, while the right panel shows a spurious detection.}
\label{fig:13}
\end{figure}

As pointed out earlier, we describe the lensing effect of the matter
contained in the light cone with a stack of lens planes. Cluster halos
are localised structures, i.e.~their signal originates from a single
lens plane. Thus, any detection should disappear when its plane is
removed from the stack. Conversely, spurious detections are not caused
by localised structures and should remain even after removing an
individual lens plane. This is illustrated by the $S/N$ maps shown in
Fig.~\ref{fig:5}. The map in the left panel includes all lens
planes, while one plane was removed for the right panel. Both maps were
obtained with the OAPT estimator with a filter size of $20'$ and a
source redshift of $z_s=1$. Clearly, the highest peak in the left panel, which
is in fact produced by a massive halo, disappears in the right panel, after
removing the lens plane from the stack which contains the halo. All other
features in the left upper map remain unchanged.

This allows us to verify the reliability of detections associated with
some halo in the catalogue. For each positive match, we estimate the
lensing signal before and after removing the plane containing the
candidate lensing halo from the lens-plane stack. If this causes a
significant decrease in the $S/N$ ratio, we classify the detection as
true, and otherwise as spurious. We estimate through several checks of
detections associated to the halos that $S/N$ fluctuations of order $25\%$ of
the initial value are possible due to different properties of the noise. Thus,
we set this limit as our threshold for discriminating between true and
spurious detections.

This method also shows its power when pixels identifying a true detection
are compared with pixels associated to a spurious detection. This is shown in
Fig.~\ref{fig:13}. The map in the left panel represents a true detections,
while the map on the right panel shows a spurious detections. The maps refer
to different regions of a $S/N$ map created with the APT estimator with a
filter size of $11'$ and a source redshift of $z_s=2$. As it is clearly
seen, it is impossible a priori to distinguish which of the two is spurious.

\subsection{Statistical analysis of the detections}
\label{sect:totdet}

\begin{figure}[!ht]
  \includegraphics[height=0.49\hsize,angle=-90]{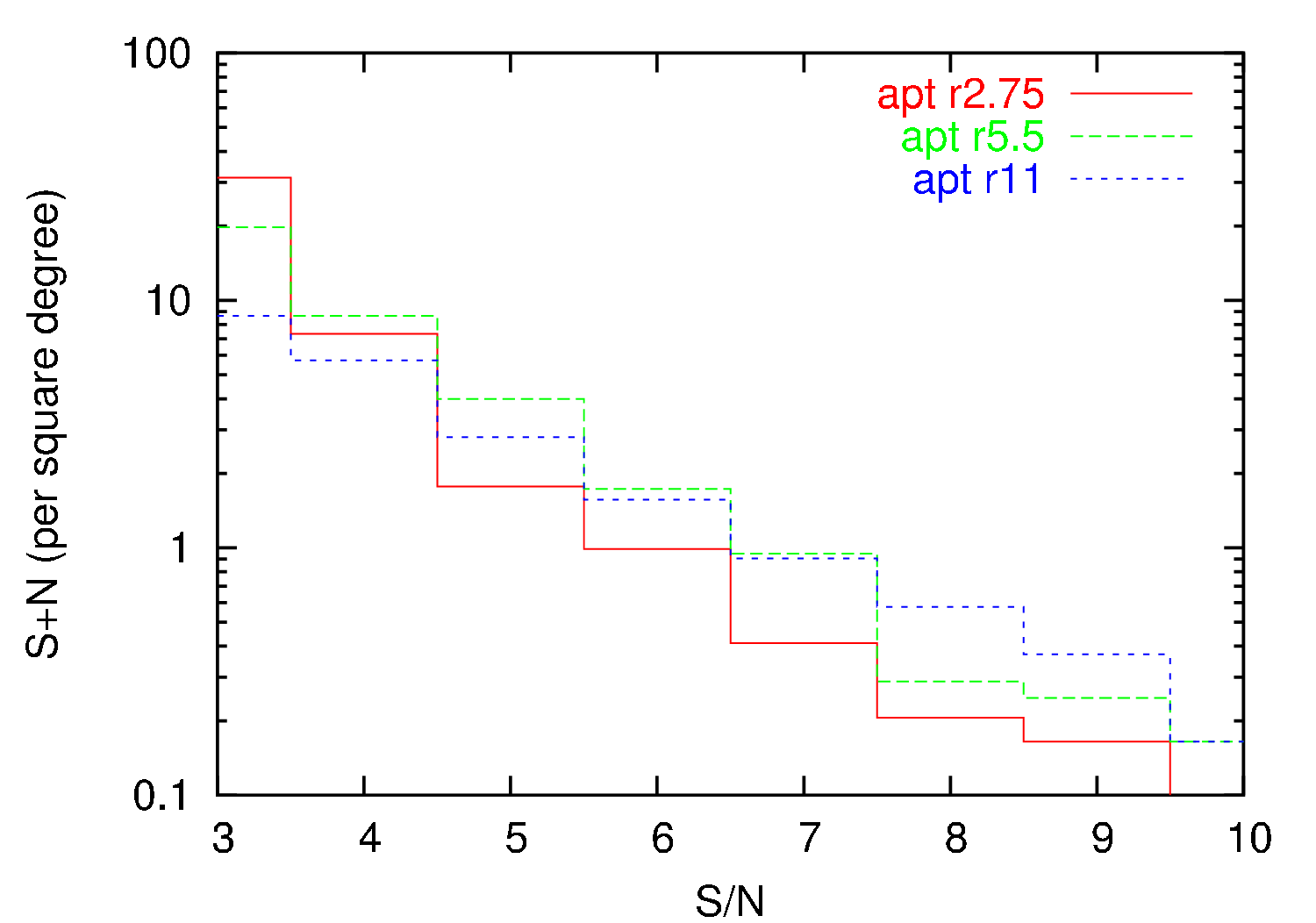}\hfill
  \includegraphics[height=0.49\hsize,angle=-90]{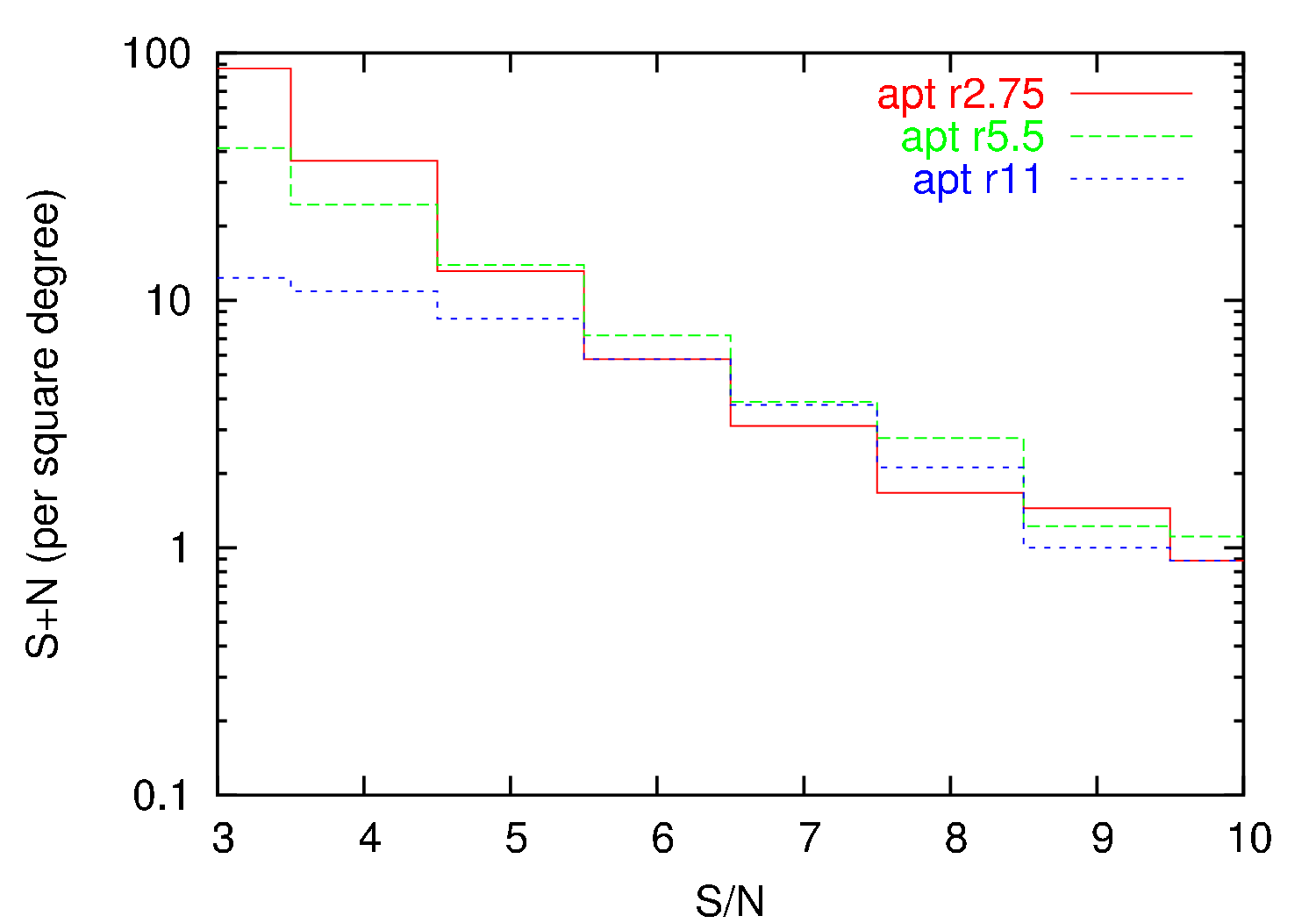}
  \includegraphics[height=0.49\hsize,angle=-90]{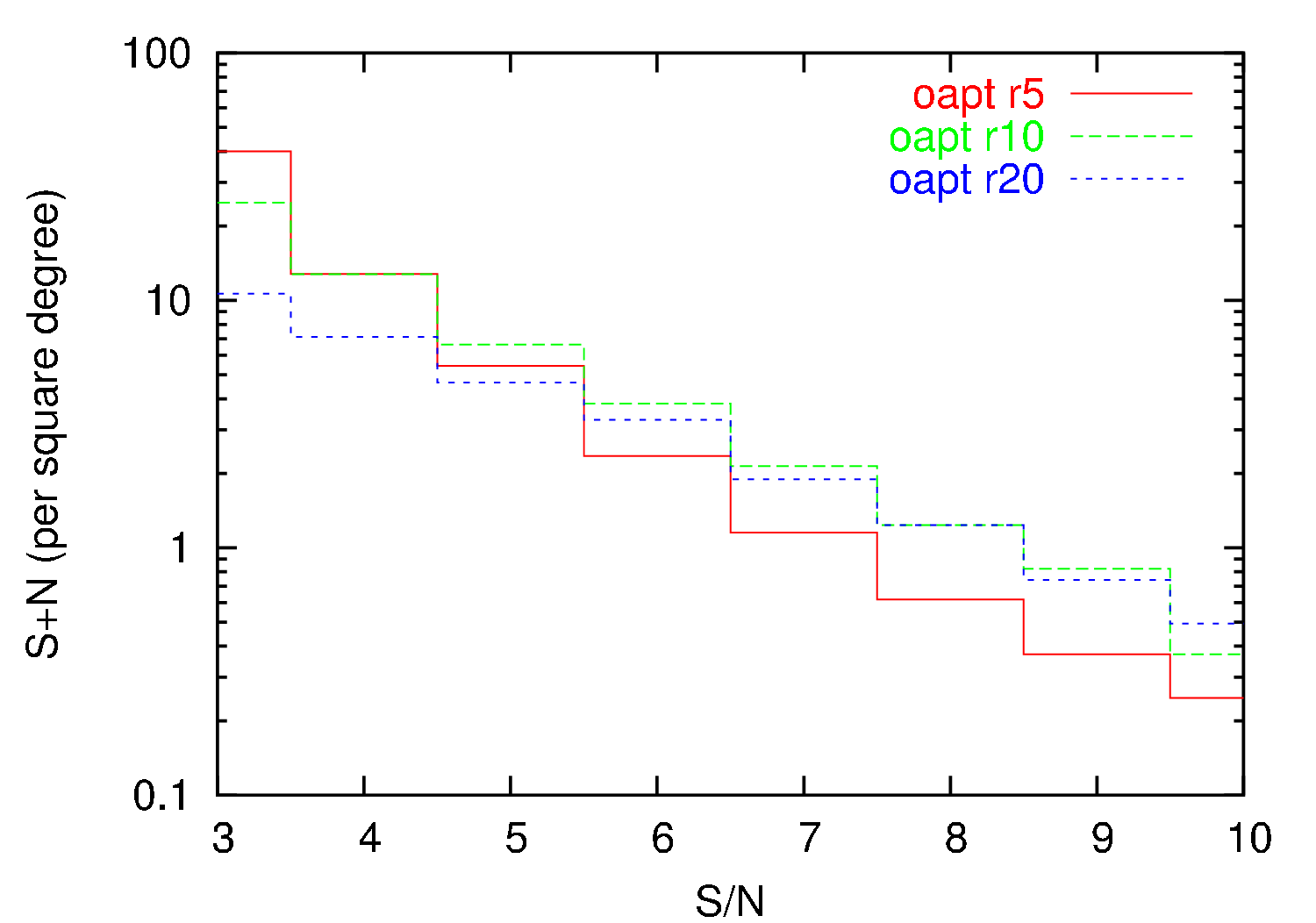}\hfill
  \includegraphics[height=0.49\hsize,angle=-90]{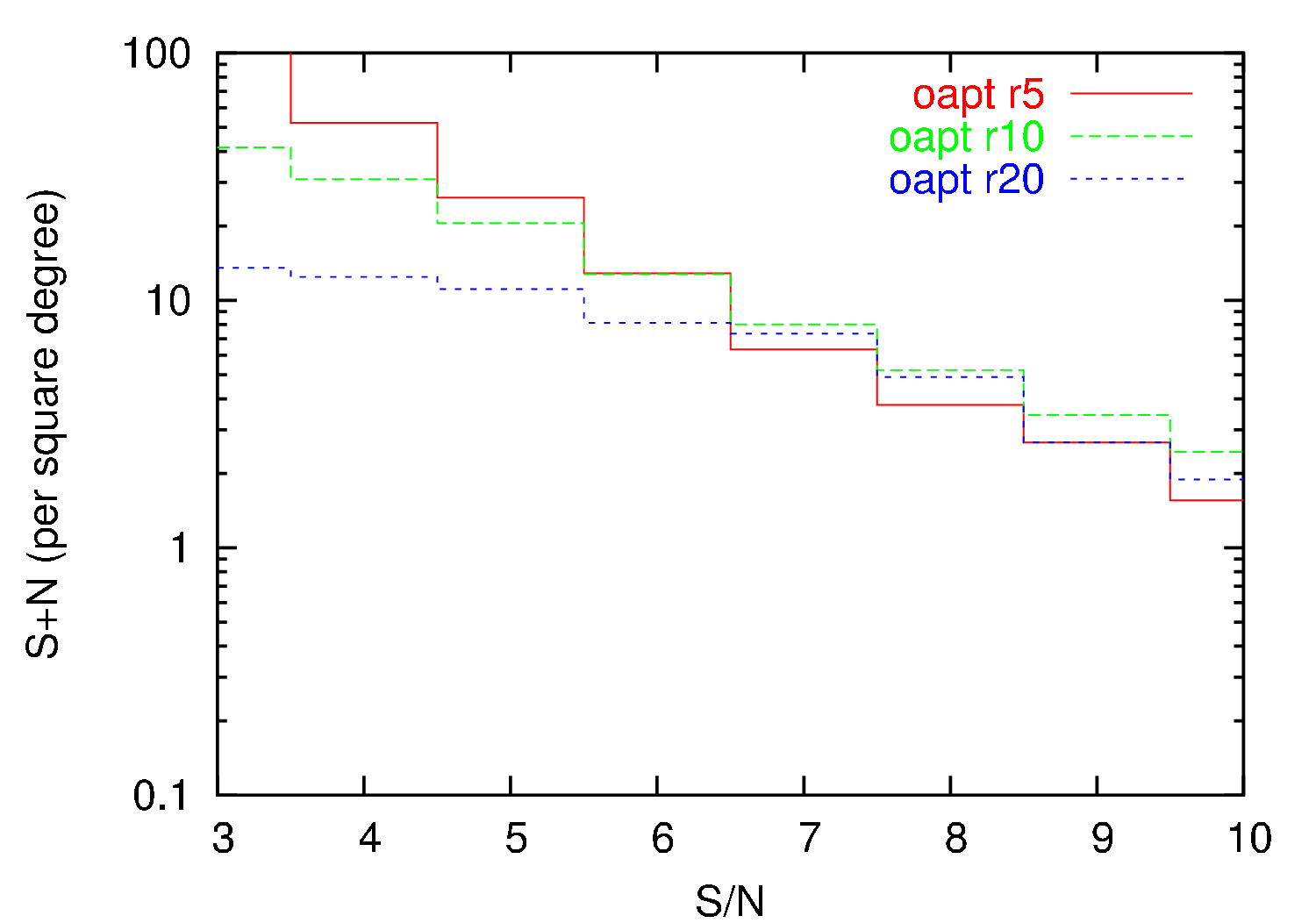}
  \includegraphics[height=0.49\hsize,angle=-90]{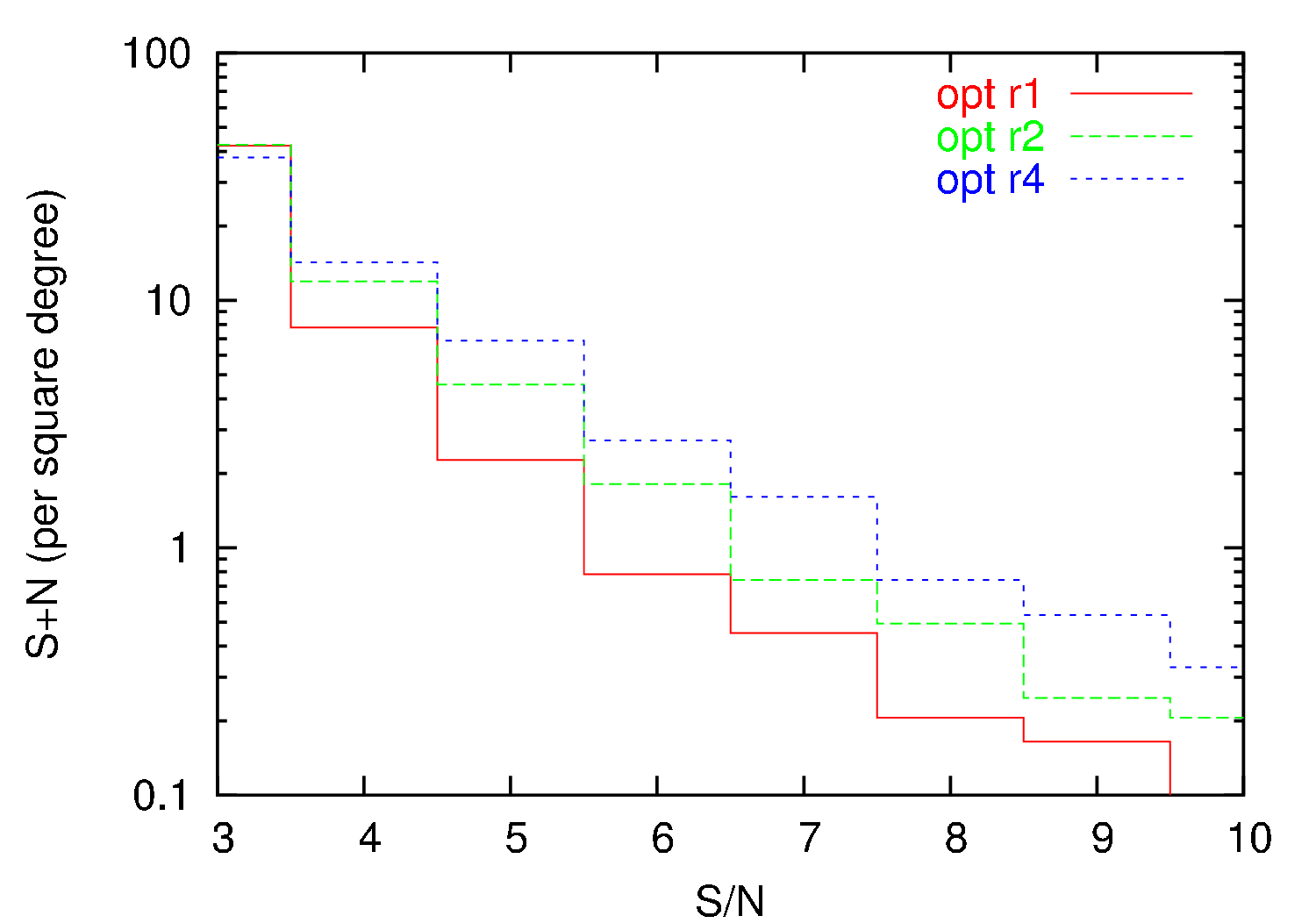}\hfill
  \includegraphics[height=0.49\hsize,angle=-90]{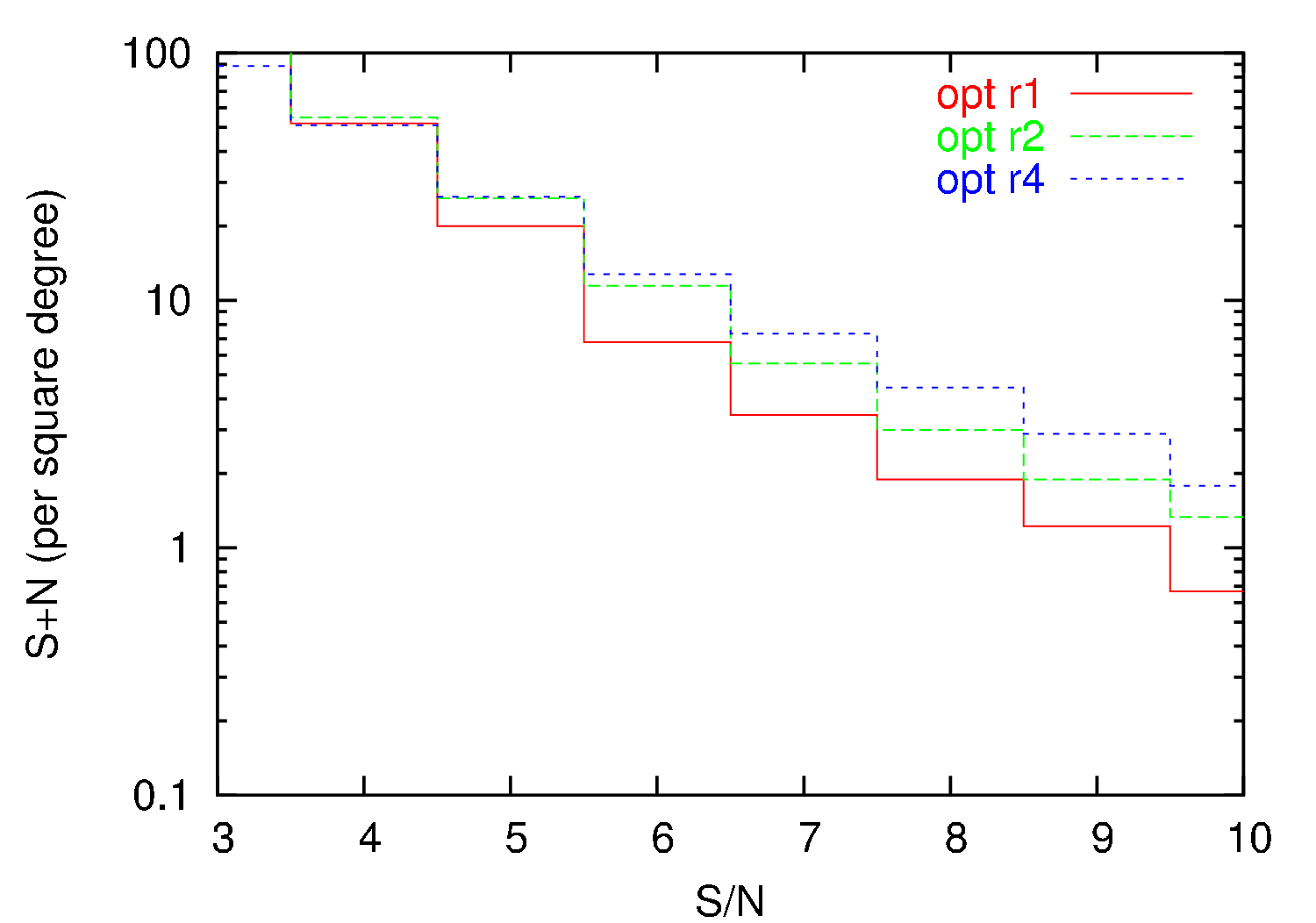}
\caption{Number of detections as a function of the $S/N$ ratio obtained
by using the APT (top panels), the OAPT (middle panels) and the $OPT$
weak lensing estimators. Results for sources at redshift $z_s=1$ and
$z_s=2$ are shown in the left and in the right panels, respectively.
Different line styles refer to three different filter sizes. For the
OPT, these are $1'$, $2'$ and $4'$. They correspond to $2.75'$, $5.5'$,
and $11'$ for the APT and to $5'$, $10'$ and $20'$ for the OAPT.}
\label{fig:6}
\end{figure}

In Fig.~\ref{fig:6}, we show the number of detections per square
degree in $S/N$ ratio bins, ignoring for now the distinction between
true and spurious detections. Left and right panels refer to simulations
with sources at redshifts $z_s=1$ and $z_s=2$, respectively. From top to
bottom, we show the results for the APT, the OAPT and the OPT
estimators. In each panel, we use solid, dashed and dotted lines to
display the histograms corresponding to increasing filter sizes.

For low source redshifts and small filter sizes, the APT and the OPT
estimators lead to similar numbers of detections. Instead, for the OAPT,
the number of detections is larger by up to a factor of two for
$S/N\gtrsim4$. Increasing the filter size, the number of detections
generally increases for all estimators, especially for large $S/N$
ratios and in particular for the OPT.

We notice, however, that for small $S/N$ ratios, larger filters produce
lower numbers of detections for the APT and for the OAPT. This behaviour
is more evident for sources at higher redshifts. For example, we find
that the number of detections with $S/N=4$ drops by a factor of $4$ for
the APT and by a factor of $\sim7$ for the OAPT, when increasing the
filter size from $2.5'$ to $11'$ and from $5'$ to $20'$, respectively.
Increasing the filter size, the weak-lensing signal is estimated by averaging
over more background galaxies. Thus, high $S/N$ peaks are smoothed, and some
detections may be suppressed. This affects mainly the detections with the APT
and the OAPT filters. On the other hand, the OPT filter shrinks in response to
the noise introduced by the large scale structure, largely reducing this
effect compared to the APT and the OAPT.

\begin{figure}[!ht]
  \includegraphics[height=0.49\hsize,angle=-90]{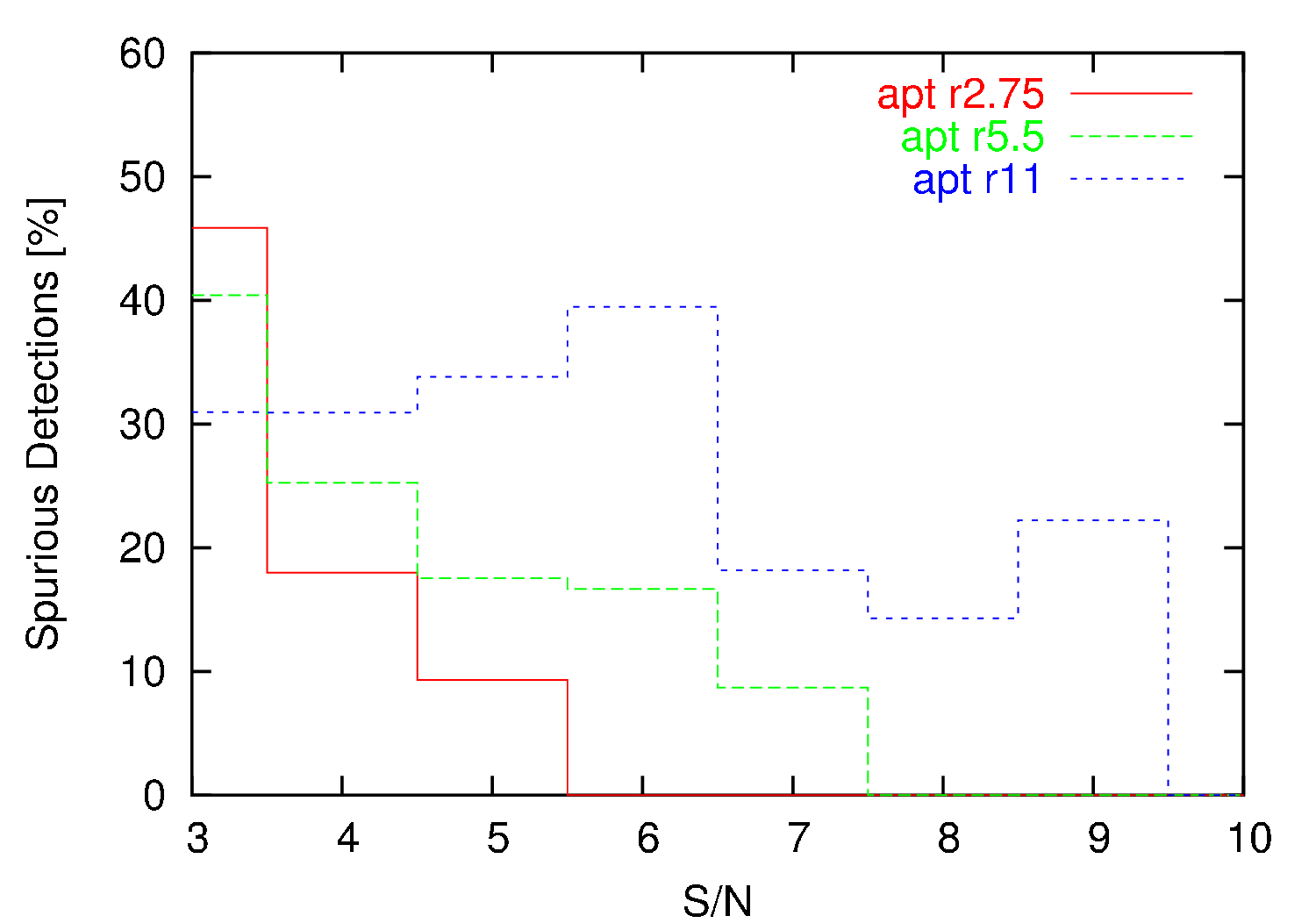}\hfill
  \includegraphics[height=0.49\hsize,angle=-90]{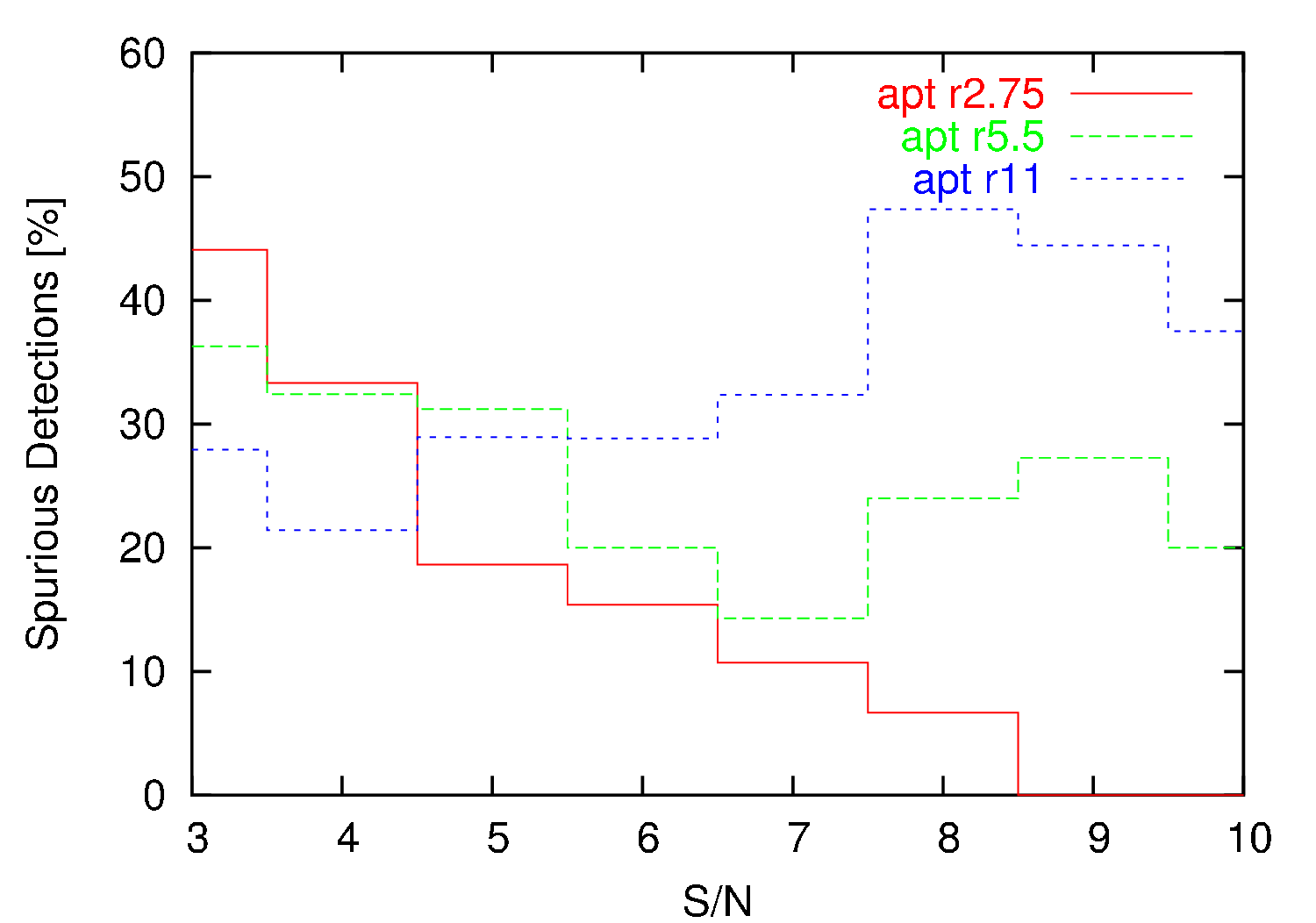}
  \includegraphics[height=0.49\hsize,angle=-90]{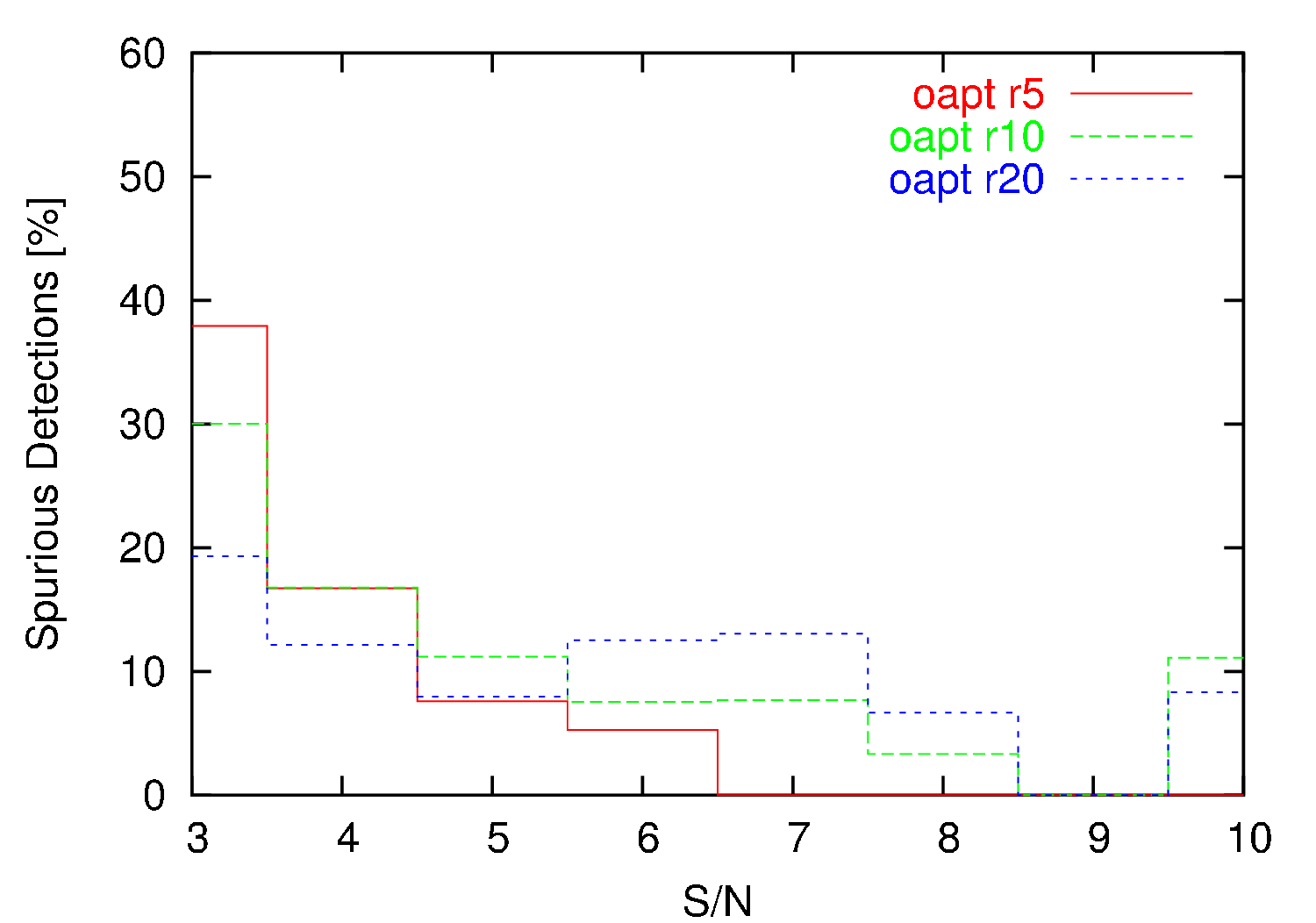}\hfill
  \includegraphics[height=0.49\hsize,angle=-90]{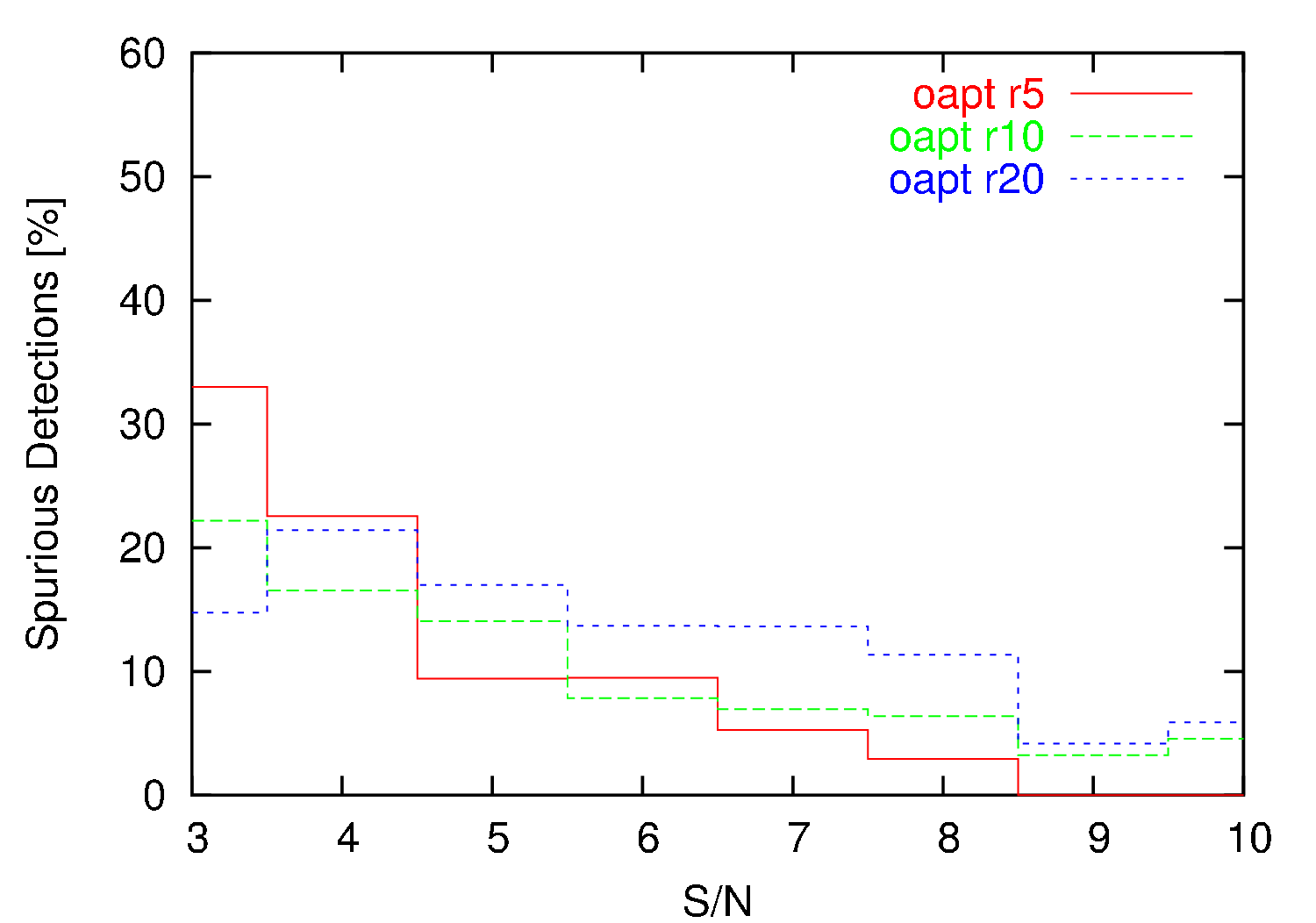}
  \includegraphics[height=0.49\hsize,angle=-90]{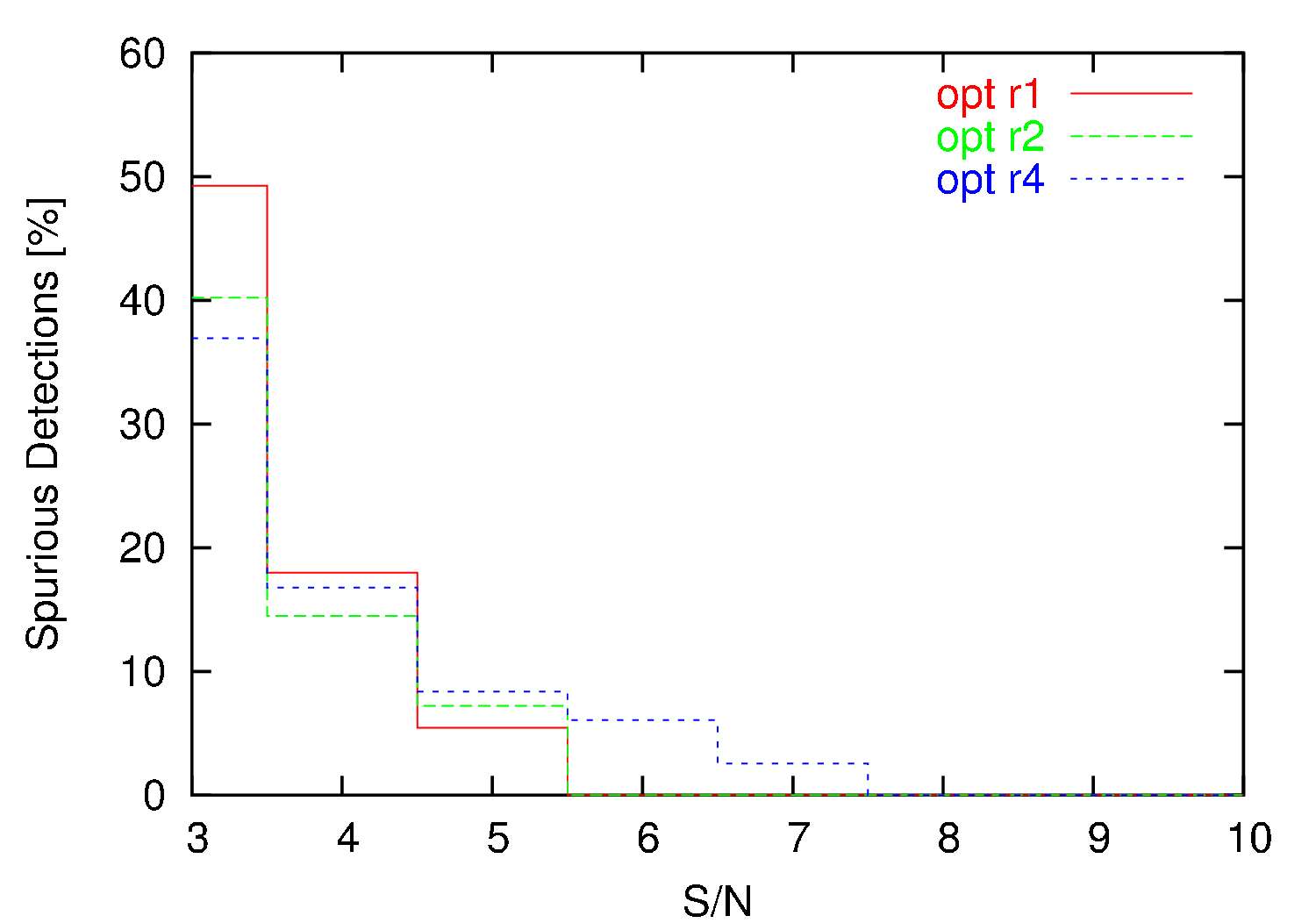}\hfill
  \includegraphics[height=0.49\hsize,angle=-90]{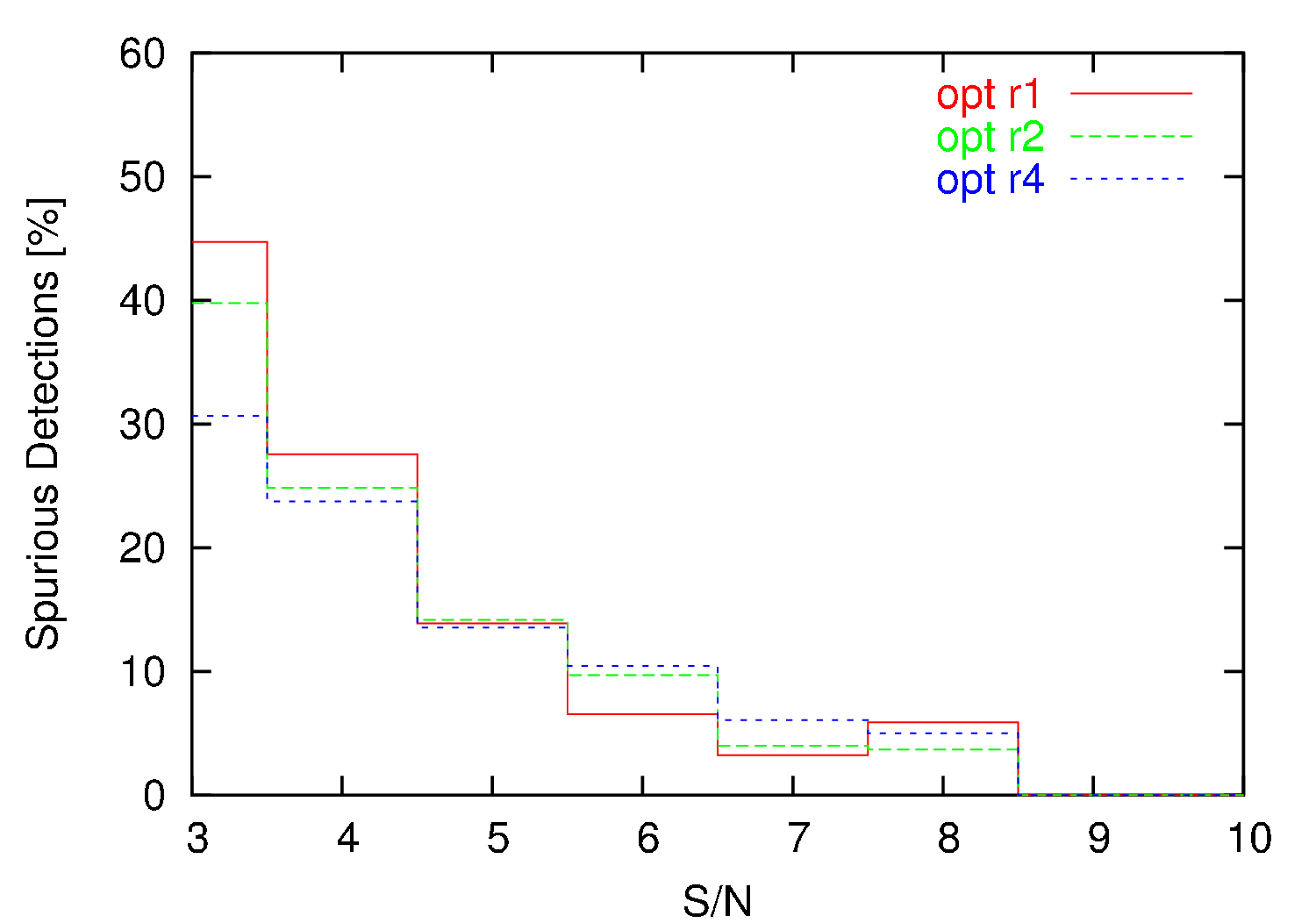}
\caption{Fraction of spurious detections as a function of the $S/N$
ratio obtained by using the APT (top panels), the OAPT (middle panels)
and the $OPT$ weak-lensing estimators. Results for sources at redshift
$z_s=1$ and $z_s$=2 are shown in the left and the right panels,
respectively. Different line styles refer to three filter sizes. For the
OPT these are $1'$, $2'$ and $4'$. They correspond to $2.75'$, $5.5'$,
and $11'$ for the APT and to $5'$, $10'$ and $20'$ for the OAPT.}
\label{fig:7}
\end{figure}

The fractions of spurious detections are shown in
Fig.~\ref{fig:7}. Clearly, the OPT estimator performs better than
the APT and the OAPT. For sources at redshift $z_s=1$ and $z_s=2$, the
fraction of spurious detections with the OPT is less than $20\%$ and
$30\%$ at $S/N\sim 4$. This fraction decreases below $10\%$ for
$S/N\gtrsim5$ and drops rapidly to zero for higher $S/N$ ratios. Results
are very stable against changes in the filter size. Conversely, the APT
and the OAPT estimators yield similarly low fractions of false
detections only for the smallest apertures.

Depending on the filter shape, its size and on the source redshift, a
$S/N$ threshold can be defined above which there are no spurious
detections and thus all detections are reliable. For the OPT estimator,
this minimal signal-to-noise ratio is between $5$ and $8$. It increases
above $10$ for the APT and the OAPT estimators if large filter sizes are
used. These results agree with the results of \cite{MA05.1}, using
numerical simulations, and of \cite{MA06.1}, regarding the analysis of
the GaBoDS survey.

Here, we studied the contaminations by the LSS, the intrinsic
ellipticity and the finite number of background galaxies all together.
To gain an idea which of those is the main source for spurious
detections, we used the APT with $r_s=11'$ to analyse a catalog of
galaxies with intrinsic ellipticities set to zero. In this case, the
S/N ratio is enhanced by a factor of four uniformly across the whole
field, but the morphology of the map is not affected. The same should
apply to the finite number of background sources. We thus conclude
that the main source of spurious detections is the LSS, as already
noted by \cite{RE99.1} and \cite{WH02.1}.

\subsection{Sensitivity}
\label{sect:sens}

We shall now quantify which halo masses the weak-lensing estimators are
sensitive to.

\begin{figure}[!ht]
  \includegraphics[height=0.49\hsize,angle=-90]{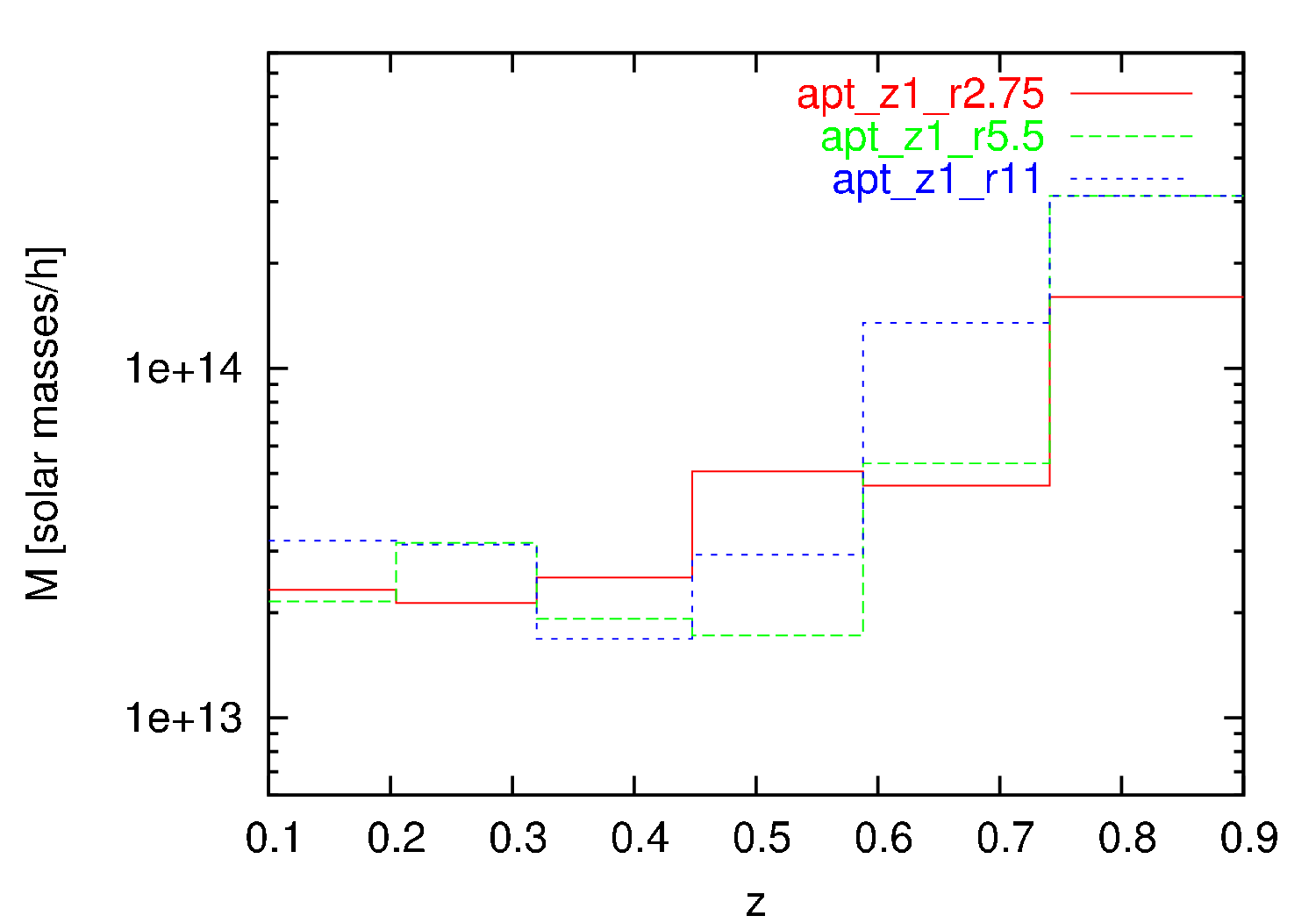}\hfill
  \includegraphics[height=0.49\hsize,angle=-90]{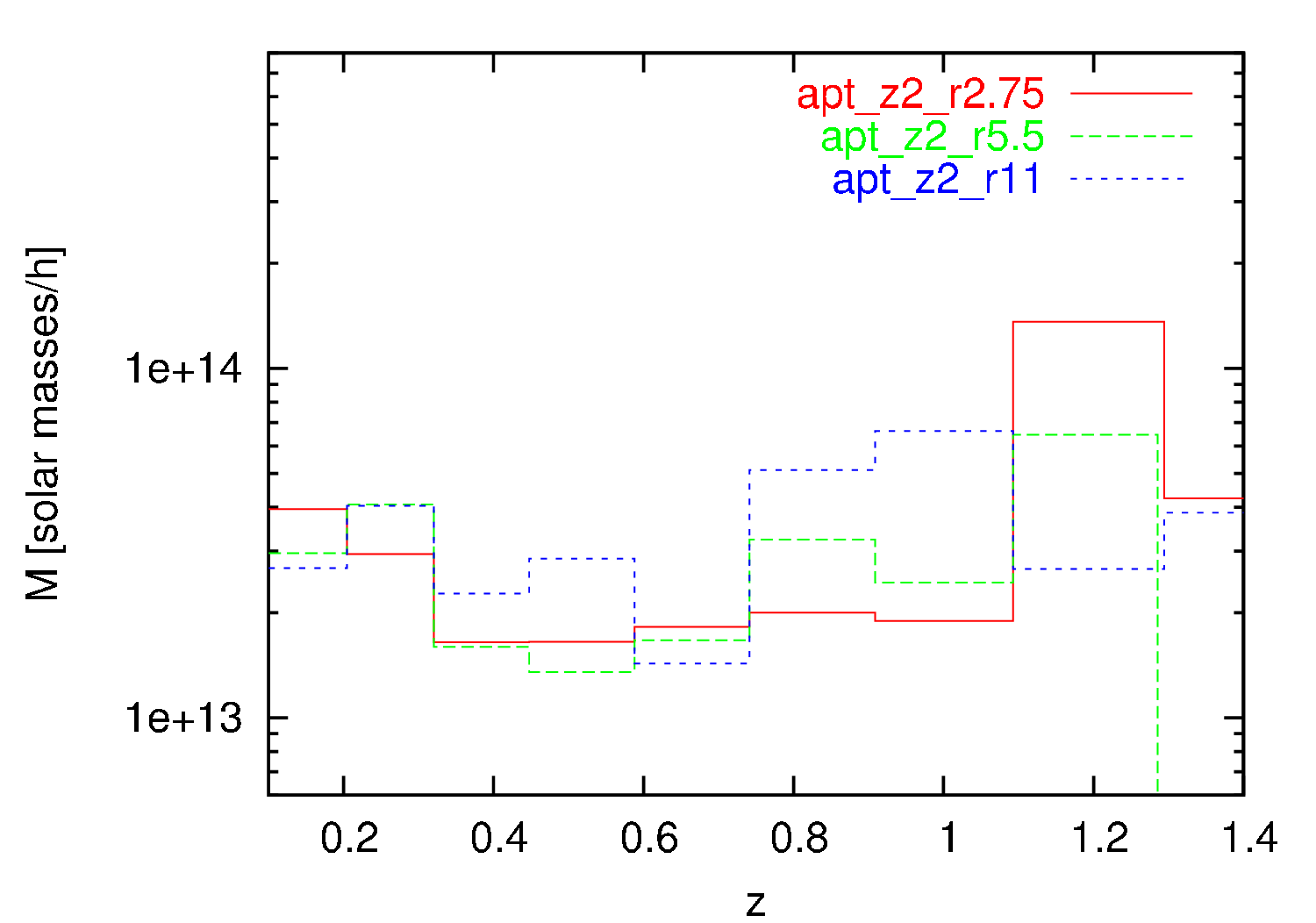}
  \includegraphics[height=0.49\hsize,angle=-90]{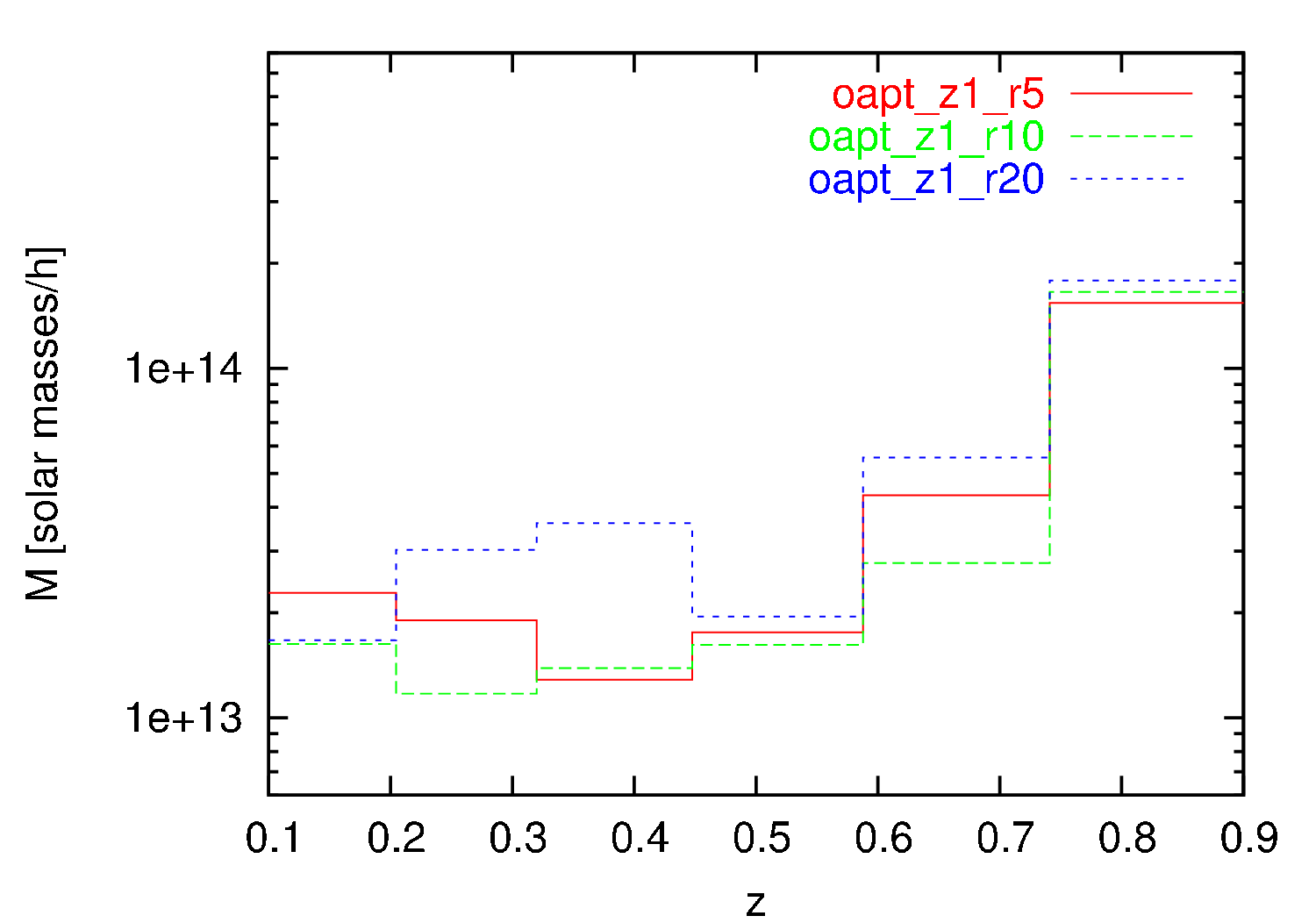}\hfill
  \includegraphics[height=0.49\hsize,angle=-90]{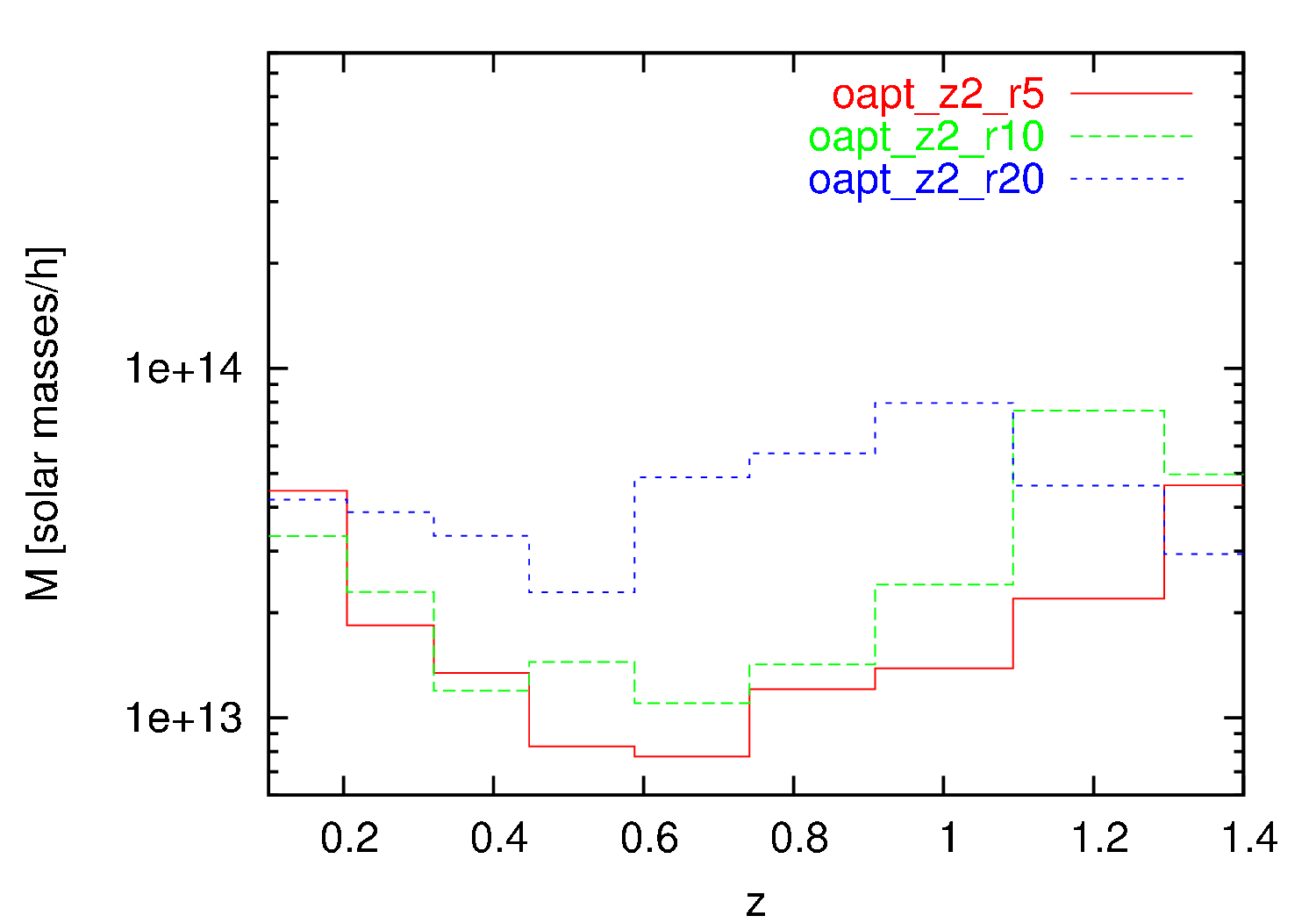}
  \includegraphics[height=0.49\hsize,angle=-90]{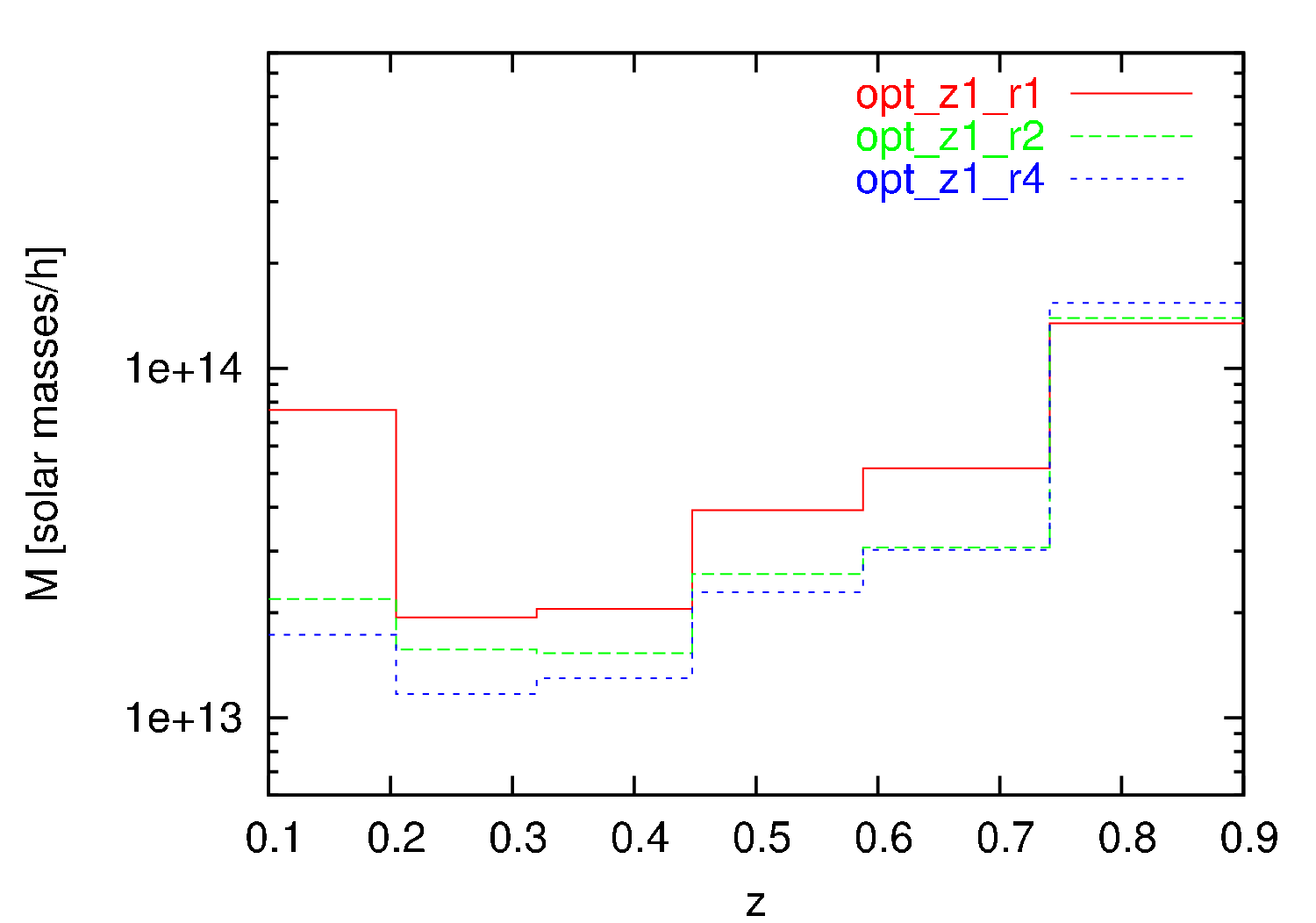}\hfill
  \includegraphics[height=0.49\hsize,angle=-90]{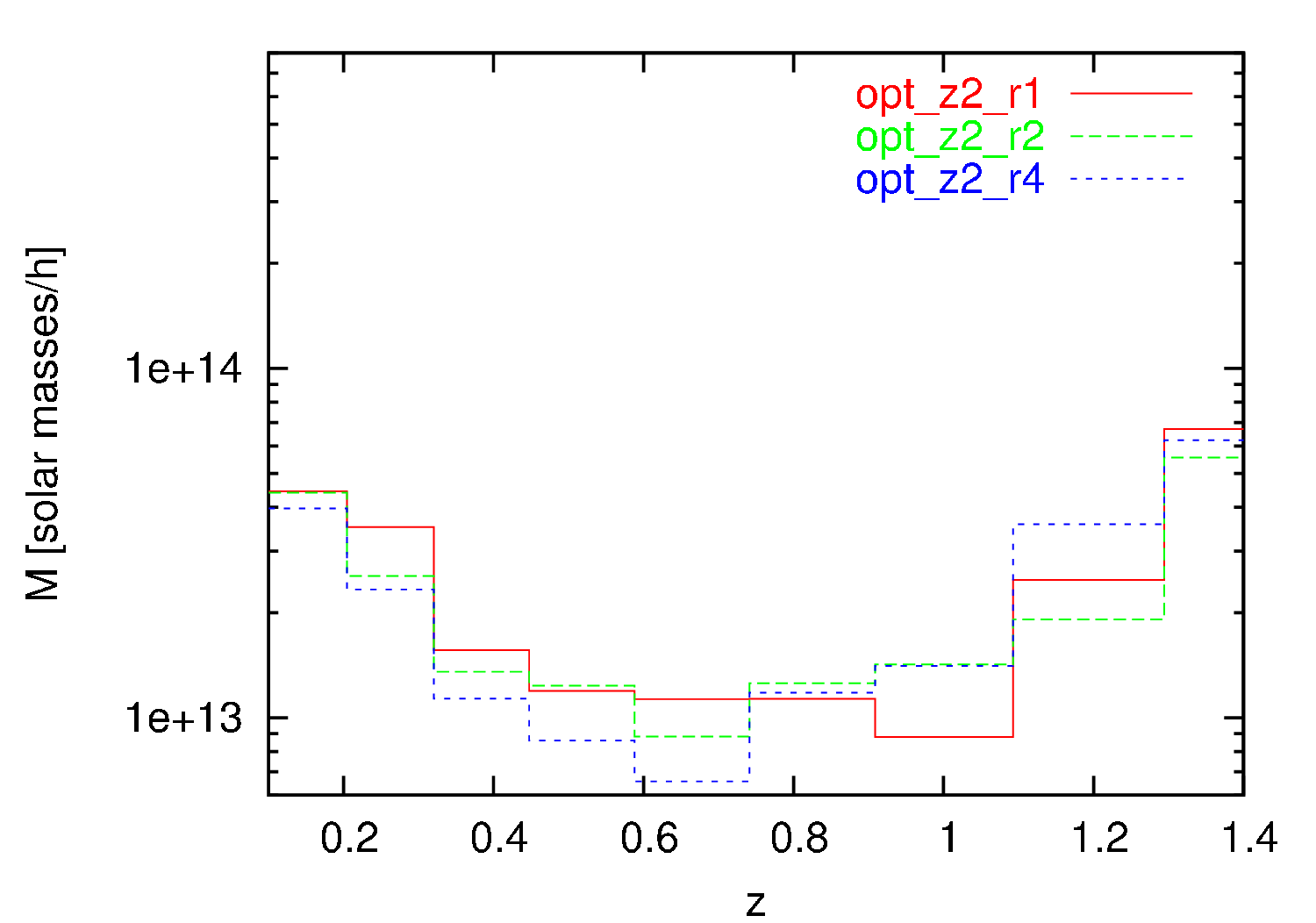}
\caption{Minimum detected halo mass as a function of redshift for the
APT (top panels), the OAPT (middle panels) and for the OPT (bottom
panels) estimators. Results for sources at redshift $z_s=1$ and $z_s$=2
are shown in the left and in the right panels, respectively. Different
line styles refer to three filter sizes. For the OPT, these are $1'$,
$2'$ and $4'$. They correspond to $2.75'$, $5.5'$, and $11'$ for the APT
and to $5'$, $10'$ and $20'$ for the OAPT. Results for each redshift bin are
averaged between two planes.}
\label{fig:8}
\end{figure}

Figure~\ref{fig:8} shows the lowest mass detected in each
redshift bin. This is defined as the mean mass of the ten least massive
halos detected in this bin. Again, results are displayed for all
weak-lensing estimators, for different filter sizes and for two source
redshifts.

We note that the performance of the three filters is very similar for
sources at redshift $z_s=1$ (left panels). The OPT (bottom panels) is
only slightly more efficient in detecting low-mass halos than the APT
(top panels) and the OAPT (middle panels). The minimal mass detected
depends on the lens redshift. All filters allow the detection of
low-mass halos more efficiently if these are at redshifts between $0.2$
and $0.5$, i.e.~at intermediate distances between the observer and the
sources. This obviously reflects the dependence of the geometrical
lensing strength on the angular-diameter distances between the observer
and the lens, the lens and the sources, and the observer and the
sources. The lowest detected masses fall within $\sim
10^{13}h^{-1}M_\odot$ and $\sim 10^{14}h^{-1}M_\odot$ for the OPT
estimator.

For sources at higher redshift, the region of best filter performance
shifts to higher lens redshift, between $0.5$ and $0.8$. We note that
due to the increasing importance of lensing by large-scale structures,
the differences between the estimators are more significant. The OPT
estimator allows the detection of halos with masses as low as $\lesssim
10^{13}h^{-1}M_\odot$, almost independently of the filter size. Similar
masses are detected with the OAPT only for the smallest apertures. With
the APT and the OAPT, the results are indeed much more sensitive to the
filter size than with the OPT. Increasing the filter size pushes the
detectability limit to larger masses. Again, as discussed in
Sect.~\ref{sect:totdet}, this is due to the fact that the signal from
low-mass halos is smeared out by averaging over an increasing number of
galaxies entering the aperture. For example, the minimal mass detected
with the OAPT filter at $z\sim0.8$ changes by one order of magnitude by
varying the filter scale from $5'$ to $20'$.

\subsection{Completeness}

We now discuss the completeness of a synthetic halo catalogue selected
by weak lensing.

\begin{figure}[!ht]
  \includegraphics[height=0.49\hsize,angle=-90]{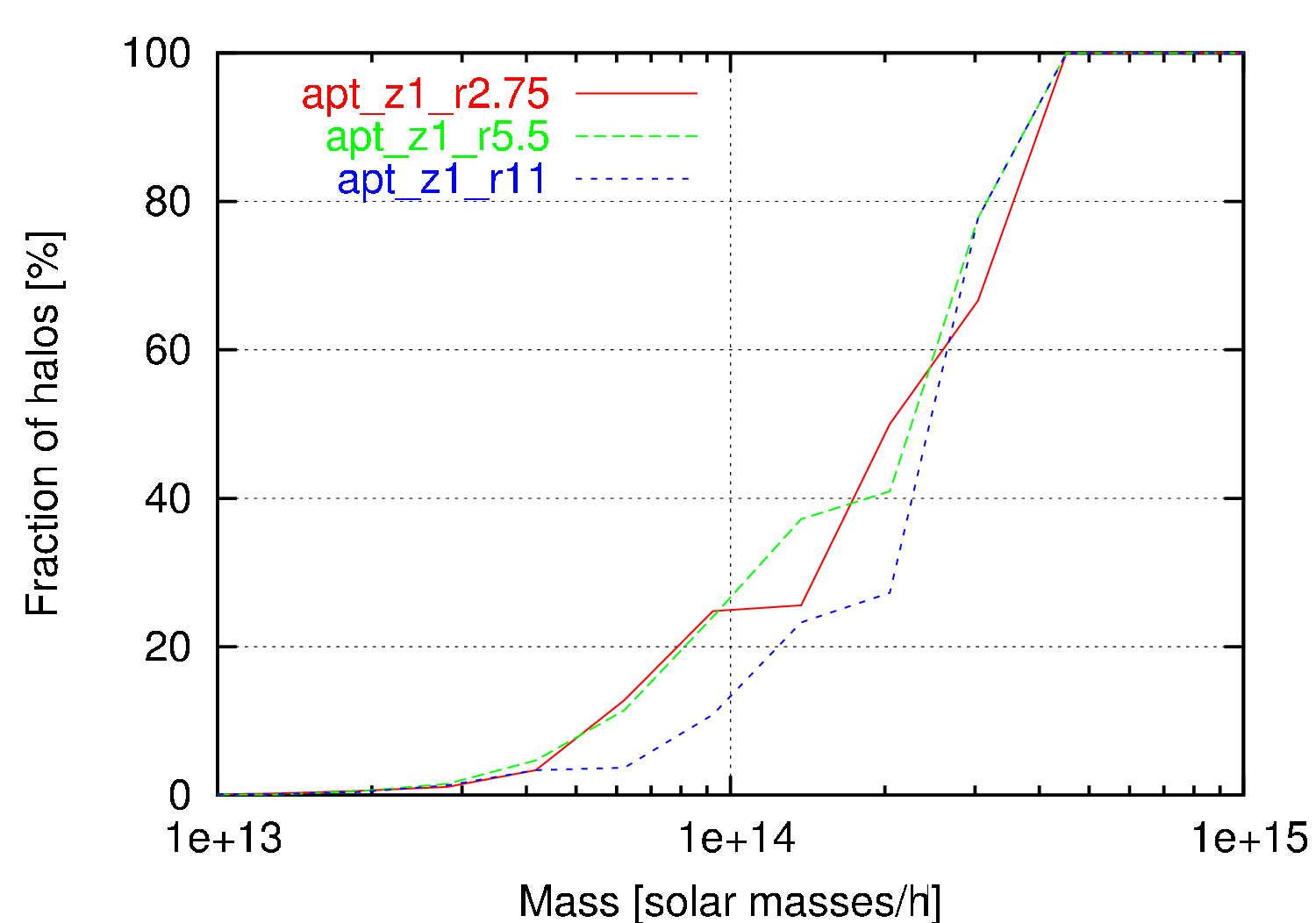}\hfill
  \includegraphics[height=0.49\hsize,angle=-90]{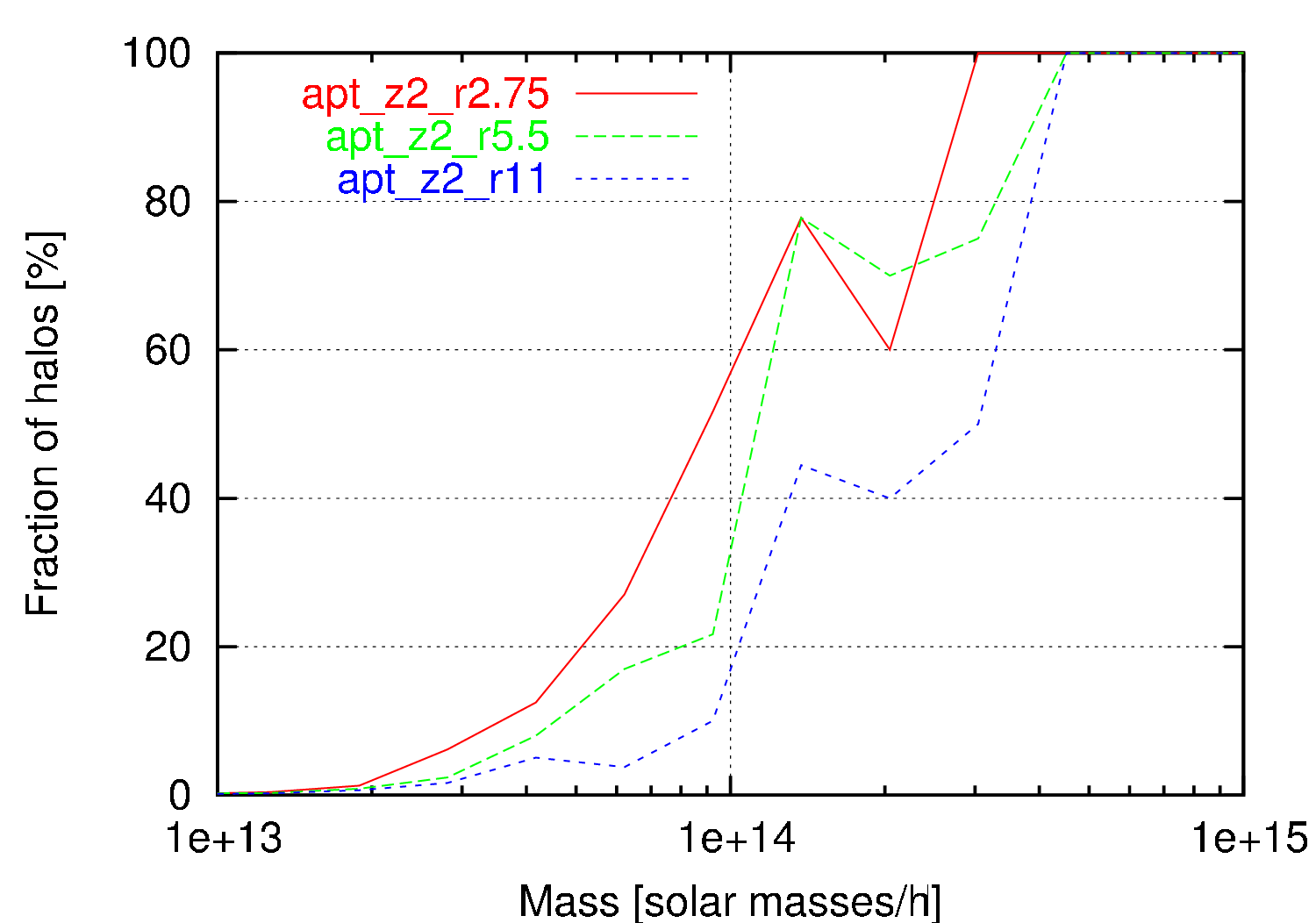}
  \includegraphics[height=0.49\hsize,angle=-90]{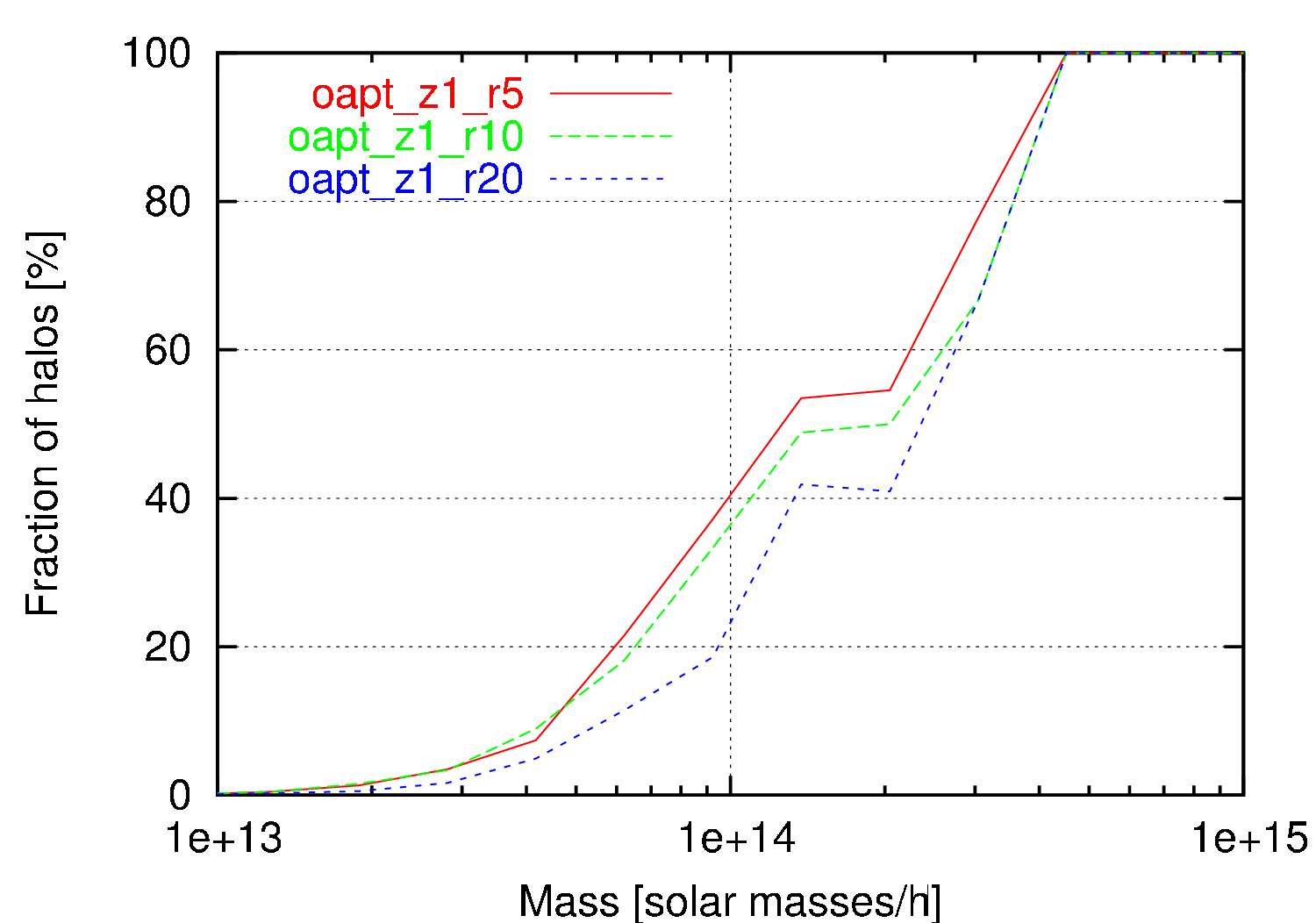}\hfill
  \includegraphics[height=0.49\hsize,angle=-90]{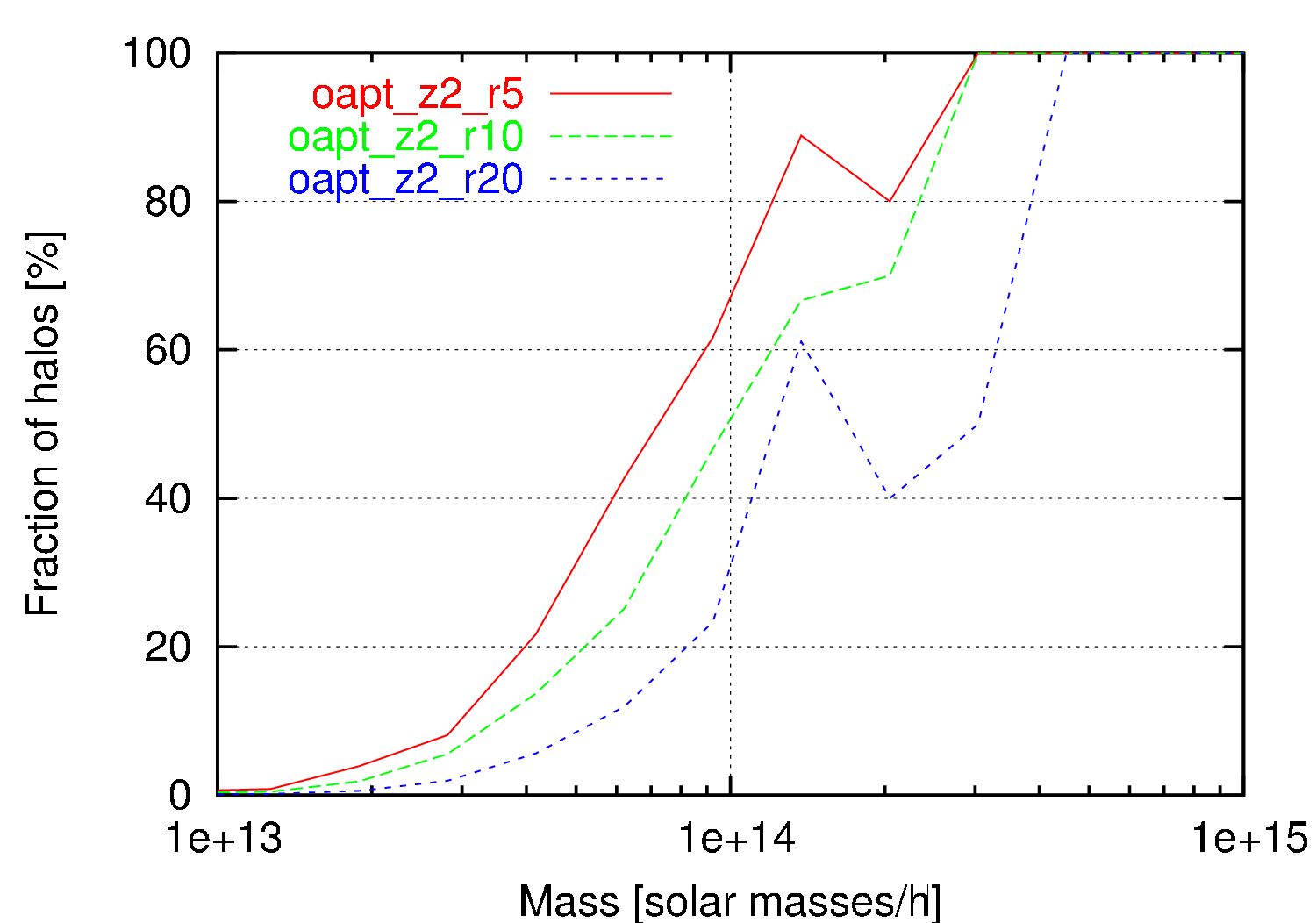}
  \includegraphics[height=0.49\hsize,angle=-90]{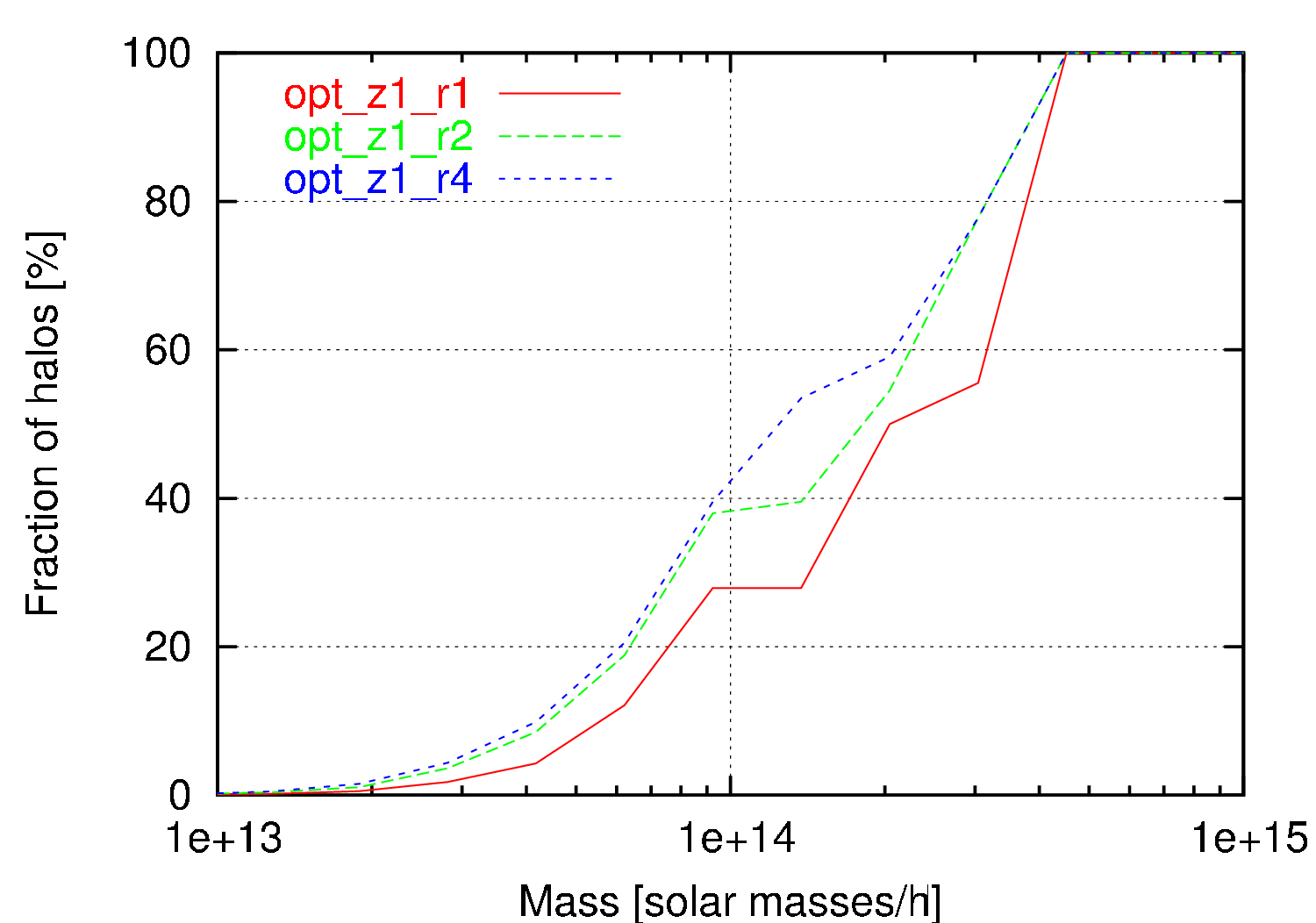}\hfill
  \includegraphics[height=0.49\hsize,angle=-90]{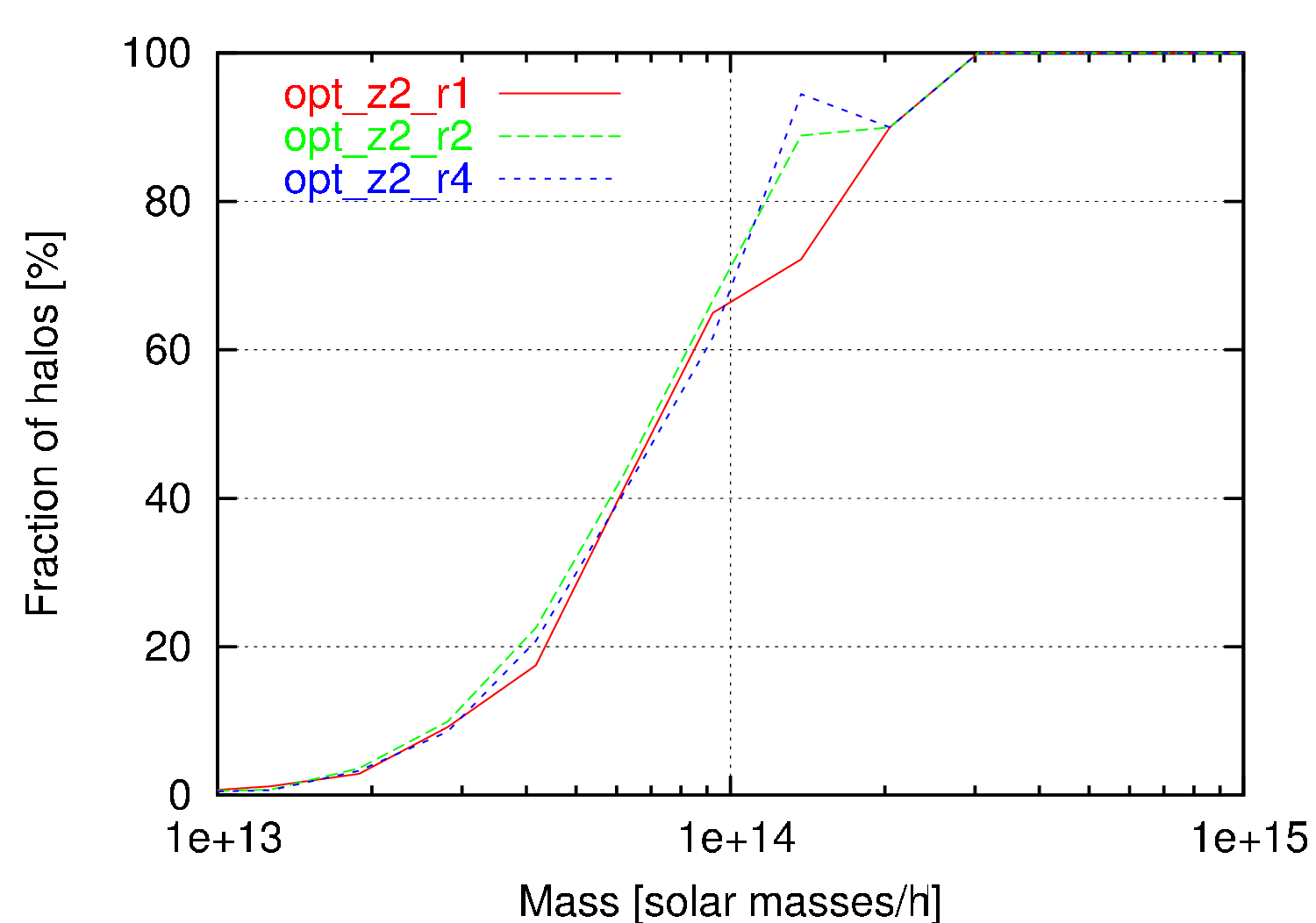}
\caption{Fraction of detections as a function of the halo mass. Each
plot contains results obtained with the three filter radii used in this
work. The panels on the left show curves for sources at $z_s=1$, the
panels on the right for sources at $z_s=2$. From top to bottom we have
the APT, the OAPT and the OPT.}
\label{fig:9}
\end{figure}

Figure~\ref{fig:9} shows the fraction of halos contained in the
light cone that are detected with different weak lensing estimators as a
function of their mass. Again, we find that the OPT filter yields the
most stable results with respect to changes in the filter size. This is
particularly evident for sources at redshift $z_s=2$ (right panels),
while the differences are smaller for $z_s=1$ (left panels). As discussed
earlier, the APT and the OAPT become less efficient in detecting
low-mass halos when the filter size is increased.

For the OPT estimator, the completeness reaches $100\%$ for masses
$M\gtrsim 3\times 10^{14}h^{-1}M_\odot$ and $M\gtrsim 2\times
10^{14}h^{-1}M_\odot$ for sources at redshift $z_s=1$ and $z_s=2$,
respectively. For lower masses, the completeness drops quickly, reaching
$\sim 50\%$ already at $M\sim 2\times 10^{14}h^{-1}M_\odot$ for
low-redshift sources, and at $M\sim 7\times 10^{13}h^{-1}M_\odot$ for
high-redshift sources. Similar results are obtained with the APT and the
OAPT only for small apertures.

\begin{figure}[!ht]
  \includegraphics[height=0.49\hsize,angle=-90]{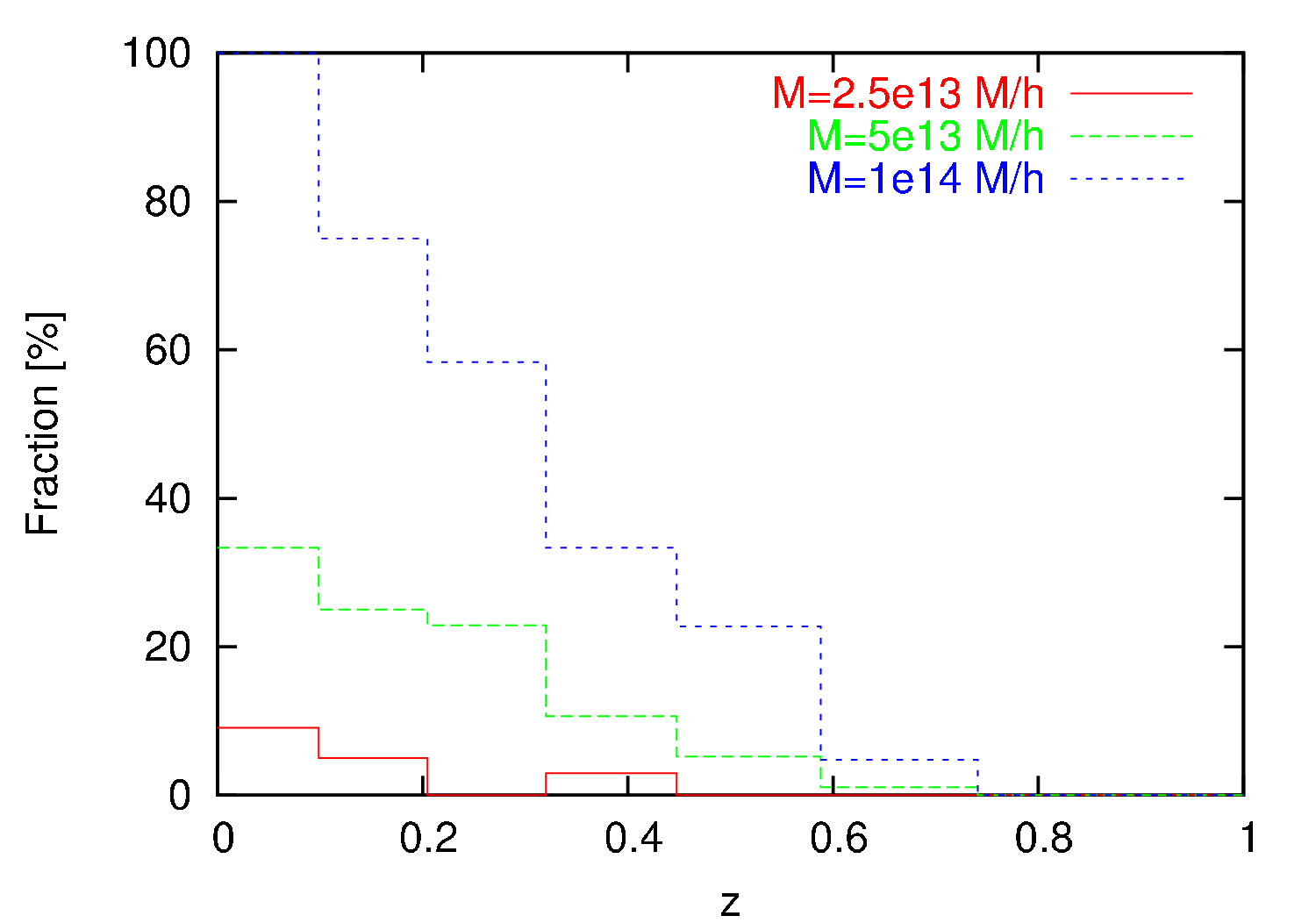}\hfill
  \includegraphics[height=0.49\hsize,angle=-90]{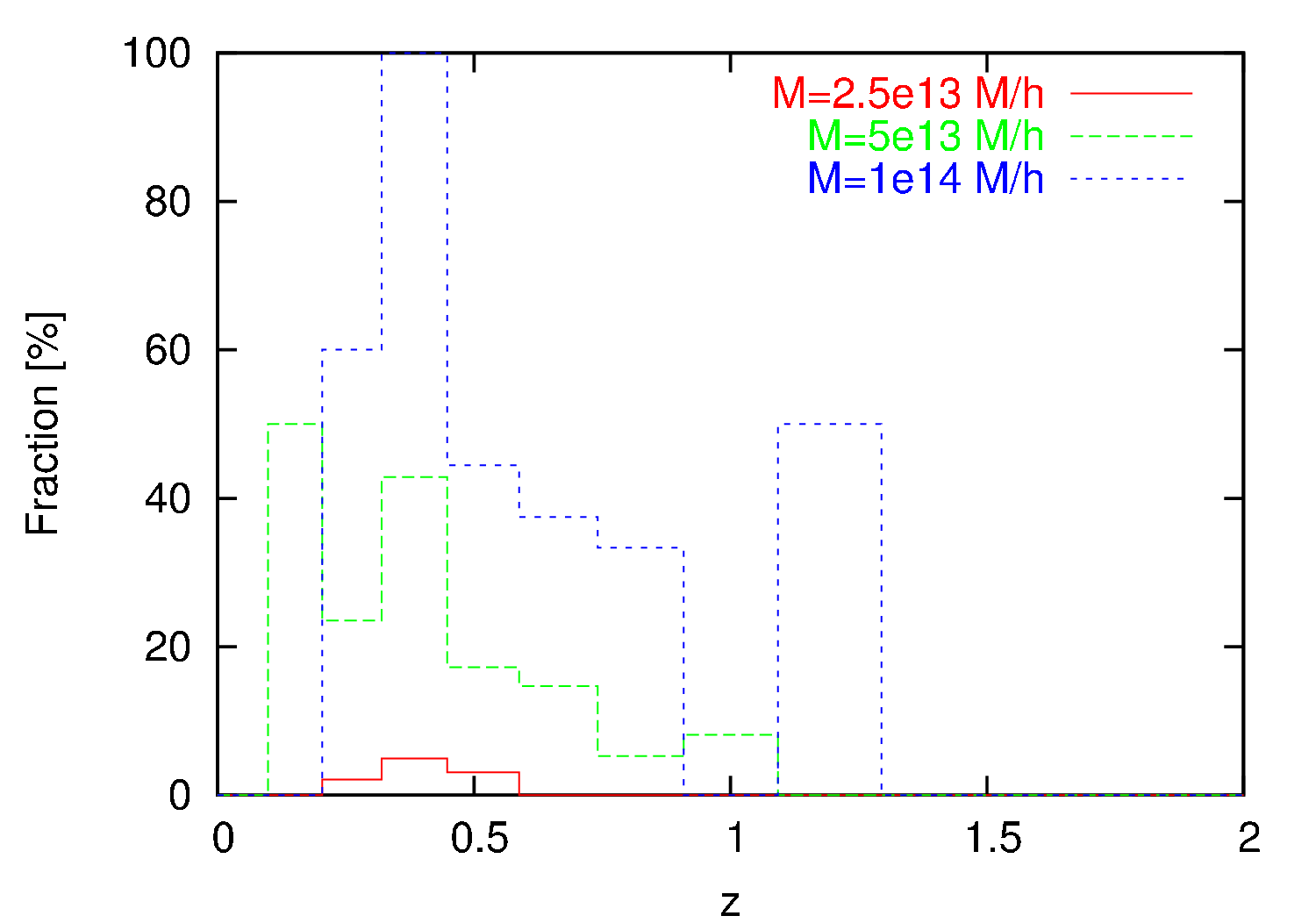}
  \includegraphics[height=0.49\hsize,angle=-90]{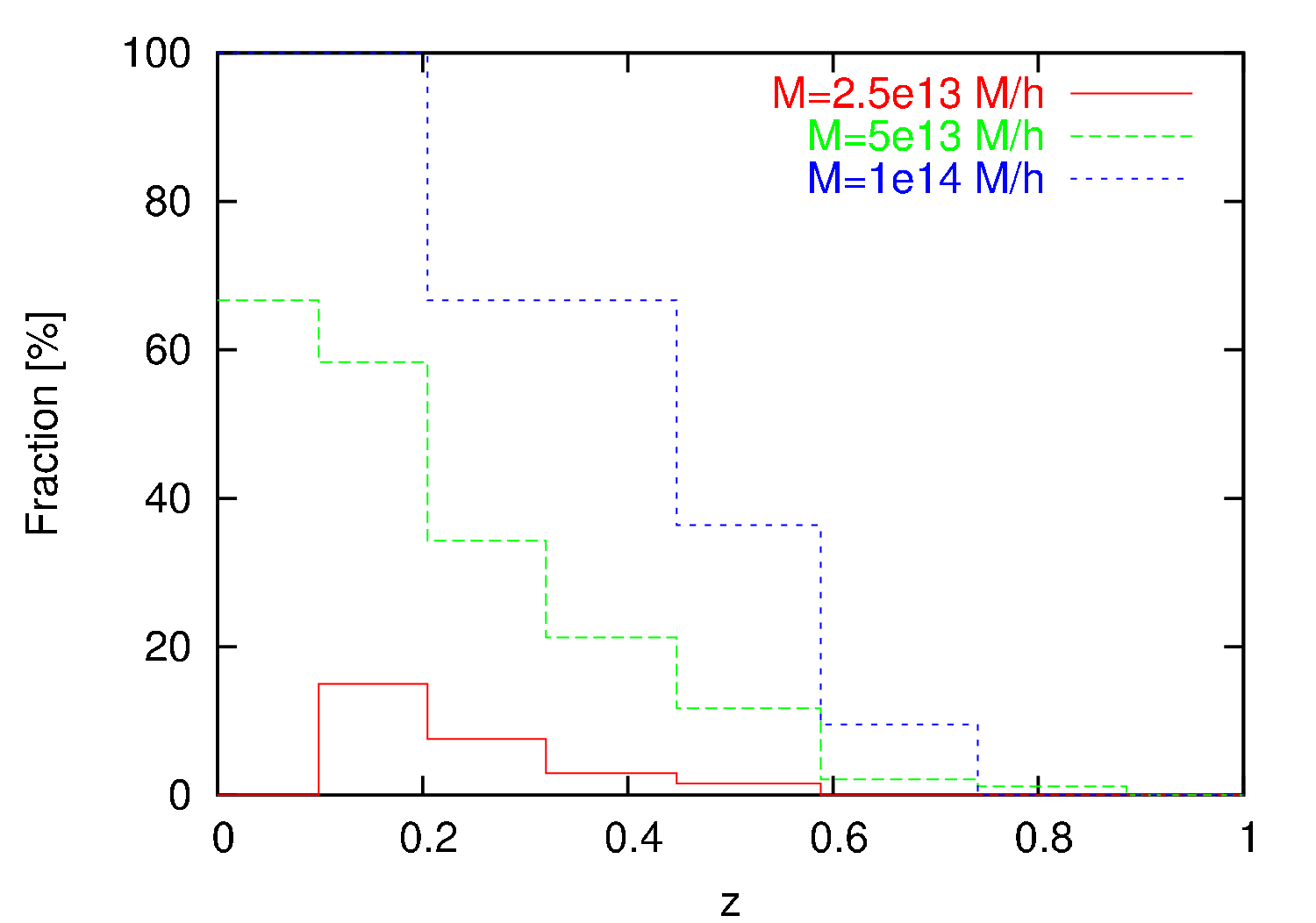}\hfill
  \includegraphics[height=0.49\hsize,angle=-90]{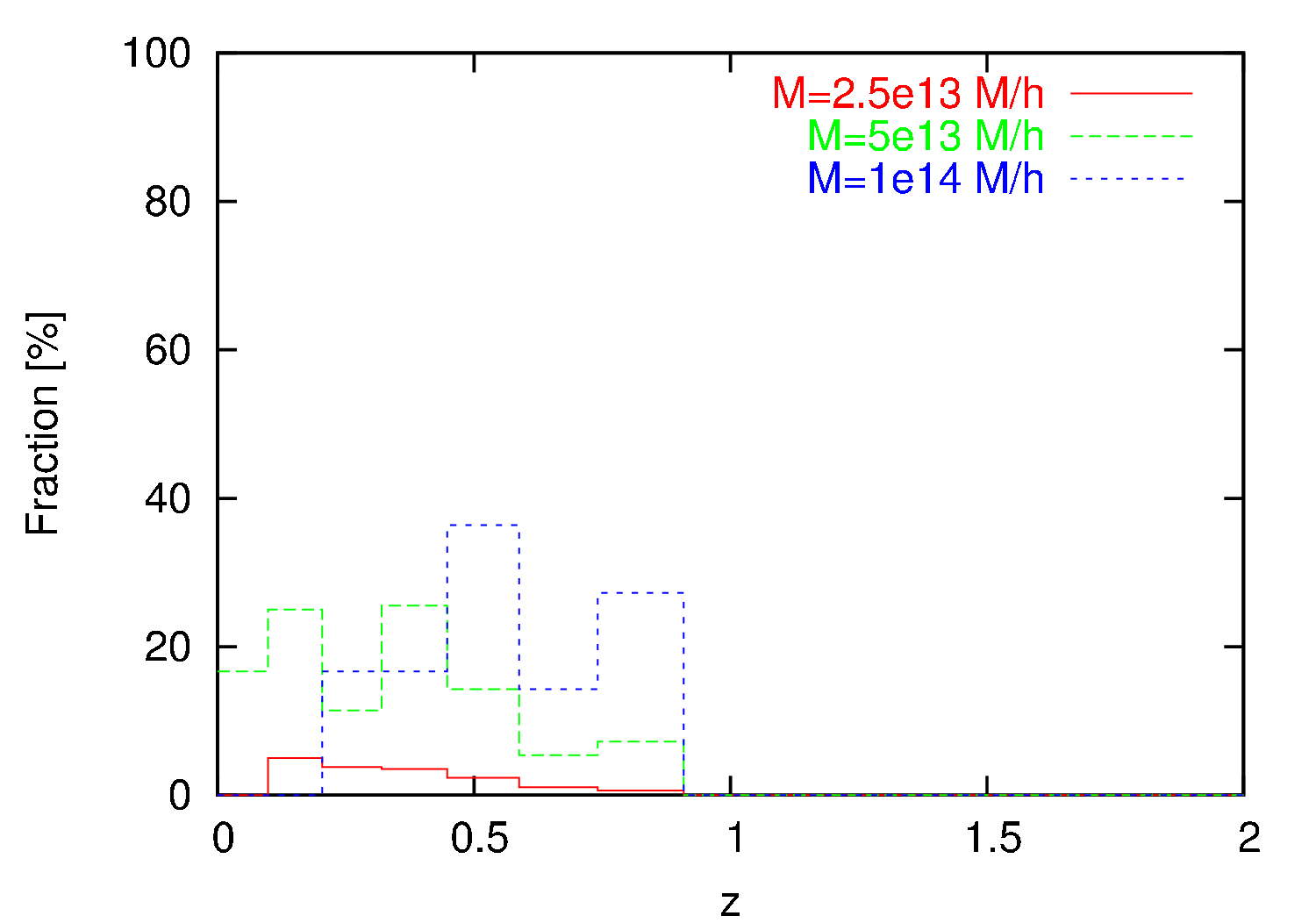}
  \includegraphics[height=0.49\hsize,angle=-90]{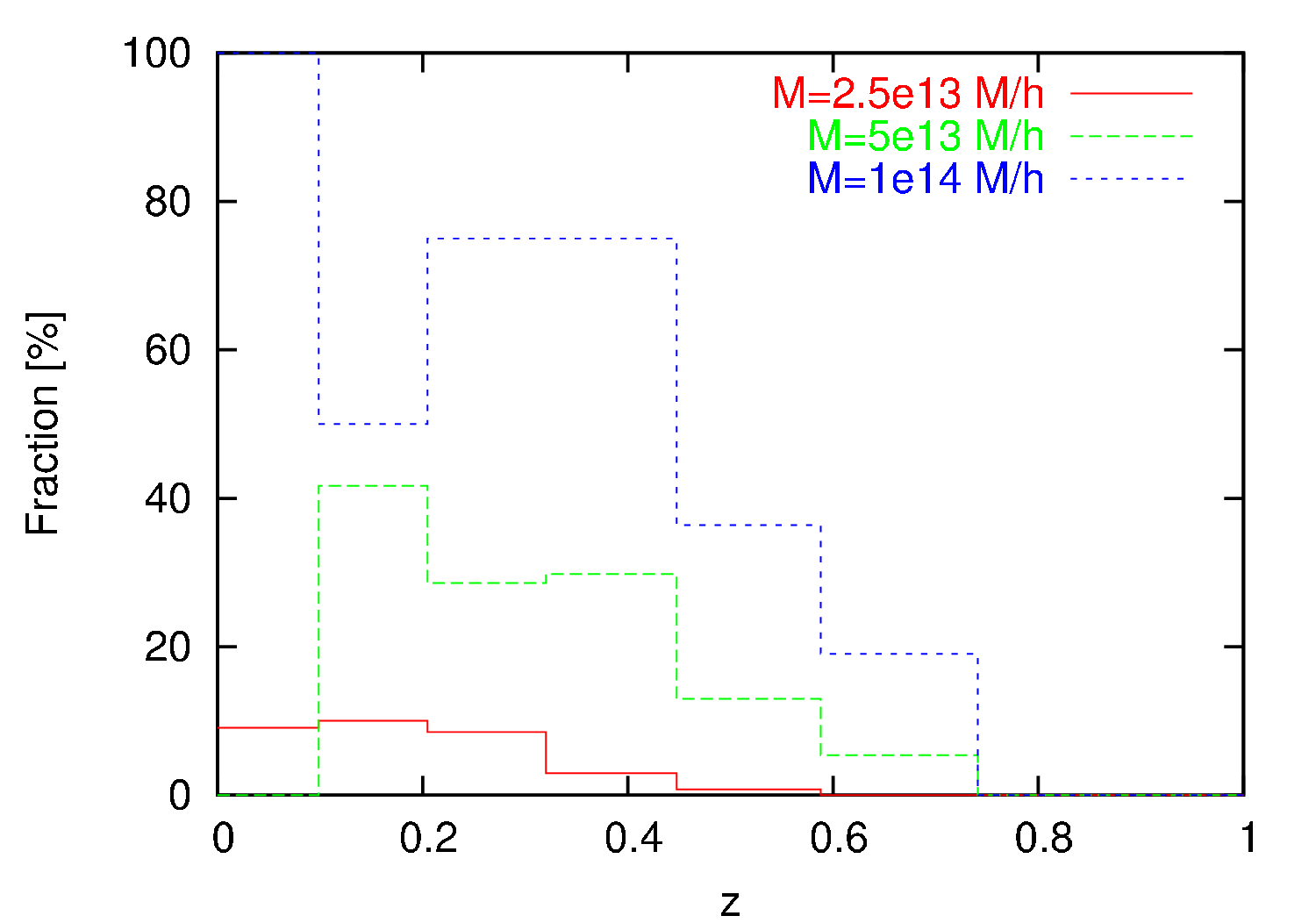}\hfill
  \includegraphics[height=0.49\hsize,angle=-90]{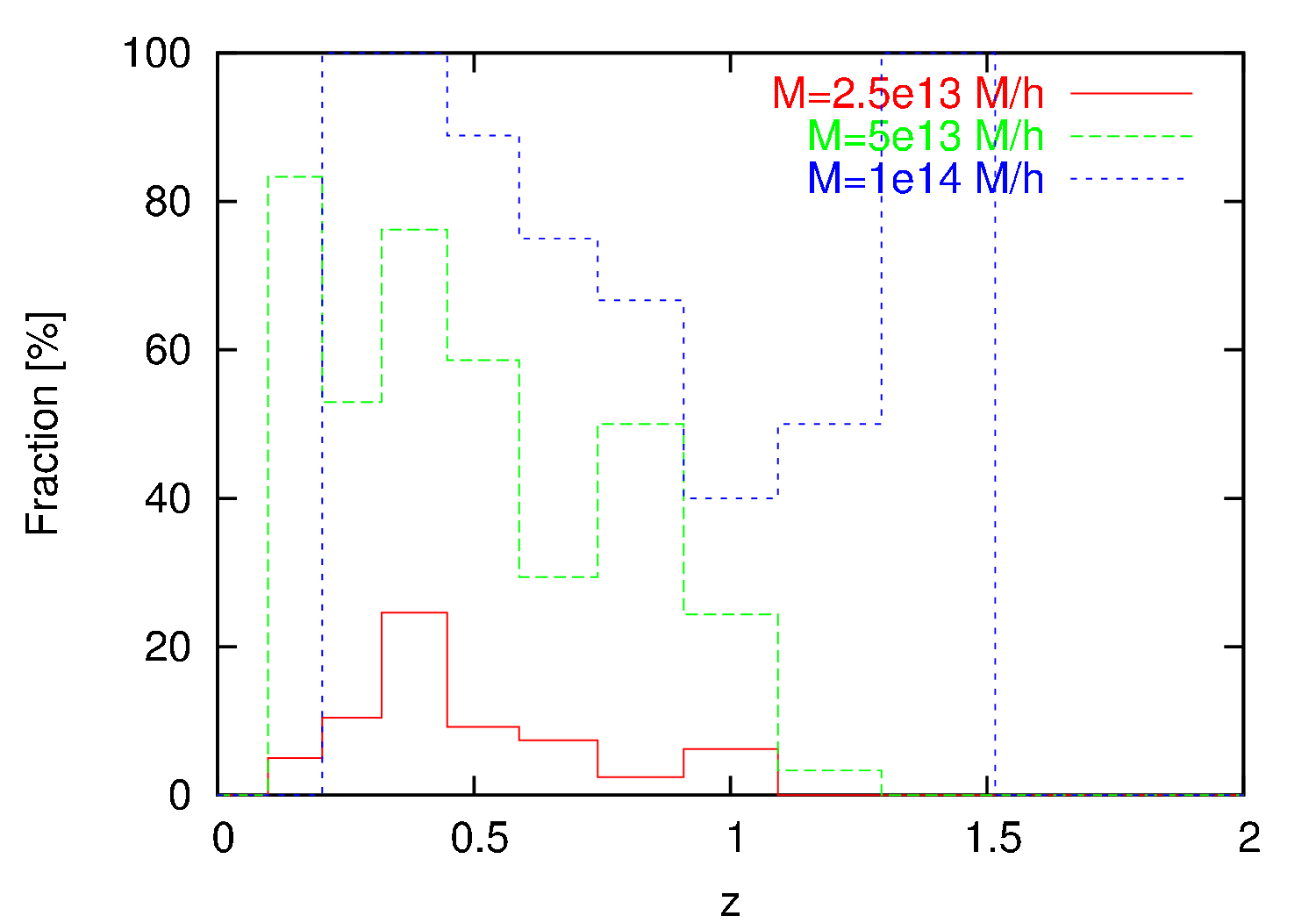}
\caption{Fraction of halo detections with the APT, OAPT and OPT (from
top to bottom) as a function of the halo redshift for three particular
masses. The red line corresponds to a mass of $M=2.5\times 10^{13}
M_{\odot}/h$, the green line to $M=5\times 10^{13} M_{\odot}/h$ and the
blue line to $M=10^{14} M_{\odot}/h$. The panels on the left show the results
for sources at $z_s=1$ and those on the right for sources at $z_s=2$. From top
to bottom, the filter radii are $r=5.5'$ (for APT), $r=10'$ (for OAPT) and
$r=2'$ (for OPT). Results for each redshift bin are averaged between two
planes.}
\label{fig:12}
\end{figure}

Figure \ref{fig:9} gives a global view of the halos detected, regardless
of the their redshift. In Fig.~\ref{fig:12}, we selected three mass bins
($M=2.5\times 10^{13} M_{\odot}/h$, $M=5\times 10^{13} M_{\odot}/h$,
$M=10^{14} M_{\odot}/h$) and determined the fraction of halos detected
as a function of the redshift.

To reduce the noise, we binned together two lens planes as in
Fig.~\ref{fig:8}. Yet, the results are still noisy, there is much
variation for all the filters when the filter radius is changed, and the
performance of the filters is quite similar in this respect. We see from
the figure that the detected halos are preferentially located at low and
moderate redshifts, due, as already said, to the geometry of the lensing
strength.

\subsection{Comparison with the peak statistics}

The peak statistic counts peaks in convergence maps, e.g. obtained with
the Kaiser-Squires inversion \citep[see][]{KA93.1,KA95.1}, usually smoothed
with a Gaussian kernel. Even though they used a different set of numerical
simulations, we can safely compare our results with the peak-statistic analysis
by \cite{HA04.1}, whose Gaussian kernel has a FWHM of 1~arcmin.

Fixing a detection threshold of $S/N> 4$ ($5$), \cite{HA04.1} found
$N\approx 6$ $(2.5)$ detections per square degree, $60\%$ $(76\%)$ of
which correspond to real haloes with masses larger than
$10^{13}\,h^{-1}M_\odot$. In our simulations, with the same S/N
threshold and the optimal filter by \cite{MA05.1}, we found $N\approx
10$ ($7$), with an efficiency in detecting real haloes of $85\%$ ($95\%$).
For halos with masses $M>2\times 10^{14}\,h^{-1} M_\odot$ ($M\approx
1\times 10^{14}\,h^{-1} M_\odot$), the \cite{HA04.1} sample is
complete at the $70\%$ ($50\%$) level, which is virtually identical to
the completeness of $70\%$ ($50-60\%$) achieved with the optimal filter.

\section{Comparison with observations}

The results outlined above show interesting differences between the
performances of the filter functions. The discrepancies are
particularly significant for high-redshift sources, indicating that
the noise due to the LSS should become important only for deep
observations. We can now attempt a quick comparison of our simulations
with the observational results existing in the literature. In
particular, we focus here on the searches for dark matter
concentrations in the GaBoDS survey \citep{SC03.3,MA06.1}.

To this goal, we perform a new set of ray-tracing simulations, where a
realistic redshift distribution of the sources is assumed. In
particular, we draw the sources from the probability distribution
function
\begin{equation}
P(z)=N\exp[-(z/z_0)^{\beta}] \;,
\end{equation} 
where $N$ is chosen such that 
\begin{equation}
\int_0^{\infty} P(z)\d z = 1 \;.
\end{equation}
We adapt $P(z)$ to the redshift distribution of the sources in the
GaBoDS survey by setting $z_0=0.4$ and $\beta=1.5$ \citep{SC03.3}. In
order to mimic the number density of galaxies in the GaBoDS
observations, we assume $n_g=10\,\mathrm{arcmin}^{-1}$.

\begin{figure}[!ht]
  \includegraphics[height=0.49\hsize,angle=-90]{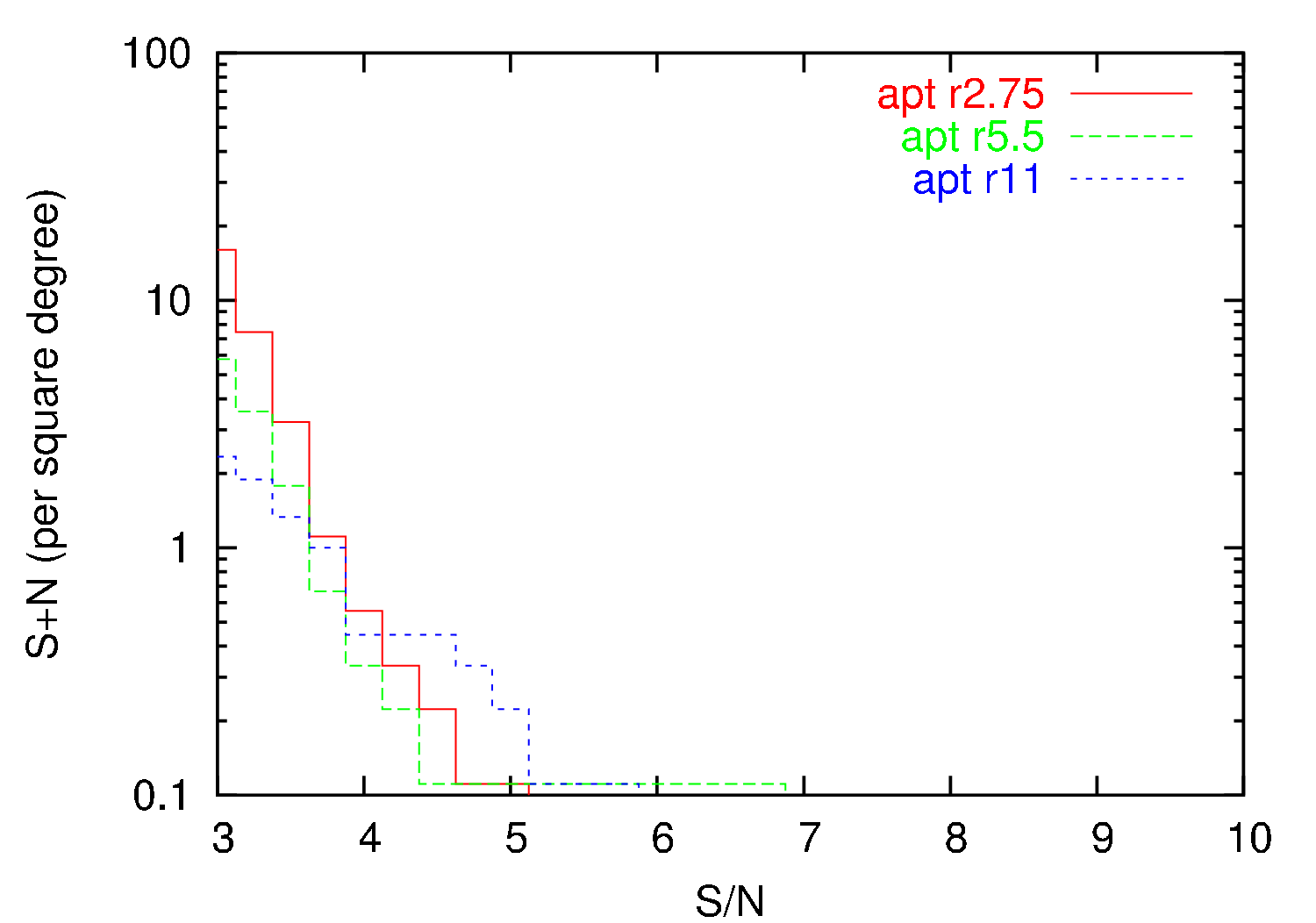}\hfill
  \includegraphics[height=0.49\hsize,angle=-90]{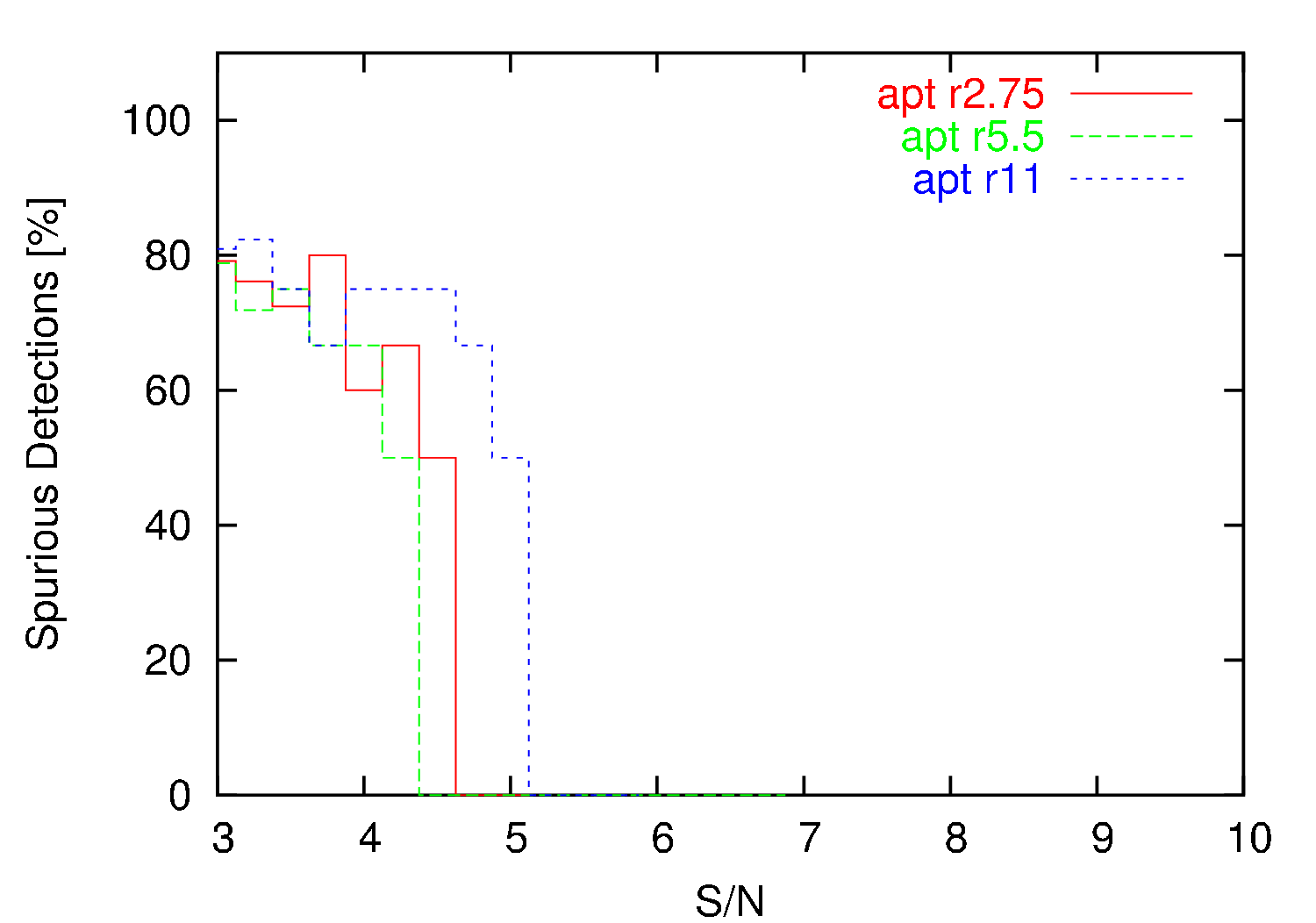}
  \includegraphics[height=0.49\hsize,angle=-90]{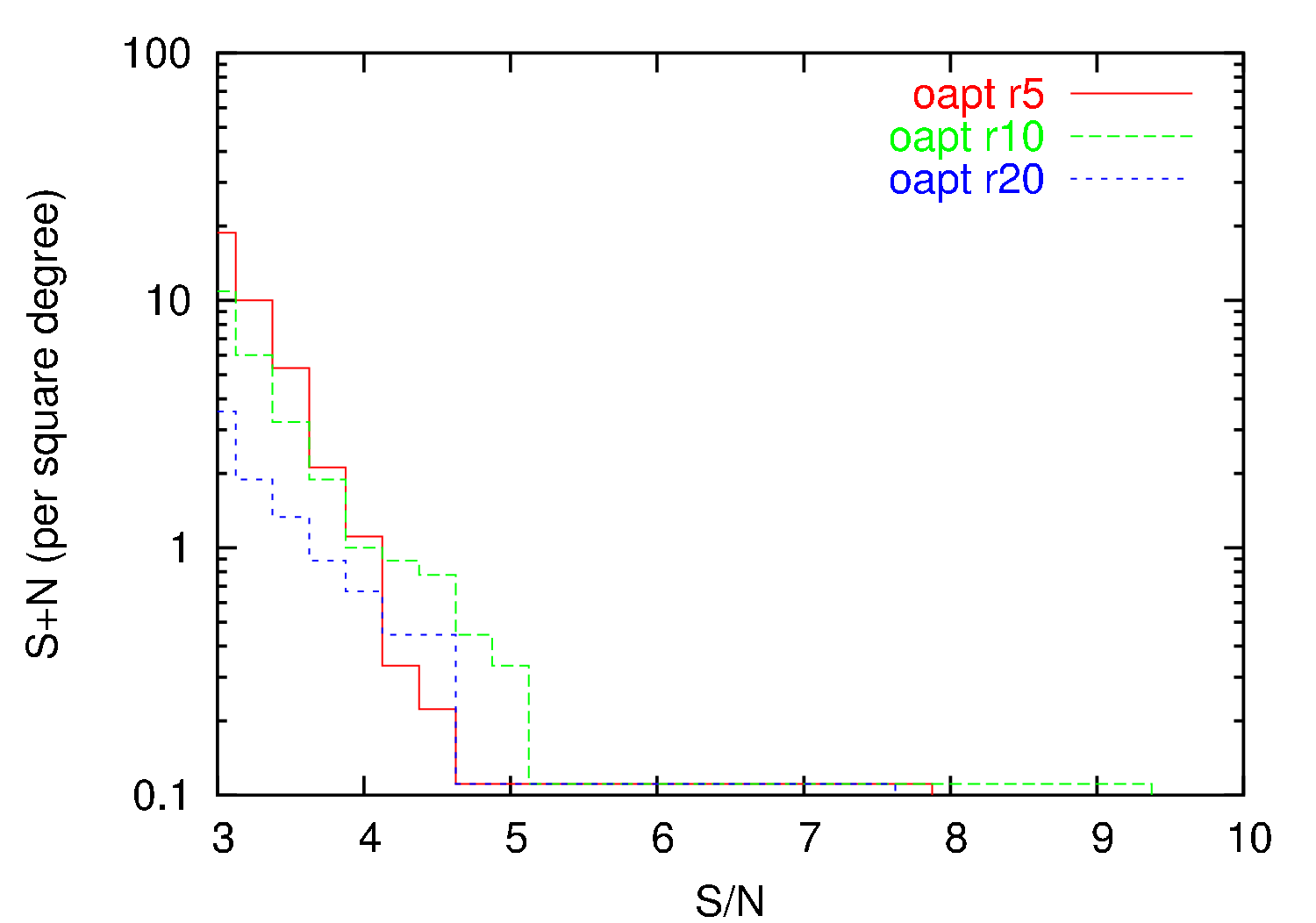}\hfill
  \includegraphics[height=0.49\hsize,angle=-90]{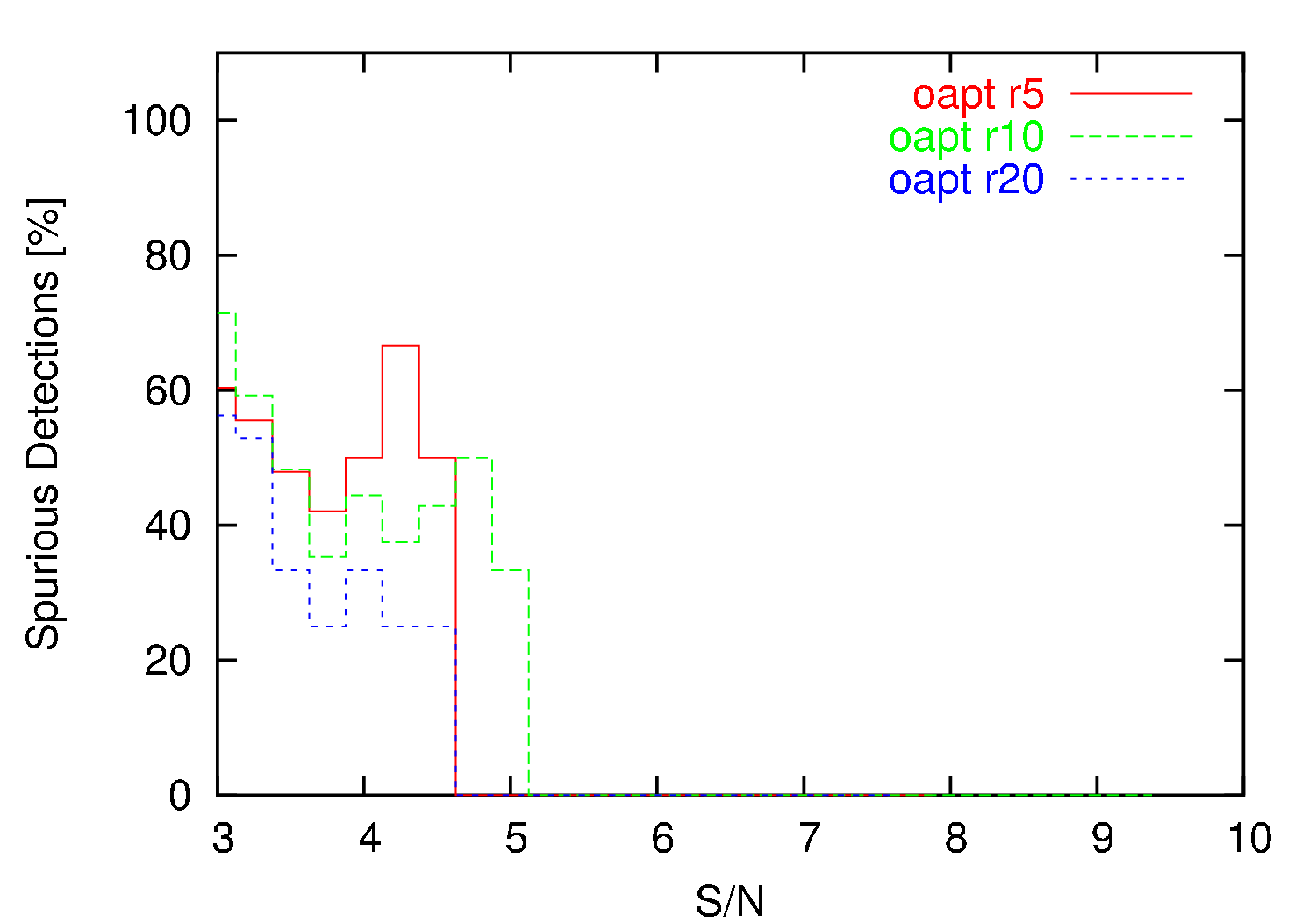}
  \includegraphics[height=0.49\hsize,angle=-90]{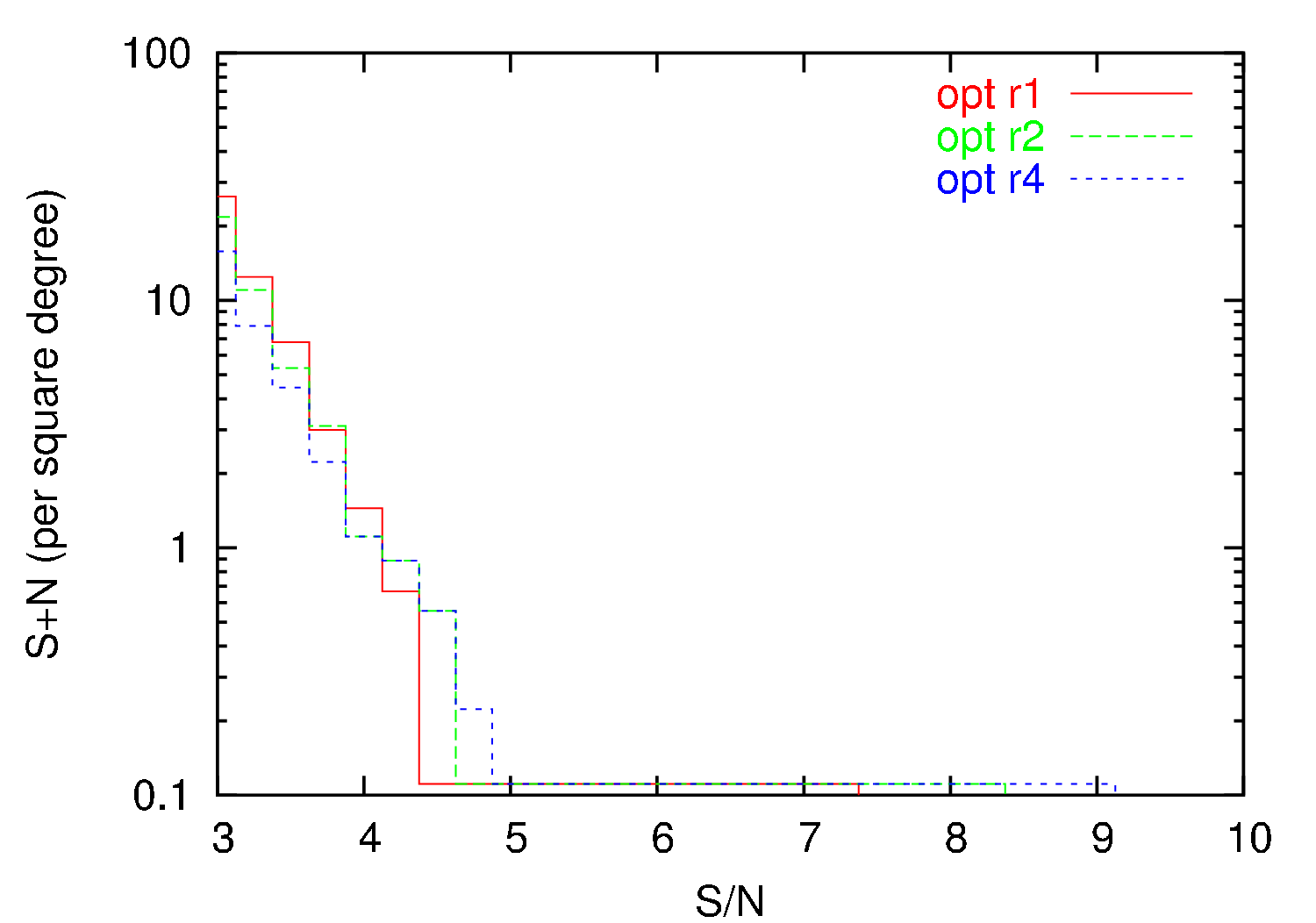}\hfill
  \includegraphics[height=0.49\hsize,angle=-90]{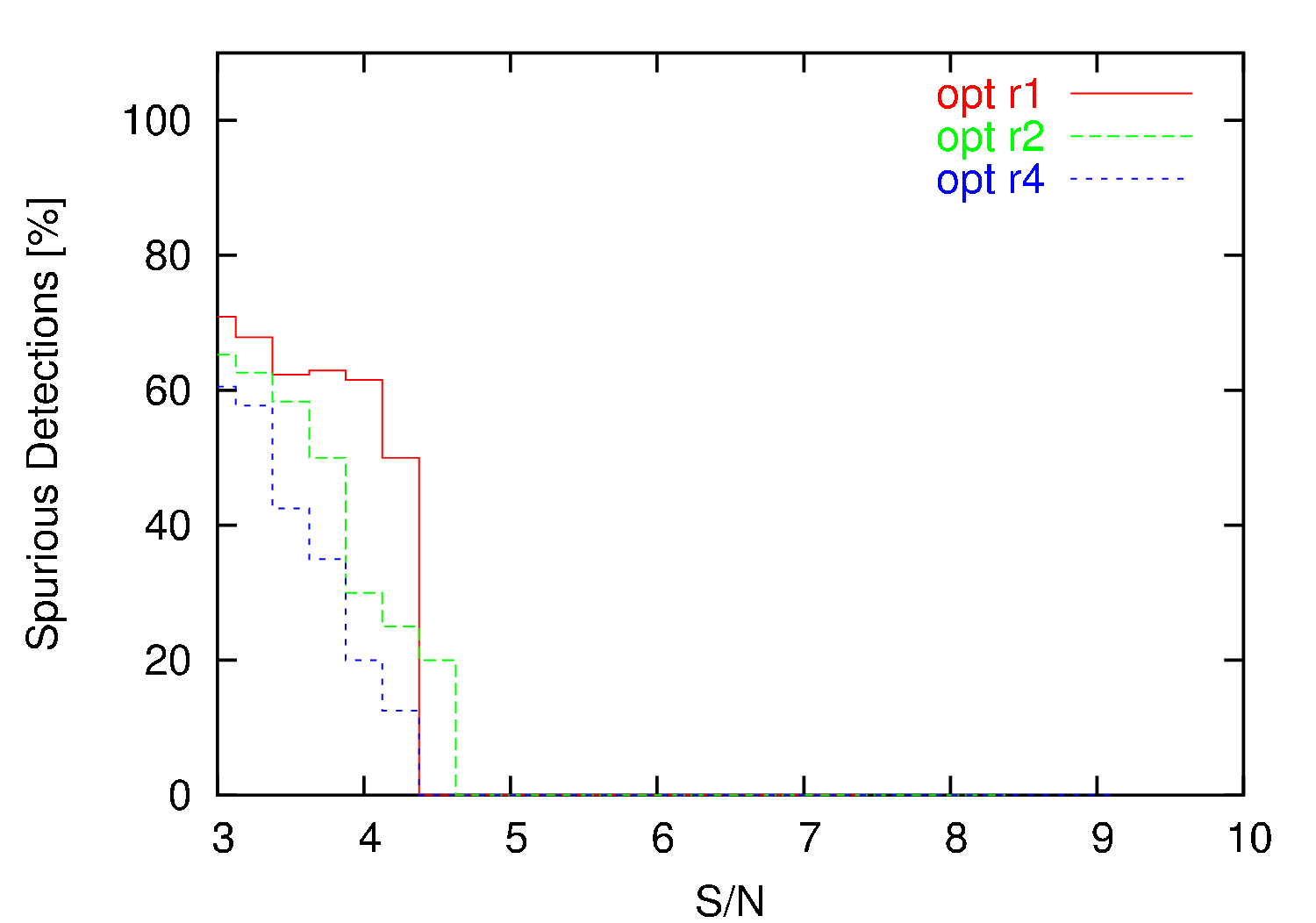}
\caption{Total number of detections per square degree (left panels)
and fraction of spurious detections (right panels) for sources
distributed in redshift as in the GaBoDS survey \citep{SC03.3}. From
top to bottom, we show the APT (for $r=2.75'$, $r=5.5'$ and $r=11'$),
the OAPT ($r=5'$, $r=10'$ and $r=20'$) and the OPT ($r=1'$, $r=2'$ and
$r=4'$).}
\label{fig:10}
\end{figure}

By repeating the same analysis outlined above, we find results that
are compatible with the results of \cite{MA06.1}. In particular,
the number of detections with $S/N=3.5$ per square degree in our
GaBoDS simulations (in GaBoDS data) are $\simeq 5$ ($\simeq 4$) for
the OPT with $r=2'$, $\simeq 3$ ($\simeq 3$) for the OAPT with $r=10'$
and $\simeq 1.5$ ($\simeq 2$) for the APT with $r=5.5'$ ($r=4'$)
respectively. A comparison between the detections with different weak
lensing estimators is shown in Fig.~\ref{fig:10}.

The fraction of spurious detections is large for all filters, but it
is generally smaller for the OAPT and the OPT. As expected, the OAPT
and the OPT estimators have similar performances, because of the small
density of background galaxies. Indeed, the noise due to the intrinsic
shape of the sources is dominant with respect to that due to the LSS
and thus, according to Equation~(\ref{eq:opt_filter}), the two filter
functions have a very similar shape.

A more detailed statistical analysis, including simulations
representative of different current surveys (both ground- and
space-based), will be presented in a future paper.

\section{Conclusions}

We studied the performance of dark-matter halo detection with three
different linear filters for their weak-lensing signal, the aperture
mass (APT), the optimised aperture mass (OAPT), and a filter optimised
for distinguishing halo signals from spurious signals caused by the
large-scale structure (OPT). In particular, we addressed the questions
how the halo selection function depends on mass and redshift, how the
number of detected halos and of spurious detections depends on
parameters of the observation, and how the filters compare.

To this end, we used a large $N$-body simulation, identified the halos
in it and used multiple lens-plane theory to determine the lensing
properties along a fine grid of light rays traced within a cone from the
observer to the source redshift. Halos were then detected as peaks in
the filtered cosmic-shear maps. By comparison with the known halo
catalog, spurious peaks could be distinguished from those caused by real
halos.

Our main results are as follows:

\begin{itemize}

\item We confirm that among those tested, the optimised filter (OPT)
proposed by \cite{MA05.1} performs best in the sense that its results
are least sensitive to changes in the angular filter scale, it produces
the least number of spurious detections, and it has the lowest mass
limit for halo detection (cf.\ Figs.~\ref{fig:7}, \ref{fig:8} and
\ref{fig:9}).

\item With the OPT filter, the fraction of spurious detections is
typically $\lesssim10\%$ for a signal-to-noise threshold of
$S/N\approx5$. It increases with source redshift due to the larger
contamination by large-scale structure lensing (cf.\ Fig.~\ref{fig:7}).

\item The number of halos detected per square degree by the OPT filter
with $S/N\gtrsim5$ is a few if the sources are at redshift $z_s=1$, and
$\sim20$ for $z_s=2$ (cf.\ Fig.~\ref{fig:6}).

\item The minimum detectable halo mass starts at a few times
$10^{13}\,h^{-1}\,M_\odot$ at redshifts $\sim0.1$, drops to
$\sim10^{13}h^{-1}\,M_\odot$ near the optimal lensing redshift and
increases towards $\sim10^{14}h^{-1}\,M_\odot$ approaching the source
redshift (cf.\ Fig.~\ref{fig:8}).

\item The fraction of halos detected reaches $\sim50\%$ at
$\sim2\times10^{14}h^{-1}\,M_\odot$ and $100\%$ at
$\sim4.5\times10^{14}h^{-1}\,M_\odot$ with sources at $z_s=1$. With
more distant sources at $z_s=2$, half of the halos with
$\sim7\times10^{13}h^{-1}\,M_\odot$ are found, and all halos above
$\sim3\times10^{14}h^{-1}\,M_\odot$ (cf.\ Fig.~\ref{fig:9}).

\item Adapting parameters to the GaBoDS survey \citep{SC03.3}, and
distributing sources in redshift, our simulation yields a number of
significant detections per square degree which is in good agreement
with what was found applying the OPT filter to the real GaBoDS data
\citep{MA06.1}.

\end{itemize}

Thus, the OPT filter, optimised for suppressing contaminations by
large-scale structures, allows the reliable detection of dark-matter
halos with masses exceeding a few times $10^{13}\,h^{-1}\,M_\odot$ with
a low contamination by spurious detections.

\acknowledgements{We are grateful to Stefano Borgani and to Giuseppe Murante
  for helpful discussions, and to the anonymous referee whose
  comments helped to improve the paper. Computations have been
  performed using the IBM-SP4/5 at Cineca (Consorzio Interuniversitario del 
  Nord-Est per il Calcolo
  Automatico), Bologna, with CPU time assigned under an INAF-CINECA grant. We
  acknowledge financial contribution from contract ASI-INAF I/023/05/0 and
  INFN PD51. This work was supported by the Deutsche Forschungsgemeinschaft
  (DFG) under the grants BA 1369/5-1 and 1369/5-2 and by DAAD and Vigoni
  programme.}

\end{document}